\documentclass[11pt,tightenlines,eqsecnum,floats,aps,amssymb,nofootinbib,prd,shownopacs,floatfix]{revtex4-2}

\usepackage{graphicx}
\usepackage{epstopdf}
\usepackage{latexsym}
\usepackage{amssymb}
\usepackage{amsmath}
\usepackage{color}
\usepackage{mathrsfs}
\usepackage{xparse}
\usepackage{float}
\usepackage[colorinlistoftodos,prependcaption,textsize=tiny]{todonotes}
\usepackage{dsfont}
\usepackage{mathtools}
\usepackage[toc,title,page]{appendix}
\usepackage[colorlinks=false, pdfborder={0 0 0}]{hyperref}

\usepackage{tensor}
\usepackage{braket}
\renewcommand{\t}{\tensor} 
\newcommand{\ft}[1]{\underset{\smile}{#1}} 

\renewcommand{\thesection}{\Roman{section}}
\renewcommand{\thesubsection}{\thesection.\Alph{subsection}}
\renewcommand{\thesubsubsection}{\thesubsection.\arabic{subsubsection}}

\makeatletter
\renewcommand{\p@subsection}{}
\renewcommand{\p@subsubsection}{}
\makeatother
\begin{document}
\begin{abstract}
We consider the coupling of a scalar field to linearised gravity and derive a relativistic gravitationally induced decoherence model using Ashtekar variables. The model is formulated at the gauge invariant level using suitable geometrical clocks in the relational formalism, broadening existing gauge invariant formulations of decoherence models. For the construction of the Dirac observables we extend the known observable map by a kind of dual map where the role of clocks and constraints is interchanged. We also discuss a second choice of geometrical clocks existing in the ADM literature. Then we apply a reduced phase space quantisation on Fock space and derive the final master equation choosing a Gibbs state for the gravitational environment and using the projection operator technique. The resulting master equation is not automatically of Lindblad type, a starting point sometimes assumed for phenomenological models, but still involves a residual time dependence at the level of the effective operators in the master equation due to the form of the correlation functions that we express in terms of thermal Wightman functions. Furthermore, we discuss why in the model analysed here the application of a second Markov approximation in order to obtain a set of time independent effective system operators is less straightforward than in some of the quantum mechanical models.
\end{abstract}

\title{A gravitationally induced decoherence model using Ashtekar variables}
\author{Max Joseph Fahn}
\email{max.j.fahn@gravity.fau.de}
\author{Kristina Giesel}
\email{kristina.giesel@gravity.fau.de}
\author{Michael Kobler}
\email{michael.kobler@gravity.fau.de}
\affiliation{Institute for Quantum Gravity, Theoretical Physics III, Department of Physics,  Friedrich-Alexander-Universit\"at Erlangen-N\"urnberg, Staudtstr. 7, 91058 Erlangen, Germany.}

\maketitle
\newpage
\tableofcontents
\newpage
\section{Introduction}
\label{sec:Intro}

Decoherence models play an important role in the semiclassical sector of quantum models in various areas of physics from condensed matter, quantum optics and quantum information \cite{tannoudji1992atom,carmichael1999statistical,dattagupta2004dissipative,heiss2008fundamentals,orszag2016quantum} to cosmology \cite{Halliwell:1989,Polarski:1995jg,LaFlamme:1990kd,Boyanovsky:2015tba, Hollowood:2017bil}, for a mathematical perspective see \cite{Bahns:2019}. These models are formulated in the framework of open quantum systems \cite{breuer2002theory,breuer2003concepts,Rivas:2012} and allow for instance to address the question of the quantum-to-classical transition \cite{Zurek:2003zz,Giulini:1996nw,Schlosshauer:2014pgr}. Because such models involve an interaction between the quantum system and its environment the choice of the latter is a crucial ingredient for any decoherence model. A very common choice is to couple the system to a bath of harmonic oscillators \cite{Caldeira+Leggett:1981,breuer2002theory,Xu:2020lhc}. In this work we want to analyse gravitationally induced decoherence for which the environment is taken to be gravity. This provides a platform to formulate and analyse decoherence models based on existing models for quantum gravity. Compared to the situation in closed quantum models where the semiclassical sector is usually described by means of semiclassical states, see for instance \cite{Klauder+Skagerstam:1985,Perelomov:1986,Zhang+Feng+Gilmore:1990,Ali+Antoine+Gazeau:2014}, the semiclassical sector in the context of quantum gravity can be investigated from a different angle within decoherence models with a gravitational environment.

There already exist several models for gravitationally induced decoherence in the literature \cite{ellis1984search,ellis1996precision,Benatti:1999sa,Benatti:2000ph,lisi2000probing,Morgan:2004vv,guzzo2016quantum,coelho2017decoherence,carpio2018revisiting,gomes2019quantum, hellmann2022searching,anastopoulos2013master,Blencowe:2012mp,oniga2016quantum,lagouvardos2021gravitational}, see also the reviews in \cite{Bassi:2017szd,Anastopoulos:2021jdz} and the review \cite{Donadi:2022szl} for further models in which quantum matter is coupled to gravity. Some of them have for instance been used to study the influence of decoherence on standard model matter such as neutrinos and their oscillations, see for instance  \cite{lisi2000probing,Benatti:2000ph,lisi2000probing,Morgan:2004vv,guzzo2016quantum,coelho2017decoherence,carpio2018revisiting,gomes2019quantum, hellmann2022searching}. All models have in common that the system's dynamics is encoded in a so-called master equation that is an effective first order differential equation for the density matrix of the system, where this density matrix is obtained once the environmental degrees of freedom in the total dynamics have been traced out.  One can group the existing models in two classes, that we will denote as bottom-up \cite{ellis1996precision,Benatti:1999sa,lisi2000probing,guzzo2016quantum,gomes2019quantum} and top-down approaches \cite{anastopoulos2013master, Blencowe:2012mp, oniga2016quantum, lagouvardos2021gravitational, Kok:2003mc, Breuer:2008rh, Asprea:2021jag}, respectively.  In the first approach one takes as a starting point a given master equation, as for instance the well known Lindblad equation \cite{gorini1976completely,lindblad1976generators}, and then constrains existing free parameters phenomenologically for a given model under consideration. In the top-down approach one starts with a given classical model, performs a quantisation of the total system and afterwards derives the master equation within that quantum model. If one aims at formulating a decoherence model based on a specific quantum gravity approach then the top-down approach is advantageous, since an interesting question in this context is whether there exist characteristic features in the decoherence model that will distinguish one quantum gravity model from another. Furthermore, in the derivation of the Lindblad equation, generally several assumptions such as the first and second Markov approximation are applied and one needs to check carefully under which circumstances these can be applied in gravitationally induced decoherence models. Compared to decoherence models derived in a quantum mechanical context, choosing quantum gravity as the environment complicates the situation in two aspects. On the one hand we have to deal with a quantum field theory where also the coupling between the system and the environment is more complex and on the other hand we consider a gauge theory and both facts need to be addressed appropriately in these models.

In this article we are interested in deriving a gravitationally induced decoherence model inspired by loop quantum gravity, see \cite{rovelli_2004,thiemann_2007,rovelli_vidotto_2014} for textbooks. Most works on the semiclassical physics of loop quantum gravity rather focus on closed quantum systems with semiclassical states \cite{Thiemann:2002vj,Freidel:2010tt,Stottmeister:2015tua,Calcinari:2020bft} or coarse graining in spin foam models \cite{Dittrich:2014ala,Steinhaus:2020lgb}. There exist a few exceptions see for instance \cite{Feller:2016zuk} where a decoherence toy model for surface states in the context of quantum black holes is introduced or the work in \cite{Ansel:2021pdk} for a coupling of a spin foam model to an environment using the adiabatic elimination method \cite{Azouit:2016} to obtain an effective Lindblad equation.

As a first step in this direction we consider a top-down approach in which we couple a scalar field to linearised gravity as it has been done in \cite{anastopoulos2013master,Blencowe:2012mp,oniga2016quantum}.  In contrast to \cite{anastopoulos2013master,Blencowe:2012mp,oniga2016quantum} where ADM variables have been used we construct the model using Ashtekar variables \cite{ashtekar1986new,ashtekar1989new}. The latter allow in principle a quantisation using techniques from loop quantum gravity along the lines of \cite{ashtekar1991gravitons,Varadarajan:2002ht}. However, as a first step due to the complexity of the model and to be able to better compare our results to the existing literature, we apply a Fock quantisation to the linearised gravitational sector.

Already at the classical level the model considered in this work requires further generalisations to the already existing results. The models in \cite{anastopoulos2013master,Blencowe:2012mp,oniga2016quantum} perturb the gravitational degrees of freedom only and we consider the coupling of the scalar field in the framework of a Post-Minkowski approximation scheme that allows to consider a setup where only the gravitational degrees of freedom are perturbed and still obtain a consistent formulation of the involved constraints in perturbation theory. Because we work with Ashtekar variables the model involves an additional Gau\ss{} constraint that needs to be taken into account for the gauge invariance. The route we follow here to deal with the gauge freedom is that we consider a reduced phase space quantisation. For this purpose we first construct a canonical transformation that separates the gauge and physical degrees of freedom into two sets. The latter are obtained by constructing Dirac observables in the relational formalism applying techniques from \cite{Rovelli:1990ph,Rovelli:1990pi,Rovelli:2001bz,Dittrich:2004cb,Dittrich:2005kc,Vytheeswaran:1994np} where suitable reference fields need to be chosen. In our case, these will be so-called geometrical clocks that only depend on the gravitational degrees of freedom. These clocks differ from the ones chosen in \cite{Dittrich:2006ee}, where complex Ashtekar variables have been used with the corresponding gauge fixing, whereas the Dirac observables used here have a natural relation to the gauge-fixed quantities different than the one presented in \cite{ashtekar1991gravitons}, which considers the vacuum case only. Geometrical clocks have also been applied in different contexts, see for instance \cite{Dittrich:2006ee,Dittrich:2007jx,Giesel:2018opa,Frob:2021ore}. Working in the relational formalism (physical) time can be defined in a relational way and circumvents some of the issues related to fluctuating light cones mentioned in \cite{Anastopoulos:2007xv}. In order to construct the above mentioned canonical transformation we need to generalise the observable map and introduce a kind of dual observable map where the role of constraints and reference fields is interchanged. When applied to the constraints themselves it reproduces the results for the perturbative formulation of weak Abelianisation in \cite{Dittrich:2006ee} but being more general in the sense that it can be applied to any phase space function, a property crucial for our application. As we work in the linearised theory when computing the algebra of Dirac observables that provides the starting point for the reduced phase space quantisation, it is necessary to consider Dirac observables up to second order in general to formulate the model at the gauge invariant level. In former works \cite{lagouvardos2021gravitational,Blencowe:2012mp} often specific gauge fixings were chosen or Dirac observables were constructed for the matter sector only \cite{anastopoulos2013master}, where to our understanding these Dirac observables are not left invariant under the spatial diffeomorphism constraint. In \cite{oniga2016quantum} Dirac quantisation was applied and the obtained results are consistent with the reduced phase space quantisation procedure we apply here. We also consider a second choice of geometrical clocks introduced in \cite{arnowitt1962dynamics} which was carried over to Ashtekar variables in \cite{Dittrich:2006ee} and discuss its differences and similarities to the geometrical clocks chosen in this work. There exist also former work on decoherence effects caused by quantum clocks \cite{Gambini:2010ut,Gambini:2015zda,Gambini:2020bup}, whereas we use classical (perfect) clocks in the language of \cite{Gambini:2020bup}. Such decoherence effects would be present in addition to the decoherence effects that we will discuss in this work and it will be interesting to compare the size of the two types of decoherence effects. A detailed analysis, requires to work with fundamental quantum clocks and perform the construction of Dirac observables in the quantum theory for instance along the lines of \cite{Hoehn:2019fsy}. However, to the knowledge of the authors this has so far only been considered in the context of quantum mechanical systems but not a full quantum field theory setup that we will need in the case considered in this work.

The physical Hamiltonian that is itself a Dirac observable and generates the dynamics of the linearised Dirac observables of the total system is then quantised using a Fock quantisation. 
The strategy we follow in the quantisation is that we choose to normal order the entire physical Hamiltonian operator. This differs from \cite{anastopoulos2013master,oniga2016quantum,lagouvardos2021gravitational} where either individual operators in products of operators are normal ordered separately or no normal ordering is applied at all and such differences appear in the operator ordering in the self-interaction part of the Hamiltonian operator. For deriving a master equation in a top-down approach two prominent techniques exist, the influence functional approach \cite{FEYNMAN1963118} and the projection operator technique \cite{Nakajima1958OnQT, Zwanzig1960, Shibata1977AGS, chaturvedi1979time}, where we use the latter to obtain a time convolution-less (TCL) master equation for the model considered in this work. This allows us to understand in detail how certain assumptions and approximations enter into the derivation of the master equations which will be useful also for later applications and generalisations of the model. We choose a Gibbs state as the initial state of the quantum gravitational environment and use a regularisation by means of a box of finite volume in order to implement this on Fock space, similar to the case in \cite{Matsubara:1955ws}. After computing the environmental trace we remove the regulator by taking the infinite volume limit. In order to circumvent to work with regulators one can use KMS states \cite{Kubo:1957mj, Schwinger:1959many}, which is beyond the scope of this work. We express the final master equation in terms of thermal Wightman functions, which on the one hand is a general formalism other models can also be formulated in and reflects the structure of the involved correlation functions. On the other hand this gives an easy access to the special case where the environment is chosen to be in the vacuum state. For a better comparison with the already existing models in \cite{anastopoulos2013master,oniga2016quantum,lagouvardos2021gravitational} we present three equivalent but different forms of the final master equation and discuss in detail the similarities and differences to \cite{anastopoulos2013master,oniga2016quantum,lagouvardos2021gravitational}.

The paper is structured as follows: After the introduction in section \ref{sec:Intro} in section \ref{sec:CassicalSetup} we present the classical formulation of the decoherence model considered in this article. Subsection \ref{sec:ReviewLinGrav} briefly reviews the existing results on linearised gravity in terms of Ashtekar variables and we consider the coupling of a scalar field to linearised gravity in a Post-Minkowski approximation scheme in subsection \ref{sec:PostMinkowski}. Subsection \ref{sec:ConstrObs} presents the construction of Dirac observables for the linearised model using geometrical clocks. Furthermore, we introduce a canonical transformation on the full phase space in which the geometrical clocks as well as the constraints are chosen as some of the new canonical coordinates, allowing to clearly separate physical and gauge degrees of freedom in the model. This requires the application of the already existing observable map as well as its dual version introduced in this section and opens the possibility for constructing such canonical transformations in principle also in other models. Given the reduced phase space derived in subsection \ref{sec:ConstrObs} and the physical Hamiltonian, we briefly discuss the Fock quantisation of the model in section \ref{sec:QuantumModel} with a focus on the individual contributions in the physical Hamiltonian operator. The derivation of the final master equation is presented in section \ref{sec:DerivMasterEqn}. We include a brief review of the projection operator technique as well as the influence functional approach in subsections \ref{sec:ReviewPOT} and \ref{sec:ReviewIPFA}, respectively. This serves as a preparation for the application of the first technique to the model in this work, presented in subsection \ref{sec:derivMeq} where we obtain the explicit form of the final master equation. The latter is further compared to already existing master equations for similar models in the literature. Finally, in section \ref{sec:Conclusions} we summarise our results and conclude. We present more detailed computations in the appendix that we refer to in the main text of this article.

\section{Classical setup for the decoherence model}
\label{sec:CassicalSetup}

Following the usual approach for gravitationally induced decoherence models, in the model presented in this work here, we will choose as the environmental degrees of freedom the gravitational ones and consider as the system's ones a coupled Klein-Gordon scalar field and discuss the corresponding action and necessary boundary contributions in terms of Ashtekar variables in subsection \ref{sec:GRaction}.  As a simpler setup than full GR we will specialise this action to a coupling to linearised gravity only. For this purpose we consider the linearisation of vacuum gravity in Ashtekar variables in subsection \ref{sec:ReviewLinGrav} and combine this in the framework of a Post-Minkowski approximation to couple the scalar field as discussed in subsection \ref{sec:PostMinkowski}. Because we are working with a gauge theory before later quantising the system and tracing out the environmental degrees of freedom to obtain an effective reduced model for the system, we need to take care of the gauge freedom involved in the theory. Here, one can either consider a specific gauge fixing or work at the gauge invariant level by formulating the model in terms of gauge invariant variables also known as Dirac observables in the context of general relativity, where we will choose the latter strategy in this work as discussed in section \ref{sec:ConstrObs}. Once the dynamics of the model is written by means of Dirac observables we can split the total physical Hamiltonian that generates the dynamics of the Dirac observables into a system and environmental part and we will take this then as our starting point for the quantisation in section \ref{sec:QuantumModel}.

\subsection{The gravity-matter system formulated in Ashtekar variables}
\label{sec:GRaction}
As the starting point for deriving the decoherence model in this work we choose the following action involving gravity coupled to a  Klein-Gordon scalar field in the ADM decomposition  \cite{arnowitt1960consistency} expressed in terms of Ashtekar variables \cite{ashtekar1986new, barbero1994real, immirzi1997real,ashtekar1989new}:
\begin{align}\label{eq:basictotalaction}
S = \int_{\mathbb{R}} dt \int_\sigma d^3x \Big( &\frac{1}{\kappa\beta} \tensor{\dot{A}}{_a^i}(\vec{x},t)\, \tensor{E}{^a_i}(\vec{x},t)+ \dot{\phi}(\vec{x},t)\, \pi(\vec{x},t) \nonumber\\ &- \left[ \Lambda^i(\vec{x},t)\, G_i(\vec{x},t) + N^a(\vec{x},t)\, C_a(\vec{x},t) + N(\vec{x},t)\,C(\vec{x},t) \right] \Big)\,,
\end{align}
where $\beta$ is the Barbero-Immirzi parameter, $\kappa = \frac{8\pi G_N}{c^4}$ with Newton's gravitational constant $G_N$, $c$ denotes the speed of light and we work with the mostly-plus signature of the metric. From here on, we will set $c=\hbar=1$. The gravitational degrees of freedom are encoded in the Ashtekar variables consisting of an SU(2)-connection $\t{A}{_a^i}(\vec{x},t)$  and  the canonically conjugate densitised triads $\t{E}{^a_i}(\vec{x},t)$. The matter sector includes the scalar field $\phi(\vec{x},t)$ as well as its canonically conjugate momentum  $\pi(\vec{x},t)$, an overdot denotes a derivative with respect to the temporal coordinate. That general relativity is a fully constrained theory is reflected by the fact that apart from the symplectic potential in the action, there appears a sum of constraints $G_i(\vec{x},t)$, $C_a(\vec{x},t)$ and $C(\vec{x},t)$ only multiplied by different Lagrange multipliers $\Lambda^i(\vec{x},t)$, $N^a(\vec{x},t)$ and $N(\vec{x},t)$, where the latter two are usually referred to as the shift vector and the lapse function, respectively. In Ashtekar variables the spatial diffeomorphism constraint reads
\begin{equation}\label{eq:vectorconstraint}
    \t{C}{_a}(\vec{x},t) = \frac{1}{2\kappa\beta}\t{F}{_a_b^i}(\vec{x},t)\, \t{E}{^b_i}(\vec{x},t) + \pi(\vec{x},t)\, \partial_a \phi(\vec{x},t)\,,
\end{equation}
where  $\t{F}{_a_b^i}(\vec{x},t) := \partial_{a} \t{A}{_b^i}(\vec{x},t)-\partial_{b} \t{A}{_a^i}(\vec{x},t) + \t{\epsilon}{_i_j_k} \t{A}{_a^j}(\vec{x},t)\, \t{A}{_b^k}(\vec{x},t)$ denotes the curvature related to the connection variable~$A$. The Gau\ss{} constraint, which is absent if working in ADM variables, is given by
\begin{equation}\label{eq:gaussconstraint}
    \t{G}{_i}(\vec{x},t) = \frac{1}{2\kappa\beta} \left( \t{\partial}{_a} \t{E}{^a_i}(\vec{x},t) + \t{\epsilon}{_i_j^k} \t{A}{_a^j}(\vec{x},t)\, \t{E}{^a_k}(\vec{x},t)\right)
\end{equation}
with the completely antisymmetric tensor $\epsilon$.
 Finally, the Hamiltonian constraint takes the form
\begin{align}\label{eq:scalarconstraint}
    C(\vec{x},t) = &\frac{1}{\kappa}\left( \t{F}{_a_b^i}(\vec{x},t) - \frac{\beta^2 +1}{\beta^2} \t{\epsilon}{_i_l_m} (\t{A}{_a^l}(\vec{x},t)-\t{\Gamma}{_a^l}(\vec{x},t)) (\t{A}{_b^m}(\vec{x},t)-\t{\Gamma}{_b^m}(\vec{x},t))  \right) \frac{\t{\epsilon}{^i^j^k} \t{E}{^a_j}(\vec{x},t)\, \t{E}{^b_k}(\vec{x},t)}{\sqrt{\det(\t{E}{^c_n})}} \nonumber\\[0.2em] &+\frac{\pi^2(\vec{x},t)}{2\sqrt{\det(\t{E}{^c_n})}} + \frac{1}{2 \sqrt{\det(\t{E}{^c_n})}}  \delta^{ij}\t{E}{^a_i}(\vec{x},t)\, \t{E}{^b_j}(\vec{x},t)\, (\t{\partial}{_a} \phi(\vec{x},t))(\t{\partial}{_b} \phi(\vec{x},t)) \nonumber\\[0.4em]
    &+ \frac{\sqrt{\det(\t{E}{^c_n})}}{2}\; m^2\phi^2(\vec{x},t)\,.
\end{align}
 Here, $m$ denotes the mass of the scalar field and $\t{\Gamma}{_a^i}$ is the spin connection, considered a functional of the densitised triads, i.e.
 \begin{equation}
     \t{\Gamma}{_a^i}=
\frac{1}{2} \t{\epsilon}{^i^j^k} \t{E}{^b_k} \left(\t{E}{_a^j_,_b} - \t{E}{_b^j_,_a} + \t{E}{^c_j} \t{E}{_a^l} \t{E}{_c^l_,_b} +  \t{E}{_a^j} \frac{\det{(\t{E}{^c_l})_{,b}}}{\det{(\t{E}{^c_l})}} - \t{E}{_b^j} \frac{\det{(\t{E}{^c_l})_{,a}}}{2\det{(\t{E}{^c_l})}} \right)\,,
 \end{equation}
    where we abbreviated partial derivatives with a comma, that is $\t{E}{_a^j_,_b} := \t{\partial}{_b} \t{E}{_a^j}$ and suppressed the spatial and temporal arguments of the involved field variables. The elementary phase space variables satisfy the  following Poisson algebra on the classical phase space:
\begin{align}\label{eq:poissongravity}
\{\t{E}{^a_i}(\vec{x},t),\t{E}{^b_j}(\vec{y},t)\} = \{ \t{A}{_a^i}(\vec{x},t),\t{A}{_b^j}(\vec{y},t) \} &= 0\,, &\{\t{A}{_a^i}(\vec{x},t), \t{E}{^b_j} (\vec{y},t)\} &= \beta\kappa \delta_a^b\delta^i_j \delta^3(\vec{x}-\vec{y})\\
\label{eq:poissonmatter}
    \{ \phi(\vec{x},t), \phi(\vec{y},t) \} = \{ \pi(\vec{x},t), \pi(\vec{y},t) \} &= 0\,, &\{ \phi(\vec{x},t), \pi(\vec{y},t) \} &= \delta^{(3)}(\vec{x}-\vec{y})\,,
\end{align}
where all the remaining ones vanish. As the decoherence model presented in this article requires to linearise the above action around a flat Minkowski background in the course of the upcoming section, we choose asymptotically flat boundary conditions \cite{thiemann1995generalized}  also addressed in \cite{Corichi:2013zza,Campiglia:2014yja} for the case of real Ashtekar variables. With the fall-off behaviour of the Ashtekar variables discussed in section \ref{sec:ReviewLinGrav} and the linearisation of the Lagrange multipliers chosen such that both are consistent with an asymptotically flat universe, it turns out that the Gauß constraint and the gravitational part of the spatial diffeomorphism constraint do not cause any issues and need no further modification in accordance with \cite{Dittrich:2006ee}. However, the gravitational part of the Hamiltonian constraint as well as its variation is not well-defined in the asymptotically flat limit and contains divergences. To circumvent these, we introduce a suitable boundary term given by
\begin{equation}
    T[N] = -2 \int_{\partial \sigma} \t{dS}{_a} N \t{A}{_b^k} \epsilon^{ijk} \frac{\t{E}{^a_i} \t{E}{^b_j} }{\sqrt{\det E}}
\end{equation}
to the smeared Hamiltonian constraint, which makes it well-defined and functionally differentiable. Note that on the constraint hypersurface where the Gauß constraint vanishes $T[1]$ is the so-called ADM-energy \cite{regge1974role}. Further we assume, as is usually done, that the initial data of the matter variables have compact support, yielding matter contributions that are well-defined in the asymptotically flat case. Given this, the canonical Hamiltonian that we will work with in the following and that is consistent with our chosen boundary conditions reads
\begin{align}\label{eq:totalhamiltonian}
\mathbf{H}_{\rm can} 
&= C(N)+\vec{C}(\vec{N})+\vec{G}(\vec{\Lambda})+T[N] \nonumber\\
&= \int_\sigma d^3x \bigg[ \frac{1}{2\kappa\beta} N^a \t{F}{_a_b^j} \t{E}{^b_j} + N^a \pi \t{\nabla}{_a} \phi+\frac{1}{2\kappa\beta} \Lambda^i \t{\partial}{_a} \t{E}{^a_i} + \frac{1}{2\kappa\beta} \Lambda^i \t{\epsilon}{_i_k_l} \t{A}{_a^k} \t{E}{^a_l}\nonumber \\&\hspace{0.8in}- N\epsilon^{jkl} \frac{1}{\sqrt{\det E}} \t{E}{^a_k} \t{E}{^b_l}\frac{1}{2\kappa\beta^2} \t{\epsilon}{_j_m_n} \left(\t{A}{_a^m} \t{A}{_b^n} + (\beta^2+1) \t{\Gamma}{_a^m} \t{\Gamma}{_b^n} - 2 \t{\Gamma}{_a^m}\t{A}{_b^n}\right) \nonumber\\  &\hspace{0.8in}+ \frac{1}{\sqrt{\det E}} \bigg(-\frac{1}{\kappa} \epsilon^{jkl} \t{E}{^a_k} \t{E}{^b_l} \t{A}{_b^j} \t{\partial}{_a} N + N \frac{\pi^2}{2} + \frac{N}{2}  \t{E}{^a_i} \t{E}{^b_i} (\t{\partial}{_a} \phi)(\t{\partial}{_b} \phi)\bigg) \nonumber\\&\hspace{0.8in}+ N \frac{\sqrt{\det E}}{2} \;m^2\phi^2\bigg]\,,
\end{align}
where ${C}({N}) := \int_\sigma d^3x\; C(\vec x,t) N(\vec x,t)$, $\vec{C}(\vec{N}) := \int_\sigma d^3x\; C_a(\vec x,t) N^a(\vec x,t)$ and $\vec{G}(\vec{\Lambda}) := \int_\sigma d^3x\; G_i(\vec x,t) \Lambda^i(\vec x,t)$.\\
In the following two subsections we will consider a linearisation of the gravity-matter system. For this purpose we first briefly review the linearisation of the vacuum case and afterwards discuss how the scalar field can be coupled to linearised gravity in the framework of a Post-Minkowski approximation scheme.

\subsection{Brief review of linearised gravity in Ashtekar variables}
\label{sec:ReviewLinGrav}

A quantisation of the full theory of general relativity is a highly complex task. As outlined in the introduction, one of the motivations for the work presented in this article comes from scenarios that involve weak couplings between the matter system and the gravitational field. Therefore, we will consider a linearisation of general relativity around a Minkowski spacetime and then consider a perturbation for the gravitational degrees of freedom using $\kappa$ as an expansion parameter. Here we will focus on vacuum gravity using Ashtekar variables first, analogous to the work in for instance \cite{ashtekar1991gravitons} or \cite{dittrich2006testing}. We denote the linear perturbations with a prefix $\delta$. The Ashtekar variables as well as the Lagrange multipliers then become
\begin{align}
\t{E}{^a_i} &= \overline{E}^a_i + \kappa\, \delta\t{E}{^a_i}=\delta^a_i + \kappa\, \delta\t{E}{^a_i} && \delta E : O(r^{-1}) \text{ even} \label{eq:FallOffdeltaE}\\
\t{A}{_a^i} &=\overline{A}^i_a + \kappa\, \delta\t{A}{_a^i}= 0 + \kappa\, \delta\t{A}{_a^i}&& \delta A : O(r^{-2}) \text{ odd}\\
N &=\overline{N} + \kappa\, \delta N=1 + \kappa\, \delta N && \delta N : O(r^{-1})\\
N^a & = \overline{N}^a + \kappa\, \delta N^a=0 + \kappa\, \delta N^a && \delta N^a : O(r^{-1})\\
\Lambda^i &= \overline{\Lambda}^i + \kappa\, \delta\Lambda^i=0 + \kappa\, \delta\Lambda^i && \delta \Lambda : O(r^{-2}) \text{ even}\,,
\end{align}
where the fall-off behaviour in the limit of spatial infinity of the first order perturbations has been chosen in accordance with the asymptotically flat boundary condition \cite{thiemann1995generalized}. The split into background and perturbed quantities is chosen such that the background corresponds to a Minkowski spacetime that remains unchanged by the action of the canonical Hamiltonian of the system, i.e.
\begin{align}
    \overline{\{\t{E}{^a_i},\mathbf{H}_{\rm can}\}} &= 0 \\ 
    \overline{\{\t{A}{_a^i},\mathbf{H}_{\rm can}\}} &= 0\,,
\end{align}
where the overline stands for evaluation with respect to the background Minkowski spacetime. In the case of vacuum gravity (after computation of the Poisson brackets) this amounts to  $\overline{E}^a_i=\delta^a_i$, $\overline{A}_a^i=0$, $\overline{N}=1$ and $\overline{N}^a=0$. As one can easily compute, the spin connection vanishes in the background:
\begin{eqnarray}
\t{\Gamma}{_a^i} &=& \overline{\Gamma}^i_a +\kappa\, \delta\t{\Gamma}{_a^i}
\approx 0+\kappa\, \delta\t{\Gamma}{_a^i} \nonumber\\
&=&- \frac{\kappa\, }{2} \epsilon^{ijk} \delta_k^b \left[- \delta_a^l \delta_c^j \t{\partial}{_b}  (\delta\t{E}{^c_l}) + \delta_c^j \delta_b^l \t{\partial}{_a} (\delta\t{E}{^c_l}) - \delta_{ac}\t{\partial}{_b} (\delta\t{E}{^c_j}) + \delta_a^j \delta_c^l \t{\partial}{_b} (\delta\t{E}{^c_l}) \right]\,.
\end{eqnarray}
The Poisson algebra of the linearised gravitational variables can be inherited from \eqref{eq:poissongravity}:
\begin{equation}\label{eq:poissongravitylin}
\{\delta \t{E}{^a_i}(\vec{x},t),\delta \t{E}{^b_j}(\vec{y},t)\} = \{ \delta \t{A}{_a^i}(\vec{x},t),\delta \t{A}{_b^j}(\vec{y},t) \} = 0\,, \hspace{0.2in} \{\delta \t{A}{_a^i}(\vec{x},t), \delta \t{E}{^b_j} (\vec{y},t)\} = \frac{\beta}{\kappa} \delta_a^b\delta^i_j \delta^3(\vec{x}-\vec{y})\,.
\end{equation}
In the next subsection we will address how the matter contributions can be included consistently in the linearised framework.

\subsection{Post-Minkowski approximation scheme}
\label{sec:PostMinkowski}
In the context of general relativistic perturbation theory, one usually chooses a background solution and then considers perturbations of the gravitational and matter sector around it assuming that the perturbations are small compared to the chosen background quantities. In this work we linearise around a flat Minkowski spacetime, which is a vacuum solution of general relativity. Hence, any considered perturbation in the matter sector will not be small compared to vanishing matter degrees of freedom in the vacuum case. However, we can formulate a model that involves the coupling between matter and linearised gravity in a Post-Minkowski approximation scheme. For the model presented in this article we will need to consider the zeroth and first order in the Post-Minkowski formalism only and this will be the underlying guiding principle of how the matter sector will be included into the Hamiltonian formulation and in particular into the constraints. We will briefly sketch the main steps that lead to the linearised action that we take as our starting point. The Post-Minkowski formalism is based on the Landau-Lifshitz formulation of general relativity, see for instance the book \cite{PoissonBook:2014} for an introduction to the subject, and one starts by introducing the so-called gothic metric defined as 
\begin{equation}\label{4:dgm}
\mathfrak{g}^{\mu\nu} := \sqrt{-\det(g_{\rho\sigma})}\, g^{\mu\nu},
\end{equation}
by means of which we can introduce the following tensor density
\begin{equation}
\label{eq:DefHTensor}
H^{\mu\nu\rho\sigma} := \mathfrak{g}^{\mu\rho}\mathfrak{g}^{\nu\sigma}-\mathfrak{g}^{\mu\sigma}\mathfrak{g}^{\nu\rho}\,,
\end{equation}
where $H^{\mu\nu\rho\sigma}$ carries the same symmetries as the Riemann tensor. Following the presentation in \cite{PoissonBook:2014} one can use \eqref{eq:DefHTensor} to rewrite Einstein's equations $G_{\mu\nu}=\kappa T_{\mu\nu}$ in the following form:
\begin{equation}\label{4:gothfeq}
\partial_\nu\partial_\sigma H^{\mu\nu\rho\sigma} = 2\kappa (-\det(g)) \left[ T^{\mu\rho} + t^{\mu\rho}_{LL}\right]\,.
\end{equation}
Here, $T^{\mu\rho}$ denotes the energy-momentum tensor of the matter degrees of freedom and $t^{\mu\rho}_{LL}$ is the so-called Landau-Lifshitz pseudotensor which consists of a sum of (contractions of) terms $\partial \mathfrak{g}\, \partial \mathfrak{g}$ and does not transform as a tensor under coordinate transformations. This quantity corresponds to the distribution of energy of the gravitational field in spacetime. If one aims at formulating the corresponding linearised theory, as a first step, one applies a partial gauge fixing on the gothic metric by the harmonic coordinate condition
\begin{equation}
\partial_\mu \mathfrak{g}^{\mu\nu}=0\,.
\end{equation}
For later convenience it is useful to introduce the following quantity also known as the metric potentials $\mathfrak{h}^{\mu\nu}$ that have the following relation to the gothic metric $\mathfrak{g}^{\mu\nu}$:
\begin{equation}
\mathfrak{g}^{\mu\nu}=\eta^{\mu\nu}-\mathfrak{h}^{\mu\nu}\,.
\end{equation}
The harmonic coordinate condition then carries over to $\partial_\mu \mathfrak{h}^{\mu\nu}=0$. In this gauge, Einstein's equations in \eqref{4:gothfeq} have the form, see e.g. equation (6.51) and (6.52) in \citep{PoissonBook:2014}:
\begin{equation}\label{4:weq}
\Box \mathfrak{h}^{\mu\nu} = -2\kappa (-\det(g)) \left[ T^{\mu\nu}[\Phi,g] + t^{\mu\nu}_{LL}[\mathfrak{h}] + t_H^{\mu\nu}[\mathfrak{h}]\right],
\end{equation}
with the d'Alembertian $\Box$ in flat spacetime and 
\begin{equation}
t^{\mu\nu}_H[\mathfrak{h}] := \frac{1}{2\kappa (-\det(g))} \left( \partial_\rho \mathfrak{h}^{\mu\sigma} \; \partial_\sigma \mathfrak{h}^{\nu\rho} - \mathfrak{h}^{\rho\sigma} \; \partial_{\rho} \partial_\sigma \mathfrak{h}^{\mu\nu}\right) \,.
\end{equation}
Among the individual contributions in \eqref{4:weq} the energy-momentum tensor $T^{\mu\nu}$ depends on the matter variables here denoted by $\Phi$ and on the metric $g$, while the other two contributions depend on the modified gothic metric $\mathfrak{h}$ only\footnote{Later we consider a Klein-Gordon scalar field for the matter sector but here $\Phi$ is understood symbolically for all possible matter choices that one would like to couple in a given model}. The idea of the Post-Minkowski approximation scheme is to construct an iterative solution for \eqref{4:weq}. For this purpose one chooses that in lowest order the gothic metric agrees with the Minkowski metric and for higher orders considers an expansion of $\mathfrak{h}^{\mu\nu}$ in terms of powers of $\kappa$ according to
\begin{equation}
\label{eq:GothicPertAnsatz}
\mathfrak{g}^{\mu\nu}=\eta^{\mu\nu}+ \sum\limits_{n=1}^\infty \kappa^n \, \mathfrak{h}^{\mu\nu}_{(n)}\,.
\end{equation}
As in this work we linearise our entire system around a flat Minkowski background and are therefore only interested in weak gravity, for the model in this article we can consider truncations of this perturbative expansion\footnote{ Note also the formal role of the expansion parameter $\kappa$ that is further explained in box 6.4 in \citep{PoissonBook:2014}: Depending on the chosen system of units, $\kappa$ could also be equal to one.}.

An iterative solution of Einstein's equations is constructed by the following procedure. Let us assume one has a solution for the potentials $\mathfrak{h}$ of order $\kappa^k$. Then one is able to compute the contributions of the quantities on the right hand side of \eqref{4:weq} with it. Due to the prefactor $\kappa$, the right hand side will then be of order $\kappa^{k+1}$, so the entire equation will yield a solution for $\mathfrak{h}_{(k+1)}$. As can be seen from \eqref{eq:GothicPertAnsatz}, the perturbative ansatz in the Post-Minkowski approximation scheme is constructed in a way that in the zeroth order where $\mathfrak{h}=0$ one has $\mathfrak{g}=\eta$, from which one directly obtains $g=\eta$. With this, one can construct the right hand side of \eqref{4:weq} using this zeroth order solution. As $t_{LL}$ and $t_H$ both contain $\mathfrak{h}$ in each summand, they vanish for $\mathfrak{h}=0$ and the only non-trivial contribution comes from the energy-momentum tensor $T^{\mu\nu}$ on the right hand side. As in zeroth order $g=\eta$, it simplifies to the energy momentum tensor $T^{\mu\nu}$ on a flat Minkowski spacetime. Given the right hand side of \eqref{4:weq}, the left hand side amounts to the first order modified gothic metric, that is
\begin{equation}\label{4:foeq}
\kappa\Box \mathfrak{h}^{\mu\nu}_{(1)} = -2\kappa T^{\mu\nu}[\Phi,\eta]\,.
\end{equation}
Finally, one needs to relate $\mathfrak{h}^{\mu\nu}_{(1)}$ to the (inverse) metric perturbations $\delta h^{\mu\nu}$. Since in this work we are interested in the linearised theory only, we can use 
 $g^{\mu\nu}=\eta^{\mu\nu}-\kappa \delta h^{\mu\nu}+O(\kappa^2)$ and obtain
 \begin{equation}
\mathfrak{h}_{(1)}^{\mu\nu} = \delta h^{\mu\nu} - \frac{1}{2} \eta^{\mu\nu} \, \eta_{\rho\sigma} \delta h^{\rho\sigma}\,.
\end{equation}
The last equations allows us to rewrite the linearised equations in a first order Minkowski approximation as
\begin{equation}\label{eq:LinEinsteineq}
-\frac{\kappa}{2}\left( \Box (\delta h^{\mu\nu}) - \frac{1}{2} \eta^{\mu\nu} \, \eta_{\rho\sigma}\Box (\delta h^{\rho\sigma}) \right)= \kappa T^{\mu\nu}[\Phi,\eta] \,. 
\end{equation}
From \eqref{eq:LinEinsteineq} we can directly read off how the interaction term in the action needs to look like, namely
\begin{equation}
S_{I} = \frac{\kappa}{2} \int_{M} d^4x \; \delta h_{\mu\nu} T^{\mu\nu}[\Phi,\eta] \,.
\end{equation}
The overall factor $\frac{\kappa}{2}$ takes into account that we start with an overall factor of $\frac{1}{2\kappa}$ and then consider linear perturbations of the metric $g_{\mu\nu}=\eta_{\mu\nu}+\kappa\delta h_{\mu\nu}$ yielding a factor $\kappa^2$ for second-order perturbations in the vacuum case. Note that, as expected, the same interaction contribution to the  action can be obtained by minimally coupling matter to gravity and then linearising the metric degrees of freedom around a Minkowski spacetime \cite{Giddings:2019wmj,anastopoulos2013master,oniga2016quantum,Blencowe:2012mp}. Let us denote the matter Lagrangian density by ${\cal L}_\Phi$, then we have
\begin{align}
\int_M d^4x \; \sqrt{-\det(g)}\, \mathcal{L}_\Phi &\approx \int_M d^4x \; \mathcal{L}_\Phi \rvert_{g=\eta} + \int_M d^4x \; (-\kappa \delta h^{\mu\nu})\left(\frac{\delta (\sqrt{-\det(g)}  \, \mathcal{L}_\Phi)}{\delta g^{\mu\nu}}\right) \Big|_{g=\eta} \nonumber\\
&= S^\eta_{\Phi} +\frac{\kappa}{2} \int_M d^4x\;\delta h^{\mu\nu} T_{\mu\nu}[\Phi,\eta] \,,
\end{align}
where $S^\eta_{\Phi}$ denotes the matter action on flat spacetime. For the model discussed in this work we choose the matter Lagrangian of a Klein-Gordon scalar field
\begin{equation}
{\cal L}_\phi=-\frac{1}{2}g^{\mu\nu}D_\mu\phi D_\nu \phi-\frac{m^2}{2}\phi^2,
\end{equation}
where $D_\mu$ denotes the torsion-free, covariant metric-compatible derivative with respect to $g$. Performing an ADM decomposition as we did for the vacuum case we end up with
\begin{equation}
S_\phi = \int dt \int_\sigma d^3x \left( \dot{\phi} \pi - \left[ N^a C_a^\phi + N C^\phi \right] \right)
\end{equation}
with
\begin{align}
C_a^\phi &:= \pi \nabla_a \phi, \\ \label{3:hcmat}
C^\phi &:= \frac{\pi^2}{2\sqrt{\det q}} + \frac{\sqrt{\det q}}{2} q^{ab} (\nabla_a \phi)(\nabla_b \phi) + \sqrt{\det q}\; \frac{m^2}{2}\phi^2\,,
\end{align}
where $\pi$ denotes the canonically conjugate momentum associated to $\phi$. If we linearise the model including the scalar field as discussed above and  take into account all contributions up to first order in $\kappa$, we obtain the following constraints in the linearised theory:
\begin{align}\label{eq:linGauss}
\delta G_i &= \frac{\kappa}{2\beta} \left( \t{\partial}{_a} (\t{\delta E}{^a_i}) + \t{\epsilon}{_i_k^l} \delta_l^a\, \t{\delta A}{_a^k} \right), \\
\delta C_a &= \frac{\kappa}{2\beta} \delta_j^b \left( \t{\partial}{_a} (\t{\delta A}{_b^j}) - \t{\partial}{_b} (\t{\delta A}{_a^j}) \right) + \kappa \underbrace{\pi\, \t{\partial}{_a}\phi}_{=: p_a(\phi,\pi)}, \\
 \delta C &= \kappa \t{\epsilon}{_j^k^l} \delta_k^a \delta_l^b \,\t{\partial}{_a} (\t{\delta A}{_b^j}) + \kappa \underbrace{\frac{1}{2} \left[ \pi^2 + \t{\partial}{_a} \phi \,\partial^a \phi + m^2 \phi^2 \right]}_{=:\epsilon(\phi,\pi)}  \,.
\end{align}
These constraints are also consistent with the $00$ and $0a$ components of the linearised Einstein's equations that one obtains in the Post-Minkowski approximation scheme. Now these include gravitational- as well as matter contributions, where we introduced the momentum density $p_a(\phi,\pi)$ and the energy density $\epsilon(\phi,\pi)$ of the scalar field. Note that these constraints are Abelian up to first order in $\kappa$. The corresponding background constraints of the zero order Post-Minkowski approximation scheme all vanish trivially, since $\overline{C}=\overline{C}^{\rm geo}=0, \overline{C}_a=\overline{C}^{\rm geo}_a=0$ and $\overline{G}_j=0$. 
The action of the linearised theory is given by
\begin{align}
S_{\rm lin} = \int dt \Bigg(&\int_\sigma d^3x\left[\frac{\kappa}{\beta} \t{\delta \dot{A}}{_a^i} \t{\delta E}{^a_i} +\pi \dot{\phi} \right]\nonumber\\&-\underbrace{\int_\sigma d^3x\bigg[ \epsilon(\phi,\pi) + \delta N \delta C + \delta N^a \delta C_a + \delta \Lambda^i \delta G_i + \delta \mathcal{H}_I +\frac{1}{\kappa} \delta^2 C^{geo} \bigg]}_{= \delta \mathbf{H}_{\rm can}}\Bigg)
\end{align}
with the interaction part $\delta \mathcal{H}_I:=\frac{\kappa}{2}\delta h_{ab}T^{ab}$ with $\delta h_{ab}$ understood as a function of the densitised triads and the second order of the geometrical Hamiltonian constraint $\delta^2 C^{geo}$. Note that the overall $\frac{1}{\kappa}$ in front of the geometrical term has been partly cancelled by the $\kappa$ involved in the linearised quantities for $N, N^a$ and $\Lambda^i$, while the constraints and the interaction term are still linear in $\kappa$. In the above form of the linearised action it was already used that in the linearised framework with asymptotically flat boundary conditions the boundary term precisely cancels the first order of the (geometrical) Hamiltonian constraint in accordance with \cite{Dittrich:2006ee}, i.e. $T= -\overline{N}\delta C^{geo}=-\delta C^{geo}$ where we used that $\overline{N}=1$. Hence, neither the linearised Hamiltonian constraint with background lapse nor the boundary term do appear in the action and canonical Hamiltonian anymore.

The linearised total canonical Hamiltonian, which follows from \eqref{eq:totalhamiltonian}, is given by:
\begin{align}\label{eq:totalhamlin}
\delta \mathbf{H}_{\rm can} =\int_\sigma d^3x \bigg[ &\epsilon(\phi,\pi) + 
\kappa \, \delta N^a\, \pi\, \partial_a \phi +\kappa\, \delta N\, \epsilon(\phi,\pi) +\kappa\, (\t{\partial}{_a}\phi) \,(\t{\partial}{^b}\phi)\, \t{\delta E}{^a_i}\, \delta^i_b \nonumber\\&+\frac{\kappa}{2} m^2 \phi^2\, \t{\delta E}{^a_i}\,\delta^i_a - \frac{\kappa}{2} \t{\delta E}{^a_i}\,\delta_a^i\, \epsilon(\phi,\pi)+ 
\frac{\kappa}{2\beta} \delta N^a\, \delta_j^b \left(\t{\partial}{_a} (\t{\delta A}{_b^j}) - \t{\partial}{_b} (\t{\delta A}{_a^j})\right) \nonumber\\&+ \frac{\kappa}{2\beta} \delta \Lambda^i\, \t{\partial}{_a} (\t{\delta E}{^a_i}) + \frac{\kappa}{2\beta} \delta \Lambda^i\, \t{\epsilon}{_i_k^l}\, \delta^a_l\, \t{\delta A}{_a^k} - \kappa\, \t{\epsilon}{_j^k^l}\, \delta_k^a\, \delta_l^b\, \t{\delta A}{_b^j}\, \t{\partial}{_a} (\delta N)\nonumber\\\ &- \frac{\kappa}{2\beta^2} \epsilon^{jkl} \t{\epsilon}{_j_m_n}\, \delta^a_k\, \delta^b_l \left( \t{\delta A}{_a^m} \t{\delta A}{_b^n}+ (\beta^2+1) \t{\delta \Gamma}{_a^m} \t{\delta \Gamma}{_b^n} -2 \t{\delta \Gamma}{_a^m} \t{\delta A}{_b^n} \right) \bigg].
\end{align}
Let us briefly comment on the individual contributions in different orders of $\kappa$. The $\kappa^0$-order encodes the equations of motion for the uncoupled system and environment. In our case these are the Klein-Gordon equation and the equations of motion for the background connection and densitised triad variables. Since the connection vanishes in the background and the triad is just given by $\overline{E}^a_i=\delta^a_i$ the corresponding equations of motion trivially vanish and this is again consistent with the vanishing of the background constraints that generate this trivial dynamics for the gravitational degrees of freedom. 
Note that the energy density of the scalar field $\epsilon(\phi,\pi)$ in $\kappa^0$-order is not part of the background constraints but just contributes to the non-vanishing part of the Hamiltonian in this order. Without the boundary term a further term that contributes to the $\kappa^0$ is the linearised Hamiltonian given by $\overline{N}\delta C^{\rm geo}=\overline{N}(\delta C-\kappa\epsilon)$. On the linearised phase space this term generates the background equations. If we compute the Poisson brackets of $\delta\t{A}{_a^i},\delta \t{E}{^a_i}$ with $\overline{N}\delta C^{\rm geo}$ they both vanish for $\overline{N}=1$ demonstrating again that the background equations are trivially fulfilled.
In linear order in $\kappa$ in the covariant case we obtain a part that is quadratic in the perturbations of all metric components. Here, this corresponds to the terms that are either quadratic in the linearised Ashtekar variables or involve them linearly together with the linearised lapse and shift variables. These terms together in $\delta \mathbf{H}_{\rm can}$ will generate the left hand side of all linearised Einstein's equations including the $00$ and $0a$ component.  The right hand side of the linearised Einstein's equations will be obtained from all contributions that involve gravitational as well as matter variables in the $\kappa^1$-order. These will be $\delta\mathcal{H}_I$ together with the parts where the matter variables occur in combination with the linearised lapse and shift. 

As can easily be seen, most of the twenty phase space variables of the linearised theory,
\begin{equation}\label{eq:oldphasespacevar}
(\phi,\pi) \hspace{0.4in} (\delta \t{A}{_a^i},\delta \t{E}{^b_j})\,,
\end{equation}
are not observable, as they transform non-trivially under the linearised constraints. In the next section \ref{sec:ConstrObs} we use the relational formalism to construct a set of independent gauge invariant quantities.

\subsection{Dirac observables in the linearised model using geometrical clocks}
\label{sec:ConstrObs}
 As mentioned above in a gauge theory we can either apply a specific gauge fixing or formulate the model in terms of gauge invariant quantities. In case we work at the level of perturbation theory, as done here, one often takes the approach that gauge invariance is guaranteed up to possible corrections that are of higher order in perturbation theory than one is truncating at, we will follow the same strategy here. To construct gauge invariant variables on the linearised phase space we consider the relational formalism \cite{Rovelli:1990ph,Rovelli:1990pi,Rovelli:2001bz} together with an observable map that maps the elementary variables of the linearised phase space to their corresponding gauge invariant quantities \cite{Dittrich:2004cb,Dittrich:2005kc,Vytheeswaran:1994np,Thiemann:2004wk}. The formalism is based on a choice of reference fields, one for each constraint in the system, and then the observable map  returns for a given tensor field of certain rank on phase space its corresponding gauge invariant extension. Explicitly, it can be written as a power series in the reference fields weighted by  contributions that involve nested Poisson brackets of the tensor fields and the constraints. The physical interpretation of these gauge invariant quantities, also known as Dirac observables in the framework of general relativity, is that they give the value of the tensor field at those values where the reference fields take specific values. For general relativity and its spatial diffeomorphism and Hamiltonian constraints this particularly means that we can formulate the dynamics of a given tensor field on phase space with respect to the reference fields and these provide a notion of physical spatial and temporal coordinates. The observable map introduced in \cite{Dittrich:2004cb,Dittrich:2005kc,Vytheeswaran:1994np} has the property that when applied to the chosen reference fields it maps them to phase space independent quantities and thus such a map can describe neither a canonical transformation on the full phase space one starts with nor a transformation that preserves the number of elementary phase space variables. For this reason we will slightly modify the map from \cite{Dittrich:2004cb,Dittrich:2005kc,Vytheeswaran:1994np} in two aspects. First we modify its action on reference fields and second we further introduce also the kind of dual version of this map to treat the choice of the reference fields and the constraints more on an equal footing, allowing to use the modified map to construct a canonical transformation for the variables $(\phi,\pi), (\delta \t{A}{_a^i},\delta \t{E}{^b_j})$. For the purpose of this work we can restrict our discussion to the linearised phase space. The transformation that we aim at constructing will map the set of elementary variables $(\phi,\pi), (\delta \t{A}{_a^i},\delta \t{E}{^b_j})$ to a new set of canonical variables containing the Dirac observables associated with $(\phi,\pi)$, seven chosen reference fields that are canonically conjugate to the seven linearised constraints as well as two canonical pairs of further Dirac observables in the gravitational sector. Using this kind of canonical transformation for the elementary phase space variables, the physical phase space can be more easily accessed because the constraints as well as the reference fields are among the new canonical variables. 
 
 As we will discuss below for a modification of the observable maps in \cite{Dittrich:2004cb,Dittrich:2005kc,Vytheeswaran:1994np,Thiemann:2004wk} in our work here, we add a so called dual observable map and the combination of both allow us to construct the canonical transformation on the full phase space.  Let us briefly introduce the observable map from \cite{Dittrich:2004cb,Dittrich:2005kc,Vytheeswaran:1994np,Thiemann:2004wk} as well as the dual one and then apply them to the model. We consider a system with a set of first class constraints $\{C_I\}$ and elementary phase space variables $(q^A,p_A)$. Then we choose a set of reference variables $\{T^I\}$ that satisfy $\det(\{T^I,C_J\})\not=0$ where we abbreviate $\{T^I,C_J\}:=M^{I}_{\,\,J}$. We can define an equivalent set of constraints given by 
\begin{equation}
\label{eq:NewConstraints}
C^\prime_I=\sum\limits_{J}(M^{-1})^J_{\,\,I}C_J.    
\end{equation}
Then by construction we have that the constraints $C_I^\prime$ are weakly canonically conjugate to the $T^I$'s. 
Given this, one further chooses a set of functions $\{\tau^I\}$ that can depend on the spatial and temporal coordinates by means of which one can construct an observable map which maps a function $f$ on the phase space to its corresponding observable denoted by ${\cal O}_{f,\{T\}}$ 
\begin{eqnarray}
\label{eq:ObsFormula}
{\cal O}_{f,\{T\}}(\tau^I)&=&\left[\exp( \xi^I\{C_I,\cdot\})\cdot f\right]\Big|_{\xi^I:=T^I-\tau^I} \\
&=&f+(T^I-\tau^I)\{C_I,f\}+\frac{1}{2!}(T^I-\tau^I)(T^J-\tau^J)\{C_J,\{C_I,f\}\}+\cdots\, ,\nonumber 
\end{eqnarray}
where the label $\{T\}$ refers to the chosen set of reference variables. The observable ${\cal O}_{f,\{T\}}(\tau^I)$ returns the value of $f$ at those values where the reference variables $T^I$ take the values $\tau^I$. As can be shown \cite{Dittrich:2004cb,Vytheeswaran:1994np} we have $\{C_I,O_{f,T}\}\approx 0$ where $\approx$ denotes weak equivalence, that is on the constraint hypersurface defined by the $C_I$'s and thus the $O_{f,\{T\}}$ are weak Dirac observables. In case the constraints $C_I$ are already canonically conjugate to the $T^I$'s then the $O_{f,\{T\}}$ are even strong Dirac observables and weak equalities are replaced by equality signs. Let us introduce ${\cal G}^I:=T^I-\tau^I$ that we can also understand as a choice of coordinate gauge fixing condition if we require ${\cal G}^I\approx 0$. Then \eqref{eq:ObsFormula} can be understood as a power series in the ${\cal G}^I$'s with nested Poisson brackets involving the constraints $C_I$ that are in this case equal to the $C^\prime_I$'s. In case we have $\{T^I,C_J\}=\delta^I_J$ because the $\tau^I$ do not depend on phase space variables we also obtain $\{{\cal G}^I,C_J\}=\delta^I_J$. Thus, the gauge fixing conditions and the constraints build canonically conjugate pairs. If this is the case, we can construct a dual version of the observable map in \eqref{eq:ObsFormula} where the role of the gauge fixing conditions and constraints are interchanged given by
\begin{eqnarray}
\label{eq:ObsFormulaDual}
{\cal O}^{\rm dual}_{f,\{C\}}
&=&\left[\exp( -\xi_I\{{\cal G}^I,\cdot\})\cdot f\right]\Big|_{\xi_I:=C_I} \\
&=&f-C_I\{{\cal G}^I,f\}+\frac{1}{2!}C_I C_J\{{\cal G}^J,\{f, { \cal G}^I \}\}\pm\cdots\nonumber 
\end{eqnarray}
where the label $\{C\}$ denotes that we have interchanged the role of gauge fixing conditions and the constraints for the dual map. This is also the reason for the additional factor of $(-1)^n$ compared to the observable map in \eqref{eq:ObsFormula}. Note that this dual map does not depend on any functions $\tau^I$ since the constraint hypersurface is defined by $C_I\approx 0$. Similar to ${\cal O}_{f,\{T\}}(\tau^I)$ which can be understood as a family of gauge invariant extension of $f$ parameterised by $\tau^I$ for ${\cal G}^I\not=0$, the quantity ${\cal O}^{\rm dual}_{f,\{C\}}$ is an extension from the $C_I\approx0$ constraint hypersurface with the property that it commutes with all the ${\cal G}^I$ by construction.

Our strategy is to construct a canonical transformation to a new set of elementary variables. For this purpose, we apply a combination of the two observable maps yielding quantities that commute by construction with all constraints $C_I$ and all ${\cal G}_I$'s. Hence, we can choose the set $({\cal G}^I,C_I)$ as new canonical variables together with the number of independent Dirac observables obtained from the observable and its dual map associated with the set $(q^I,p_I)$. How many independent such Dirac observables exist depends on the number of constraints. 

Note that the introduction of the equivalent set of constraints in \eqref{eq:NewConstraints} is also called weak Abelianisation \cite{Dittrich:2004cb} and ensures that the corresponding Hamiltonian vector fields of the $C_I$'s weakly commute. As a consequence, the order in which we apply the constraints $C_I$ in the nested Poisson bracket is irrelevant. Now the situation relevant for us is that the constraints satisfy $\{T^I,C_J\}=\delta^I_J$ and thus the $C_I$'s can be expressed linearly in the momenta conjugate to the $T^I$'s.  Using that the set $\{C_I\}$ is first class, this is sufficient to show that the constraints are Abelian and hence their corresponding Hamiltonian vector fields commute and the order how we apply the $C_I$ also does not matter in our case. However, if we consider the entire set $({\cal G}^I,C_I)$ then the ${\cal G}^I$ and $C_I$ do not commute, not even weakly and neither is the entire set first class and in general also the subset of the ${\cal G}^I$ is not Abelian. This has the effect that the final observable that we obtain by applying the observable map in combination with its dual in general depends on the order in which we apply the two maps as well as the order in which the ${\cal G}^I$ occur in the nested Poisson brackets. This causes no problem for the model considered in this article, it just means that there exist different choices of possible coordinate transformations on phase space, but important for us is rather that we can choose one among those.

In the following we will generalise the observable map to the field theoretic setup and perturbations theory which can easily be done. In the context of perturbation theory it is sufficient to require gauge invariance or some specific form of the Poisson algebra up to corrections that are higher than the order that is considered in perturbation theory. This means we can truncate the power series for the map and its dual at some order in accordance with the desired order in perturbation theory that we consider for the linearised model. For instance if we want to compute Poisson brackets of observables up to linear order then we need to perturb both observables up to second order and collect all terms that contribute up to linear order to the final result. Such a perturbative approach for constructing observables has for instance also been used in \cite{Dittrich:2006ee,Dittrich:2007jx,Giesel:2018opa}.
Alternatively, if available we can also take the result of the corresponding Poisson bracket in full general relativity and perturb it up to linear order. But since many quantities we work with will be known at the perturbative level only the latter option is often not possible.

The first step in the relational formalism consist of choosing suitable reference fields for the given set of constraints. In our case these will be the Hamiltonian, spatial diffeomorphism and Gau\ss{} constraint. If we are interested in results up to corrections of second order in the perturbations inside the Poisson brackets we need to consider perturbations up to second order of these constraints. The properties our chosen reference fields should have are:
\begin{itemize}
    \item[(i)] Each chosen reference field consists of linear perturbations of the elementary gravitational variables and its derivatives only. These are also known as linearised geometrical clocks.
    \item[(ii)] Each chosen reference field is in lowest order canonically conjugated to one of the constraints.
    \item[(iii)] Each reference field commutes in lowest order with all constraints except the one that it is in lowest order canonically conjugated to.
    \item[(iv)] All reference fields mutually commute. 
\end{itemize}
Let us introduce the notation 
\begin{equation*}
\delta{\cal C}_I(\vec{x},t):=(\delta C(\vec{x},t), \delta C_a(\vec{x},t), \delta G_i(\vec{x},t)), \quad 
\delta^2{\cal C}_I(\vec{x},t):=(\delta^2 C(\vec{x},t), \delta^2 C_a(\vec{x},t), \delta^2 G_i(\vec{x},t))
\end{equation*}
for the set of linearised constraints  $\delta{\cal C}_I(\vec{x},t)$ and $\delta^2{\cal C}_I(\vec{x},t)$ for the set of second-order perturbations. For each of the individual constraints we have to choose one reference field and we introduce the following notation ${\cal G}^I(\vec{x},t)$ for this set  with
\begin{eqnarray}
\label{eq:DefcalG}
&& {\cal G}^I(\vec{x},t):=(\delta{\cal G}(\vec{x},t),\delta{\cal G}^a(\vec{x},t),\delta{\cal G}^j(\vec{x},t))\nonumber \\
&&\delta{\cal G}(\vec{x},t):=\delta T(\vec{x},t)-\delta\tau(\vec{x},t), \quad \delta {\cal G}^a(\vec{x},t):=\delta T^a(\vec{x},t)-\delta\sigma^a(\vec{x},t),\nonumber \\
&&\delta {\cal G}^j(\vec{x},t):=\delta\Xi^j(\vec{x},t)-\delta\xi^j(\vec{x},t).
\end{eqnarray}
where $\delta T, \delta T^a, \delta \Xi^j$ denote the individual reference fields for the Hamiltonian, spatial diffeomorphism and Gau\ss{} constraint respectively. We denote background quantities with a bar and assume that in the background the gauge fixing conditions as well as the constraints are satisfied, that is
\begin{eqnarray*}
&&\overline{\cal G}(\vec{x},t):=\overline{T}(\vec{x},t)-\overline{\tau}(\vec{x},t)=0, \overline{C}=0,\quad  \overline{\cal G}^a(\vec{x},t):=\overline{T}^a(\vec{x},t)-\overline{\sigma}^a(\vec{x},t)=0, \overline{C}_a=0,\nonumber \\
&&\overline{\cal G}^j(\vec{x},t):=\overline{T}^j(\vec{x},t)-\overline{\xi}^j(\vec{x},t)=0, \overline{G}_j=0.
\end{eqnarray*}
With the assumptions (i)-(iv) above we know that $\delta^2{\cal G}^I=0$ and further we have
\begin{eqnarray}
\label{eq:ReqPBAlgebra}
\{{\cal G}^I(x),{\cal C}_J(y)\} 
:&=&\{\delta{\cal G}^I(x),\delta {\cal C}_J(y)\} +\{\delta{\cal G}^I(x),\delta^2 {\cal C}_J(y)\} +O(\delta^2,\kappa^2)\nonumber \\
&=&\frac{1}{\kappa}\delta^I_J\delta^{(3)}(x,y) +\{\delta{\cal G}^I(x),\delta^2 {\cal C}_J(y)\} +O(\delta^2,\kappa^2),
\end{eqnarray}
where $O(\delta^2,\kappa^2)$ means that we neglect all terms that are second-order in the perturbations and/or of order $\kappa^2$ and the factor $\frac{1}{\kappa}$ has been chosen because it is also involved in the Poisson bracket of the elementary gravitational variables. This means there could be terms being of order $\kappa^n$ with $n\leq 1$ that do not contribute because they involve second or higher orders of $\delta$. That those terms can be present is caused by an asymmetry in $\delta$ and $\kappa$ due to the fact that we only perturb the gravitational degrees of freedom but not the matter ones and further that the individual terms in the action involve different powers of $\kappa$ from the beginning.
We realise that the linearised constraints $\delta{\cal C}_I$ are canonically conjugated to the reference fields ${\cal G}^I$ but in general there can be non-vanishing contribution in linear order coming from the Poisson bracket $\{\delta{\cal G}^I(x),\delta^2 {\cal C}_J(y)\}$. To ensure that we have a vanishing contribution in linear order we will use the dual observable map as discussed below. For this purpose we need to adapt the observable map and its dual to field theory and perturbation theory and for both maps we need the observable map up to second order. Taking this into account the observable formula up to second order reads
\begin{eqnarray}
\label{eq:LinObs}
&&{\cal O}_{f,\{T\}}(\delta\tau,\delta\sigma^a,\delta\xi^j)=\delta f +\delta^2f +\kappa\int\limits_{\sigma}d^3y\;\delta{\cal G}^I(y)\{ \delta {\cal C}_I(y),\delta f\}\\
&&
+\kappa\int\limits_{\sigma}d^3y\;\delta^2{\cal G}^I(y)\{ \delta {\cal C}_I(y),\delta f\}
+\kappa\int\limits_{\sigma}d^3y\;\delta{\cal G}^I(y)\Big(\{ \delta^2 {\cal C}_I(y),\delta f\}
+\{ \delta {\cal C}_I(y),\delta^2 f\}
\Big)\nonumber \\
&&+\frac{\kappa^2}{2}\int\limits_{\sigma}d^3y\;\delta{\cal G}^I(y)\int\limits_{\sigma}d^3z
\;\delta{\cal G}^J(z)\Big(\{\delta {\cal C}_J(z),\{ \delta^2 {\cal C}_I(y),\delta f\}\} 
+\{\delta {\cal C}_J(z),\{ \delta{\cal C}_I(y),\delta^2f\}\}
\Big)+O(\delta^3,\kappa^3)\nonumber,
\end{eqnarray}
where we used that $\overline{\cal G}=\overline{\cal G}^a=\overline{\cal G}^j=0$ and allowed possible non-vanishing second-order perturbations of ${\cal G}^I$ that might be present if one wants to relax the assumptions (i)-(iv) from above.
Because we consider perturbations around flat spacetime, in the background it holds that $\overline{C}=0, \overline{C}_a=0$ as well as $\overline{G}_i=0$, all trivially vanish and hence we also have $\overline{\cal G}=0, \overline{\cal G}^a=0$ and $\overline{\cal G}^j=0$. Hence, for the background there is no gauge freedom we have to deal with and thus no corresponding observables to construct. Note that if we construct the observable for elementary phase space variables then $\delta^2f=0$ and the observable formula above simplifies. This construction of observables order by order in $\kappa$ plays also a role when we consider the linearised Hamiltonian $\delta{\bf H}_{\rm can}$ which, as can be seen from \eqref{eq:totalhamlin}, has contributions in $\kappa^0$ and linear order in $\kappa$. The transformation behaviour under the linearised constraints of the matter variables can be found in the appendix \ref{app:DetailsRefFields} in \eqref{eq:PhiTrafoHam} and \eqref{eq:PhiTrafoDiffeo} and the results are an expression linear in $\kappa$. This again demonstrates that in the limit where $\kappa$ is sent to zero, which corresponds to the situation that we consider a scalar field on Minkowski only with no coupling to linearised gravity, the elementary variables $\phi,\pi$ are suitable observables. Once we consider the coupling with linearised gravity perturbations, this is no longer the case and we need the gauge invariant version of these variables. For the $\delta{\bf H}_{\rm can}$ this means that in the $\kappa^0$ term we can still work with the original $\phi,\pi$ whereas for the linear order in $\kappa$ we also need to construct Dirac observables for the matter sector, see for instance also the discussion in \cite{Giddings:2019wmj} in the context of the covariant theory.

Note that as before, the linearised phase space in the Post-Minkowski approximation scheme involves the linear perturbations of the gravitational degrees of freedom as well as the variables $(\phi,\pi)$ for the matter sector. In a similar way we obtain the linearised dual observable map given by
\begin{eqnarray}
\label{eq:LinObsdual}
&&{\cal O}^{\rm dual}_{f,\{C\}}=
\delta f +\delta^2f -\kappa\int\limits_{\sigma}d^3y\;\delta{\cal C}_I(y)\{ \delta {\cal G}^I(y),\delta f\}\\
&&
-\kappa\int\limits_{\sigma}d^3y\;\delta^2{\cal C}_I(y)\{ \delta {\cal G}^I(y),\delta f\}
-\kappa\int\limits_{\sigma}d^3y\;\delta{\cal C}_I(y)\Big(\{ \delta^2 {\cal G}^I(y),\delta f\}
-\{ \delta {\cal G}^I(y),\delta^2 f\}
\Big)\nonumber \\
&&+\frac{\kappa^2}{2}\int\limits_{\sigma}d^3y\;\delta{\cal C}_I(y)\int\limits_{\sigma}d^3z
\;\delta{\cal C}_J(z)\Big(\{\delta {\cal G}^J(z),\{ \delta^2 {\cal G}^I(y),\delta f\}\} 
+\{\delta {\cal G}^J(z),\{ \delta{\cal G}^I(y),\delta^2f\}\}
\Big)+O(\delta^3,\kappa^3)\nonumber,
\end{eqnarray}
The alternating signs compared to the observable map in \eqref{eq:ObsFormula} are needed since the order of how the constraints and the reference fields enter into the Poisson bracket is switched. Note that if we choose for instance ${\cal G}^I$ such that $\delta^2{\cal G}^I=0$ and not the non-linear ${\cal C}_I$ but only the linearised parts as the clocks then we drop all terms that involve $\delta^2{\cal G}^I,\delta^2{\cal C}_I$ in the dual observable map \eqref{eq:LinObsdual} and this will be exactly the application we need later. 

As we will work with so-called geometrical clocks, in a first step we will choose the set of reference fields among the linearised elementary gravitational degrees of freedom. As a consequence, we will work with geometrical clocks for which $\delta^2{\cal G}^I=0$ and hence they satisfy assumption (i). To also fulfil (ii) we further choose ${\cal G}^I$ such that 
 \begin{equation*}
\{\delta{\cal G}^I(x), \delta C_J(y)\} = \frac{1}{\kappa}\delta^I_J\delta^{(3)}(x,y),    
 \end{equation*}
where the factor $\frac{1}{\kappa}$ is chosen because it is also involved in Poisson bracket of the elementary geometrical variables. These requirements still allow several choices for suitable ${\cal G}^I$ and the specific choice for ${\cal G}^I$ taken here is motivated by the fact that we can relate the set of gauge invariant variables to a gauge fixing often chosen in the context of linearised gravity as will be discussed at the end of this subsection.
For the reference fields $\delta T(\vec{x},t), \delta\Xi^j(\vec{x},t)$, for the Hamiltonian constraint and the Gau\ss{} constraint respectively we choose
\begin{eqnarray}
\label{eq:GaussandHamclock}
   \delta T(\vec{x},t) &:=& -\frac{1}{\kappa\beta}  \left[\frac{1}{2} \delta_c^i \t{\epsilon}{_a^c^b} \partial_b \left(\t{\delta E}{^a_i} * G^\Delta \right)(\vec x,t) + \delta^a_i \left(\t{\delta A}{_a^i} * G^\Delta\right)(\vec{x},t)\right]\,,\\
  \delta \Xi^i(\vec{x},t)&:=&  \frac{2}{\kappa} \partial^a \left( \t{\delta A}{_a^i} * G^\Delta \right)(\vec{x},t),
\end{eqnarray}
with the three dimensional, spatial convolution
\begin{equation}
    \left(\t{\delta E}{^a_i} * G^\Delta \right)(\vec x,t) := \int d^3y \; \t{\delta E}{^a_i}(\vec y,t) \, G^\Delta(\vec{x}-\vec y)\,,
\end{equation}
the derivative of the convolution acting with respect to $\vec x$, hence only on $G^\Delta(\vec x -\vec y)$, and the Green's function of the Laplacian
\begin{equation}
    G^\Delta(\vec x-\vec y) := \int \frac{d^3k}{(2\pi)^3} \frac{1}{||\vec{k}||^2} e^{i\vec{k}(\vec{x}-\vec{y})}\,,
\end{equation}
such that $\Delta  G^\Delta(\vec x-\vec y) =-\delta(\vec x -\vec y)$.
Thus the reference fields are, as in general expected, non-local quantities because they are local in momentum space. As discussed in the appendix \ref{app:DetailsRefFields} these reference fields satisfy  
\begin{eqnarray*}
 \{\delta T(\vec{x},t),\delta C(\vec{y},t)\} &=& \frac{1}{\kappa}\delta(\vec{x}-\vec{y}) ,\quad 
\{\delta \Xi^i(\vec{x},t),\delta G_j(\vec{y},t)\} = \frac{1}{\kappa}\delta_j^i \delta(\vec{x}-\vec{y}), \quad  \{\delta T(\vec{x},t), \delta \Xi^i(\vec{y},t)\} =0\,. 
\end{eqnarray*}
Furthermore, the reference fields $\delta T,\delta \Xi^j$ commute with all remaining linearised constraints and hence they satisfy (iii) and (iv) as well. A more detailed discussion can be found in the appendix \ref{app:DetailsRefFields}. Reference fields that are canonically conjugate to the components of the spatial diffeomorphism constraint and commute with the linearised Hamiltonian as well as the linearised Gau\ss{} constraint are given by
\begin{equation}
    \delta\widetilde{T}^a(\vec{x},t):=  \frac{2}{\kappa} \left(  \delta^a_b \delta^i_c \partial^c -\frac{1}{2} \delta^i_b \partial^a + \delta^{ac}\delta_c^i \partial_b \right) \left(\delta \t{E}{^b_i} * G^\Delta\right)(\vec{x},t).
\end{equation} 
Although the  reference fields $ \delta\widetilde{T}^a(\vec{x},t)$ mutually commute, they do not have vanishing Poisson brackets with all the remaining reference fields and violate assumption (iv). To obtain reference fields for the spatial diffeomorphism constraint we can employ the dual observables map for
$\delta\widetilde{T}^a(\vec{x},t)$ yielding a quantity that by construction commutes with $\delta{\cal G}, \delta{\cal G}^a, \delta{\cal G}^j$ and hence with all reference fields. As the clock for the dual observable map we choose the geometric part of the linearised Hamiltonian constraint, that is $\delta C-\kappa\epsilon$ as well as the linearised Gau\ss{} constraint. Note that $\delta C-\kappa\epsilon\not\approx 0$ and as a consequence $\delta\widetilde{T}^a\not\approx\delta T^a$ but this causes no problems because two different choices of set of clocks need not necessarily to be weakly equivalent. In order that the observable formula can be applied we just need $\{\delta T^I,\delta{\cal C}_J\}=\{\delta {\cal G}^I,\delta{\cal C}_J\}=\delta^I_J$ which is satisfied also if we use the geometric contribution of $
\delta C$ only. 
For $ \delta\widetilde{T}^a(\vec{x},t)$ the part corresponding to the spatial diffeomorphism constraint will not contribute in the dual observable map. Since we use the linearised constraint as the clock and not ${\cal C}_I$ we can  drop all terms in involving $\delta^2C_I$. Using this together with $\delta^2{\cal G}^I=0$ the dual observable map in \eqref{eq:ObsFormulaDual} simplifies in this case to
\begin{eqnarray*}
&&{\cal O}^{\rm dual}_{\delta\widetilde{T}^a,\{\delta C\}}=:\delta T^a =
\delta \widetilde{T}^a  -\kappa\int\limits_{\sigma}d^3y\;(\delta C-\kappa\epsilon)(y)\{ \delta T(y),\delta \widetilde{T}^a\}
-\kappa\int\limits_{\sigma}d^3y\;\delta G_j(y)\{ \delta \Xi^j(y),\delta \widetilde{T}^a\}.
\end{eqnarray*}
The explicit result we obtain has the form
\begin{align}
\delta T^a(\vec{x},t) :=& \frac{2}{\kappa} \left(  \delta^a_b \delta^i_c \partial^c -\frac{1}{2} \delta^i_b \partial^a + \delta^{ac}\delta_c^i \partial_b \right) \left(\delta \t{E}{^b_i} * G^\Delta\right)(\vec{x},t) \nonumber\\ &+\frac{4\beta}{\kappa^2}\left[ \frac{1}{2} \delta^i_b \partial^a \partial^b \left( \delta G_i * G^{\Delta\Delta}\right)(\vec x,t) - \delta^{ab}\delta_b^i \left( \delta G_i * G^\Delta\right)(\vec x,t) \right]  + \frac{1}{\kappa^2} \partial^a \left[\left(\delta C-\kappa\epsilon\right) * G^{\Delta\Delta} \right](\vec{x},t)\,,
\end{align}
where $G^{\Delta\Delta}$ denotes the Green's function of the squared Laplacian, that is
\begin{equation}
    G^{\Delta\Delta}(\vec x-\vec y) = \int \frac{d^3k}{(2\pi)^3} \frac{1}{||\vec k||^4} e^{i\vec{k}(\vec x-\vec y)} = \int d^3z \; G^\Delta(\vec x -\vec z) \, G^\Delta(\vec y-\vec z)\,.
\end{equation}
Note that in \cite{Dittrich:2006ee} also geometrical clocks were used  based on the ADM clocks introduced in \cite{arnowitt1962dynamics} and \cite{kuchavr1970ground}. This set of clocks is not equivalent to ours and when requiring that all clocks vanish this corresponds to a  different gauge fixing, as is discussed at the end of appendix \ref{app:GaugeFixing}.\\
Now we are in the situation, that the seven linearised constraints as well as the seven reference fields form in lowest order two Abelian subalgebras and further obey the following Poisson algebra:
\begin{align}
\label{eq:PoissonAlgClocks}
\{\delta T(\vec{x},t),\delta C(\vec{y},t)\} &= \frac{1}{\kappa}\delta(\vec{x}-\vec{y}) & \{\delta T(\vec{x},t),\delta C_a(\vec{y},t)\} &= 0 & \{\delta T(\vec{x},t),\delta G_i(\vec{y},t)\} &= 0 \\
\{\delta T^a(\vec{x},t),\delta C(\vec{y},t)\} &= 0 & \{\delta T^a(\vec{x},t),\delta C_b(\vec{y},t)\} &= \frac{1}{\kappa}\delta_b^a \delta(\vec{x}-\vec{y}) & \{\delta T^a(\vec{x},t),\delta G_i(\vec{y},t)\} &= 0\\ 
\{\delta \Xi^i(\vec{x},t),\delta C(\vec{y},t)\} &= 0 & \{\delta \Xi^i(\vec{x},t),\delta C_a(\vec{y},t)\} &= 0 & \{\delta\Xi^i(\vec{x},t),\delta G_j(\vec{y},t)\} &= \frac{1}{\kappa}\delta_j^i \delta(\vec{x}-\vec{y})\,.
\end{align}
Since $\delta{\cal G},\delta{\cal G}^c,\delta{\cal G}^j$ differ from $\delta T,\delta T^c,\delta\Xi^j$ by some phase space independent quantity only, we can replace the reference fields by the corresponding $\delta{\cal G}^I$s and the Poisson algebra above will not change. We have defined the Poisson algebra with a factor $\frac{1}{\kappa}$ because the gravitational degrees of freedom involve a similar factor. The algebra reference fields have been defined such that the algebra of them and the linearised constraints no longer involves the Barbero-Immirzi parameter on the right hand side for the reason that $\kappa$ is the parameter labelling the order of perturbations.
Going back to the Poisson algebra of ${\cal G}^I$ and ${\cal C}_I$ in \eqref{eq:ReqPBAlgebra} so far we have chosen reference fields that satisfy this algebra. However, the in general non-vanishing contribution in linear order will prevent us from chosen ${\cal G}^I$ and ${\cal C}_J$ as new canonical coordinates. To achieve this we will apply the dual observable map to ${\cal C}_I$ and obtain an equivalent set of constraints ${\cal C}_I^\prime$ satisfying
\begin{equation}
\label{eq:ReqAlgCprime}
 \{{\cal G}^I(x),{\cal C}^\prime_J(y)\}=\{\delta{\cal G}^I(x),\delta {\cal C}^\prime_J(y)+\delta^2 {\cal C}^\prime_J(y)\}=\frac{1}{\kappa}\delta^I_J\delta^{(3)}(x,y) +O(\delta^2,\kappa^2)
\end{equation}
With this algebra given we can then choose $\delta{\cal G}^I$ as new configuration variables and $\delta{\cal C}^\prime_I$ as new momentum variables. Because in lowest order we have already the correct algebra relations we only need a modification in the linear order. To accomplish this we apply the dual observable map to all ${\cal C}_I$ and drop the linear order terms. The latter would just abelianise the lowest order and cancel the ${\cal C}_I$ contribution in the observable something we do not want. The second order that we keep will then modify the linear order in a way that we obtain the algebra shown in \eqref{eq:ReqAlgCprime}. Since we want an equivalent set of constraints here we are forced to choose as clocks the linearised constraints $\delta{\cal C}_I$ involving also the matter contributions in $\delta C$ and $\delta C_a$. Taking all this into account the equivalent constraints ${\cal C}_I^\prime$ are given by
\begin{eqnarray}
\label{eq:DefCIPrime}
&&{\cal C}_I^\prime:={\cal O}^{\rm dual}_{{\cal C}_I,\{\delta C\}}=
\delta{\cal C}_I+\delta^2{\cal C}_I
-\kappa\int\limits_{\sigma}d^3y\;\delta{\cal C}_I(y)\{ \delta {\cal G}^I(y),\delta^2 {\cal C}_I\}
\nonumber \\
&&+\frac{\kappa^2}{2}\int\limits_{\sigma}d^3y\;\delta{\cal C}_J(y)\int\limits_{\sigma}d^3z
\;\delta{\cal C}_K(z)\{\delta {\cal G}_J(z),\{ \delta{\cal G}_K(y),\delta^2{\cal C}_I\}\}
+O(\delta^3,\kappa^3),
\end{eqnarray}
where we used again that $\delta^2{\cal G}^I=0$ in our case. Up to linear order we have $\delta{\cal C}_I^\prime=\delta{\cal C}_I$ showing that as expected we have modified the constraint only in second order. This modifications become not relevant when we work with the constraints directly in the linearised theory since they are of second order but need to be considered when we compute Poisson brackets and want to consider the result up to linear order.
For a set of mutually commuting reference fields, that we consider here, in a more general context it was proven in \cite{Dittrich:2006ee} that the constraints ${\cal C}_I^\prime$ are Abelian up to corrections of second order and applying this result here we have 
\begin{equation*}
\{{\cal C}_I^\prime(x),  {\cal C}_J^\prime(y)\}=\{\delta{\cal C}_I^\prime(x)+\delta^2{\cal C}_I^\prime(x),  \delta{\cal C}_J^\prime(y)+\delta^2{\cal C}_J^\prime(y)\}=0+O(\delta^2,\kappa^2)  
\end{equation*}
and thus in the order of perturbation theory we consider here, we can treat them as Abelian constraints. Because we can just consider higher orders in the dual observable map we can always ensure that the constraints are Abelian up to a chosen order in perturbation theory and if the entire sum of the dual observable map converges even in the full unperturbed theory. Note that a similar result has already been discussed in \cite{Dittrich:2006ee} in the context of applying weak Abelianisation order by order in perturbation theory and afterwards modifying the constraints by terms that involve higher order powers of the linearised constraints. Here we can rediscover that case in the framework of the dual observable map which seems to us to be slightly more general because it cannot only be applied to the constraints as this corresponds to the case discussed in \cite{Dittrich:2006ee} but to any phase space function as we did for instance for the clock of the spatial diffeomorphism constraint before. Further, the strategy discussed in \cite{Dittrich:2006ee} separates weak Abelianisation and the addition of terms higher order in the linearised constraints where this happens all at once here using the dual observable map from second order on. 

Given the Abelian constraints ${\cal C}^\prime_I(x)$ now we can use them in the observable map in \eqref{eq:ObsFormula} and construct gauge invariant quantities that commute by construction with the constraints ${\cal C}_I^\prime(x)$ up to $O(\delta^2,\kappa^2)$ corrections. We want to apply the observable map to the elementary phase space variables of the matter and gravitational degrees of freedom and use $\delta{\cal G}^I$ as our clocks. Therefore, we can drop all terms involving $\delta^2f$ and 
$\delta^2{\cal G}^I$ and the observable map reduces to
\begin{eqnarray}
(\delta f)^{GI}(\delta\tau,\delta\sigma^a,\delta\xi^j)&:=&{\cal O}_{\delta f,\{\delta T\}}(\delta\tau,\delta\sigma^a,\delta\xi^j)\\
&=&{\cal O}^{(1)}_{\delta f,\{\delta T\}}(\delta\tau,\delta\sigma^a,\delta\xi^j)+{\cal O}^{(2)}_{\delta f,\{\delta T\}}(\delta\tau,\delta\sigma^a,\delta\xi^j)+O(\delta^3,\kappa^3)\nonumber,
\end{eqnarray}
where the label $GI$ means gauge invariant with
\begin{eqnarray}
\label{eq:ObsMapOurCase}
{\cal O}^{(1)}_{\delta f,\{\delta T\}}(\delta\tau,\delta\sigma^a,\delta\xi^j)&:=&\delta f +\kappa\int\limits_{\sigma}d^3y\;\delta{\cal G}^I(y)\{ \delta {\cal C}^\prime_I(y),\delta f\}\nonumber\\
{\cal O}^{(2)}_{\delta f,\{\delta T\}}(\delta\tau,\delta\sigma^a,\delta\xi^j)&:=&
\kappa\int\limits_{\sigma}d^3y\;\delta{\cal G}^I(y)\{ \delta^2 {\cal C}^\prime_I(y),\delta f\}
\nonumber \\
&&+\frac{\kappa^2}{2}\int\limits_{\sigma}d^3y\;\delta{\cal G}^I(y)\int\limits_{\sigma}d^3z
\;\delta{\cal G}^J(z)\Big(\{\delta {\cal C}^\prime_J(z),\{ \delta^2 {\cal C}^\prime_I(y),\delta f\}\} 
\Big)
\end{eqnarray}
where ${\cal O}^{(1)}_{\delta f,\{\delta T\}}$ and ${\cal O}^{(2)}_{\delta f,\{\delta T\}}$ denote all contributions to the observable in linear and second order respectively.
Since in this work we perturb the geometrical degrees of freedom only for the matter sector we just replace $\delta f$ by $f=\phi,\pi$ in the formula above. That this works so easily in the matter sector might not be obvious in the first place for the following reason: If we consider in  ${\cal O}^{(1)}_{\delta f,\{\delta T\}}$ the second term then for $\delta f$ chosen from geometric degrees of freedom this will be of zeroth order and hence trivially commute with the linearised or higher order constraints. In contrast, for the matter variables the corresponding term still involves the matter variables linearly and hence does not commute with the linearised constraints or higher order ones and thus yields an additional contribution compared to the geometric case. However, since the Poisson brackets of the gravitational degrees of freedom involve a factor $\frac{1}{\kappa}$, whereas the Poisson brackets of the  matter variables do not, these additional contributions come with an extra factor of $\kappa$. In the linearised theory we require for Dirac observables to commute with the constraints up to order $\delta$ and $\kappa$ and hence, these kind of contributions will only occur in higher orders in $\kappa$. A similar argument applies to these kind of additional contributions in ${\cal O}^{(2)}_{\delta f,\{\delta T\}}$ and therefore we can indeed apply the observable formula also to the matter variables. 
To obtain the explicit form of these Dirac observables we apply the observable map in \eqref{eq:ObsMapOurCase} to $(\phi,\pi,\t{\delta A}{_a^i},\t{\delta E}{^a_i})$ and obtain
\begin{align}
\phi^{GI}(\delta\sigma^c,\delta\tau,\delta\xi^j) &= \phi(\vec{x},t) - \kappa^2 (\delta T(\vec{x},t) - \delta\tau(\vec{x},t)) \pi(\vec{x},t)\label{phiGI}\\ 
& - \kappa^2 (\delta T^c(\vec{x},t) - \delta\sigma^c(\vec{x},t)) \partial_c \phi(\vec{x},t) \nonumber\\
&+{\cal O}^{(2)}_{\phi,\{\delta T\}}(\delta\tau,\delta\sigma^a,\delta\xi^j)+O(\delta^3,\kappa^3),\nonumber\\
\pi^{GI}(\delta\sigma^c,\delta\tau,\delta\xi^j) &= \pi(\vec{x},t) - \kappa^2 \partial_a [(\delta T(\vec{x},t) - \delta\tau(\vec{x},t)) \partial^a\phi(\vec{x},t)] \label{piGI} 
\\ & - \kappa^2 \partial_c[(\delta T^c(\vec{x},t) - \delta\sigma^c(\vec{x},t)) \pi(\vec{x},t)] + \kappa^2 (\delta T(\vec{x},t) - \delta\tau(\vec{x},t))\, m^2 \phi(\vec{x},t)\nonumber \\ &+{\cal O}^{(2)}_{\pi,\{\delta T\}}(\delta\tau,\delta\sigma^a,\delta\xi^j)+O(\delta^3,\kappa^3),\nonumber\\ \label{obsdelE}
(\t{\delta E}{^a_i})^{GI}(\delta\sigma^a,\delta\tau,\delta\xi^j) &= \t{\delta E}{^a_i}(\vec{x},t) - \frac{\kappa}{2}(\delta_i^a \partial_c -\delta_c^a \delta_i^b \partial_b)(\delta T^c(\vec{x},t) - \delta\sigma^c(\vec{x},t)) \\&+\kappa\beta \t{\epsilon}{_c^a^b} \delta_i^c \partial_b(\delta T(\vec{x},t) - \delta \tau(\vec{x},t))\nonumber -\frac{\kappa}{2} \t{\epsilon}{_i_j^k}  \delta_k^a (\delta \Xi^j(\vec{x},t)-\delta\xi^j(\vec{x},t))\nonumber \\ 
&+{\cal O}^{(2)}_{\t{\delta E}{^a_i},\{\delta T\}}(\delta\tau,\delta\sigma^a,\delta\xi^j)+O(\delta^3,\kappa^2), \nonumber\\ \label{obsdelA}
(\t{\delta A}{_a^i})^{GI}(\delta\sigma^a,\delta\tau,\delta\xi^j)&=\t{\delta A}{_a^i}(\vec{x},t) +\frac{\kappa}{2} \partial_a (\delta\Xi^i(\vec{x},t)-\delta\xi^i(\vec{x},t))\\ &+{\cal O}^{(2)}_{\t{\delta A}{_a^i},\{\delta T\}}(\delta\tau,\delta\sigma^a,\delta\xi^j)+O(\delta^3,\kappa^2)\nonumber \,,
\end{align}
where we displayed only the linear order in explicit form and $(\delta\tau(\vec{x},t), \delta\sigma^c(\vec{x},t), \delta\xi^j(\vec{x},t))$ denote the spacetime functions corresponding to the reference fields $(\delta T, \delta T^c, \delta\Xi^j)$, see $\delta{\cal G},\delta{\cal G}^c$ and $\delta{\cal G}^j$ in \eqref{eq:DefcalG}. The so constructed observables are Dirac observables for all values of the functions $(\delta\tau(\vec{x},t), \delta\sigma^c(\vec{x},t), \delta\xi^j(\vec{x},t))$ and thus each can be understood as a family of Dirac observables parameterised by $\delta\tau(\vec{x},t), \delta\sigma^c(\vec{x},t)$ and $\delta\xi^j(\vec{x},t)$ respectively. Now let us choose specific spacetime functions for $\delta\tau(\vec{x},t), \delta\sigma^c(\vec{x},t)$ and $\delta\xi^j(\vec{x},t)$. A specific choice for $\delta\tau(\vec{x},t), \delta\sigma^c(\vec{x},t)$ and $\delta\xi^j(\vec{x},t)$ can be chosen as follows: For the Gauß constraint we choose  vanishing parameters $\delta\xi^j(\vec{x},t)$. The remaining parameters we associate with the temporal and spatial coordinates respectively. Taking into account that the full non-linear parameters $\tau$ and $\sigma^c$ with $\tau=\overline{\tau}+\kappa \delta\tau + O(\kappa^2)$ and $\sigma^c=\overline{\sigma}^c+\kappa\delta\sigma^c + O(\kappa^2)$ should be set to $t$ and $x_\sigma^c$ respectively and that the clocks vanish in the background as well as for higher orders than the linear one, we only need to choose the linearised parameters $\delta \tau$ and $\delta \sigma^c$ yielding
\begin{equation}
\label{eq:Choicetausigma}
\delta\tau(\vec{x},t):= \frac{t}{\kappa},\quad \delta\sigma^c(\vec{x},t):=\frac{x^c_{\sigma}}{\kappa},\quad \delta\xi^j(\vec{x},t):=0,
\end{equation}
with $x^c_\sigma$ being the unique solution where $\delta T^c(\vec{x},t)=\frac{1}{\kappa}\sigma^c$ with $\sigma^c={\rm const}$. We reinsert these choices into $\delta{\cal G},\delta{\cal G}^c$ and $\delta{\cal G}^j$, respectively, in \eqref{eq:DefcalG}. Then we can understand the Dirac observables as functions of $(\vec{x}_\sigma,t)$.

We have that $(\phi^{GI},\pi^{GI})$ are already the independent Dirac observables in the matter sector. However, the 18 observables $(\t{\delta A}{_a^i})^{GI}, (\t{\delta E}{^a_i})^{GI}$ are not all independent, but only four of them are in total. For completeness we have presented the explicit forms of $(\t{\delta A}{_a^i})^{GI}$ and  $(\t{\delta E}{^a_i})^{GI}$ in Fourier space in \eqref{eq:dirobsmombE} and \eqref{eq:dirobsmombA} respectively. We realise that $\delta\tau$ and $\delta\sigma^c$ enter explicitly into these formulae. However, for the choice discussed here where $\delta\tau$ is linearly in $t$ and $\delta\sigma^c$ is constant  the contributions both vanish because they come with spatial derivatives acting on $\delta t$ and $\delta\sigma^c$. In case we choose $\delta\sigma^c$ linearly in $x^c$ then $(\t{\delta E}{^a_i})^{GI}$ has an additional term involving a Kronecker delta that also survives when we express $(\t{\delta E}{^a_i})^{GI}$ in terms of the independent gauge invariant degrees of freedom similar to what was observed in \cite{arnowitt1962dynamics}. As a consequence, considering $(\t{\delta E}{^a_i})^{GI}$ as an isolated quantity its fall-off behaviour gets modified and does no longer satisfy the requirement in \eqref{eq:FallOffdeltaE}. This, however causes no issue because we need this quantity as an intermediate step only to reinsert it into the physical Hamiltonian. The final result of the physical Hamiltonian has a suitable fall-off behaviour because these critical terms are combined with scalar field contributions for which we assumed, as usually done, that the initial data has compact support, see also our discussion below \eqref{eq:linHamredps}. The four independent degrees of freedom in the gravitational sector are encoded in the symmetric\footnote{If one considers the soldered version of the quantities, i.e. $\t{\delta \mathcal{A}}{_a_b} := \t{\delta \mathcal{A}}{_a^i} \delta_{i}^c\delta_{cb}$ and analogously for $\t{\delta \mathcal{E}}{^a^b}:= \t{\delta \mathcal{E}}{^a_i} \delta^{i}_c\delta^{cb}$.} transverse-traceless part of the the variables $(\t{\delta A}{_a^i})^{GI}, (\t{\delta E}{^a_i})^{GI}$ and therefore the 4 independent Dirac observables in the gravitational sector are given by
\begin{eqnarray}
\t{\delta{\cal A}}{_a^i}(\vec{x}_\sigma,t)&:=& \t{P}{_a^i_j^b}(\t{\delta{A}}{_b^j})^{GI}(\vec{x}_\sigma,t) \quad 
\t{\delta{\cal E}}{^a_i}(\vec{x}_\sigma,t):=\t{P}{^a_i^j_b}(\t{\delta{E}}{^b_j})^{GI}(\vec{x}_\sigma,t),
\end{eqnarray}
where $\t{P}{_a^i_j^b}$ and $\t{P}{^a_i^j_b}$ denote the projector on the transverse-traceless part in position space. Its explicit form is given by
\begin{align}\label{ttprojectorPositionSpace}
 \t{P}{_a^i_j^b} X(\vec x_\sigma,t)= & \delta^b_a \delta^i_j X(\vec x_\sigma,t) + \delta^i_j \partial_a \partial^b \left( X * G^\Delta\right)(\vec x_\sigma,t)  + \delta^b_a\delta^i_c \delta^d_j \partial^c \partial_d\left( X * G^\Delta\right)(\vec x_\sigma,t) \nonumber\\&-\frac{1}{2}\delta^b_j\delta^i_c \partial_a \partial^c\left( X * G^\Delta\right)(\vec x_\sigma,t)-\frac{1}{2} \delta_a^i \delta^b_j  X(\vec x_\sigma,t)+\frac{1}{2} \delta^i_c\delta^d_j \partial_a \partial^c \partial_d \partial^b\left( X * G^{\Delta\Delta}\right)(\vec x_\sigma,t) \nonumber\\&- \frac{1}{2} \delta^i_a\delta_j^c \partial_c \partial^b\left( X * G^\Delta\right)(\vec x_\sigma,t) + \frac{1}{2}\delta^i_d \delta_j^f \t{\epsilon}{^d_e_a} \t{\epsilon}{_c^b_f} \partial^c \partial^e\left( X * G^\Delta\right)(\vec x_\sigma,t)
\end{align}
and
\begin{equation}
    \t{P}{^a_i^j_b} X(\vec x_\sigma,t)=  \delta^{ac}\delta_{ik}\delta^{jm}\delta_{bd}\t{P}{_c^k_m^d} X(\vec x_\sigma,t)\,.
\end{equation}
As expected these projectors are non-local in position space but local in momentum space, see the appendix \ref{app:Basis} for more details. A comparison between the local projector in position space also often used in the literature in the context of the transverse-traceless gauge has been analysed in \cite{Ashtekar:2017wgq}.

 From now on we will drop the label $\sigma$ at $\vec{x}_\sigma$ and by an abuse of notation denote it by $\vec{x}$ again and always keep in mind that these $x^a$ coordinates are related to values that the reference field $T^a$ takes. By construction the Dirac observables Poisson commute with all constraints ${\cal C}_I^\prime$. Since the set ${\cal C}_I^\prime$ is weakly equivalent to the set ${\cal C}_I$ the Dirac observables are weak Dirac observables with respect to ${\cal C}_I$ where the weak equivalence refers to the constraint hypersurface defined by the linearised constraints $\delta{\cal C}_I$ that, as discussed above, agree with ${\delta C}^\prime_I$. Furthermore, as discussed in the appendix \ref{app:ConstrDiracObs} all Dirac observables truncated at linear order, that is neglecting contributions from ${\cal O}^{(2)}_{\delta f,\{\delta T\}}$, Poisson commute with all linearised constraints. What still remains to be discussed is the algebra of the independent Dirac observables. For this purpose we will consider the relation between the algebra of Dirac observables and the observable associated to the corresponding Dirac bracket. As proven in theorem 2.2 in \cite{Thiemann:2004wk}, see also \cite{Dittrich:2004cb} for an alternative proof in general we have 
\begin{equation*}
\{O_{f,\{T\}},O_{g,\{T\}} \}\approx O_{\{f,g\}^*,\{T\}}    
\end{equation*}
where $\{f,g\}^*$ denotes the Dirac bracket associated with the total set $({\cal G}^I,{\cal C}_J)$, that, when requiring ${\cal G}^I$ to be gauge fixing conditions is a second class system of constraints for which a Dirac bracket can be constructed. If we have such a relation also at the perturbative level then it provides an efficient way to compute the observable algebra because if we are interested in the algebra up to linear order then we can just expand the observable on the right hand side up to linear order whereas we need to consider the observables up to second order on the left hand side. In the following we will not consider the most general case here but just discuss the necessary result we need for the model derived in this work. 

As proven in the appendix under the assumption that we have linearised clocks, that is ${\cal G}^I=\delta{\cal G}^I$ with $\delta^k{\cal G}^I=0$ for all $k\geq 2$  and we consider observables of quantities $f=\delta f$ for which $\delta^k f=0$ with $k\geq 2$ then one can show that
\begin{equation}
\label{eq:RelDiracBPerturb}
\left\{O^{(1)}_{\delta f,\{\delta T\}}+O^{(2)}_{\delta f,\{\delta T\}}\, ,\, O^{(1)}_{\delta g,\{\delta T\}}+O^{(2)}_{\delta g,\{\delta T\}}\right \}=\{\delta f,\delta g\}^*+ O^{(1)}_{\{\delta f,\delta g\}^*,\{\delta T\}}  +O(\delta^2,\kappa^2),   
\end{equation}
where the Dirac bracket $\{\delta f,\delta g\}^*$ corresponds to the set $({\cal G}^I=\delta{\cal G}^I,{\cal C}_I^\prime={\cal C}^{\prime (1)}_I+{\cal C}^{\prime (2)}_I)$. Again as above because we do not perturb the matter degrees of freedom we can apply the same result here if we replace $\delta f$ by $f$ again with $f=\phi,\pi$. As discussed in appendix \ref{app:AlgRelDiracBracket} the case interesting for us is when $f,g$ are the elementary phase space variables of the linearised theory. Given this and using that for the reference fields we have ${\cal G}^I=\delta{\cal G}^I$ as well as $\{{\cal G}^I(x),{\cal C}^\prime_J(y)\}=\frac{1}{\kappa}\delta^I_J\delta^{(3)}(x,y)+O(\delta^2,\kappa^2)$ the explicit form of the Dirac bracket up to linear order reads
\begin{eqnarray}
\{\delta f(x),\delta g(y)\}^*&:=&
\{\delta f(x), \delta g(y)\}
-\kappa\int\limits_\sigma d^3z\{\delta f(x), \delta{\cal C}^\prime_L(z)\}\delta^L_M\{\delta{\cal G}^M(z),\delta g(y)\}\nonumber \\
&&-\kappa\int\limits_\sigma d^3z\{\delta f(x), \delta^2{\cal C}^\prime_L(z)\}\delta^L_M\{\delta{\cal G}^M(z),\delta g(y)\}
+\kappa\int\limits_\sigma d^3z\{\delta g(x), \delta{\cal C}^\prime_L(z)\}\delta^L_M\{\delta{\cal G}^M(z),\delta f(y)\}\nonumber \\
&&+\kappa\int\limits_\sigma d^3z\{\delta g(x), \delta^2{\cal C}^\prime_L(z)\}\delta^L_M\{\delta{\cal G}^M(z),\delta f(y)\}+O(\delta^2,\kappa^2).\nonumber 
\end{eqnarray}
Now for $\delta f$ and $\delta g$ that have the property that they commute with all $\delta{\cal G}^I$ the Dirac bracket reduces to the ordinary Poisson bracket. If $\delta f,\delta g$ are further elementary phase space variables that satisfy standard canonical Poisson brackets then their Poisson bracket is phase space independent and as a consequence, the Dirac observable of the Dirac bracket has only a zeroth order contributions that agrees with the original Poisson bracket. As shown in appendix \ref{app:PhysDofCommClocks} the variables $ (\phi,\pi), (\t{\delta{\cal A}}{_a^i},\t{\delta{\cal E}}{^a_i})$ all commute with all $\delta {T}^I$ which is equivalent that they commute with all $\delta{\cal G}^I$. Thus, for these set of variables the Dirac bracket agrees with the original Poisson bracket. Consequently, we can immediately conclude that the non-vanishing Poisson brackets of the set of Dirac observables $\phi^{GI},\pi^{GI},\t{\delta{\cal A}}{_a^i},\t{\delta{\cal E}}{^a_i}$ read
\begin{eqnarray}
\{\phi^{GI}(\vec{x},t),\pi^{GI}(\vec{y},t)\}&=&\delta(\vec{x}-\vec{y})+O(\delta^2,\kappa^2),\\
\{\t{\delta{\cal A}}{_a^i}(\vec{x},t),\t{\delta{\cal E}}{^b_j}(\vec{y},t)\}&=&\frac{\beta}{\kappa} \t{P}{_a^i_j^b} \delta(\vec{x}-\vec{y})+O(\delta^2,\kappa^2).\,
\end{eqnarray}
All remaining ones vanish up to corrections of order $O(\delta^2,\kappa^2)$. The matter variables satisfy the same Poisson algebra as their gauge variant counter parts and for the geometric gauge invariant degrees of freedom the Poisson bracket involves on the right hand side as well the expected projector on the transverse and traceless part. Considering this together with the Poisson algebra of the ${\cal G}^I$ and ${\cal C}_I^\prime$ in \eqref{eq:ReqAlgCprime} we realise we can perform a canonical transformation from the set of variables 
\begin{equation*}
 (\phi,\pi), \quad (\t{\delta{A}}{_a^i},\t{\delta{E}}{^a_i})
\end{equation*}
to the new set of variables
\begin{equation*}
(\delta{\cal G},C^\prime),\quad  (\delta {\cal G}^a,C^\prime_a),  \quad
(\delta {\cal G}^j,G^\prime_j), \quad (\phi^{GI},\pi^{GI}), \quad (\t{\delta{\cal A}}{_a^i},\t{\delta{\cal E}}{^a_i})
\end{equation*}
with
\begin{equation}
\label{eq:GFCondClocks}
\delta{\cal G}=\delta T -t,\quad \delta{\cal G}^a=\delta T^a-x^a_\sigma,\quad \delta {\cal G}^j=\delta\Xi^j.    
\end{equation}
Because the definition of ${\cal C}_I^\prime$ as can be seen from \eqref{eq:DefCIPrime} involves second-order contributions to implement this canonical transformation on the linearised phase space we restrict ${\cal C}_I^\prime$ to its linear part and using that in our case $\delta{\cal C}^\prime_I=\delta{\cal C}_I$ the new set of canonical variables on the linearised phase space is given by
\begin{equation*}
(\delta{\cal G},\delta C),\quad  (\delta {\cal G}^a,\delta C_a),  \quad
(\delta {\cal G}^j,\delta G_j), \quad (\phi^{GI},\pi^{GI}), \quad (\t{\delta{\cal A}}{_a^i},\t{\delta{\cal E}}{^a_i})\,,
\end{equation*}
the explicit transformation can be found in appendix \ref{app:Trafo}. An advantage of this new set of elementary phase space variables is that it allows to clearly separate the physical degrees of freedom from the remaining gauge degrees of freedom at the level of the full phase space. The physical phase space corresponds to the subspace involving the subalgebra of all Dirac observables $(\phi^{GI},\pi^{GI}), (\t{\delta{\cal A}}{_a^i},\t{\delta{\cal E}}{^a_i})$ including the expected six physical degrees of freedom. The price to pay in this context is that, as far as the position space is considered, we need to work with highly non-local clocks that become local in momentum space. This is not unexpected here considering the fact that the projector to the transverse-traceless part of the gravitational perturbations is local in momentum but non-local in position space\footnote{In \cite{Ashtekar:2017wgq} it is discussed that there are situations particularly in the case of sources where a replacement of the non-local projector in position space by the also widely used local projector can yield different physical effects.}. 

Finally, let us briefly sketch the strategy we followed here that can also be applied to a different choice of reference fields.
Suppose we are interested in perturbation theory up to linear order.
\begin{itemize}
    \item First choose a set of linearised reference fields, that is $\delta^2{\cal G}^I=0$ such that they build canonical pairs with the linearised constraints, that is $\{\delta\widetilde{\cal G}^I(x),\delta{\cal C}_J(y)\}=\delta^I_J\delta^{(3)}(x,y)$.
    \item In case not all reference fields and hence $\delta\widetilde{\cal G}^I(x)$ mutually commute then apply the dual observable map with the linearised constraints as the clocks up to second order\footnote{Note for this choice of clocks and since we apply it to the linearised clocks only the linear terms in the power series will contribute.} to those that do not to obtain ${\cal G}^I$.
    \item Next apply the dual observable map up to second order neglecting the linear order to all constraints ${\cal C}_I$ to get the constraints ${\cal C}_I^\prime$ that are Abelian and canonically conjugate to the reference fields ${\cal G}^I$ up to corrections of second order. 
    \item Define the observable map by means of the constraints ${\cal C}_I^\prime$ using as reference fields ${\cal G}^I$.
    \item Choose next to $(\delta{\cal G}^I,\delta{\cal C}^\prime_I)$ further independent phase space variables on the linearised phase space denoted by $\delta q^I,\delta p_J$. 
    \item Apply the observable observable map to $(\delta q^I,\delta p_I)$ to obtain the physical gauge invariant degrees of freedom. 
    \item Compute the algebra of the Dirac observables  of $(\delta q^I,\delta p_I)$ using \eqref{eq:RelDiracBPerturb}.
    \item If we further apply the dual observable map to these Dirac observables and they still satisfy the same Poisson algebra as their gauge variant counterparts we can also say that given the set of variables  $(\delta{\cal G}^I,\delta{\cal C}^\prime_I,\delta q^I,\delta p_I)$ then applying the observable map and its dual to the entire set with the modification that we exclude the linear order in the observable map and its dual for the set $(\delta{\cal G}^I,\delta{\cal C}^\prime_I)$ defines a canonical transformation on the entire phase space. 
\end{itemize}
In our case the final application of the dual observable map on the variables $(\delta q^I,\delta p_J)$ corresponding to $\phi^{\rm GI},\pi^{\rm GI},\delta{\cal A}^i_a,\delta{\cal E}^a_i$ acts trivially since the original gauge variant quantities already commute with our chosen geometrical clocks. We expect that one can extend this strategy without any problems to higher orders in perturbation theory and for the aspect of constructing abelian constraints this has be done in \cite{Dittrich:2006ee}. How and if also the further steps can be generalised to higher order goes beyond the scope of this work and  will be discussed more in detail elsewhere. 

The work in \cite{anastopoulos2013master} also aims at constructing Dirac observables in the framework of the ADM formalism. In appendix \ref{app:ComparisonDiracObs} we compare their construction to the one presented here. To our understanding the explicit form of the observable for the scalar field momentum presented in \cite{anastopoulos2013master} is not invariant under transformations generated by the linearised spatial diffeomorphism constraint.
~\\
~\\
We will close this subsection with discussing the dynamics of our constructed Dirac observables. This is the last ingredient missing in order to take the gauge invariant formulation as the starting point for the quantisation in the next section. As we know from full general relativity, the dynamics of Dirac observables cannot be generated by the canonical Hamiltonian because by construction, all Dirac observables commute with the constraints and $\mathbf{H}_{\rm can}$ is just a linear combination of the smeared  constraints. The generator of the dynamics of the Dirac observables is the so-called physical Hamiltonian \cite{Thiemann:2004wk}. If we rewrite the Hamiltonian constraint linearly in the clock momentum of the temporal reference field, that is $C=P_T+h$, then the physical Hamiltonian corresponds to the Dirac observable associated with the phase space function $h$. Here we need to adapt this to the framework of perturbations theory and consider the fact that in addition we have an interaction Hamiltonian as well as a non-vanishing Hamiltonian in $\kappa^0$ order. The perturbed Hamiltonian constraint up to second order is given by $C=\delta C+\delta^2C^{\rm geo}+O(\delta^2,\kappa^2)$. The reference field $\delta T$ is canonically conjugate to the linearised Hamiltonian constraint $\delta C$ and thus we can identify $\delta C$ with the momentum variable conjugate to $\delta T$ that we denote by $\delta P_T$, then we have $C=\delta P_T+\delta^2C^{\rm geo}+O(\delta^2,\kappa^2)$\footnote{Note that the boundary term $T[N]$ in $\delta H_{can}$ cancels exactly the contribution $\overline{N}\delta C=\delta C$ in $\delta H_{can}$ but since the clock momentum $\delta P_T=\delta C$ is anyway not part of the physical Hamiltonian this causes no problems.}. The Dirac observable associated with $\delta^2C^{\rm geo}$ corresponds to that part of the physical Hamiltonian that is generating the dynamics of the purely gravitational physical degrees of freedom. The Dirac observable corresponding to the interaction Hamiltonian ${\cal H}_I=\frac{\kappa}{2}\delta h_{ab}T^{ab}$ with $\delta h_{ab} = -\delta_b^i\delta_{ac} \delta\t{E}{^c_i}-\delta_a^i\delta_{bc} \delta\t{E}{^c_i}+\delta_{ab} \delta_c^i \t{\delta E}{^c_i}$ encodes the interaction between the matter and gravitational physical degrees of freedom. As we will see below when expressed in terms of the independent Dirac observables it decomposes into a term where the physical triad variables $\delta{\cal E}\t{}{^a_i}$ couple to the physical matter variables as well as a self-interaction term that involves a coupling between the physical matter variables only. Finally, in $\kappa^0$-order the physical Hamiltonian involves the matter Hamiltonian for the scalar field on Minkowski spacetime. Summarising the above discussion we denote the physical Hamiltonian of the model considered in this work by $\delta{\bf  H}$ that takes the following form
 \begin{equation}
 \label{eq:PhysHam}
 \delta{\bf  H} 
 ={\cal H}_{\phi} + O_{{\cal H}_{\rm geo},\{T\}}+O_{{\cal H}_I,\{T\}}\quad{\rm with}\quad {\cal H}_{\rm geo}:=\int\limits_\sigma d^3x\; \delta^2C^{\rm geo}(x),\quad {\cal H}_\phi:=\int\limits_\sigma d^3x\; \epsilon(\vec{x},t).
 \end{equation}
As can be seen in \eqref{eq:linHamredps} it can be written entirely in terms of the Dirac observables $(\phi^{GI},\pi^{GI})$, $(\t{\delta{\cal A}}{_a^i},\t{\delta{\cal E}}{^a_i})$ and therefore is by construction gauge invariant up to corrections of order $O(\delta^2,\kappa^2)$. Here, as discussed above, we insert the Dirac observables in all terms being linearly in $\kappa$ but keep $\phi,\pi$ in the $\kappa^0$ contribution because at zeroth order $\phi,\pi$ are suitable observables. 
Because in second-order perturbation theory the constraints ${\cal C}^\prime_I$ are only weakly equivalent to the original constraints ${\cal C}_I$ and the observable of the interaction Hamiltonian ${\cal H}_I$ is linearly in the perturbations we need to consider the Dirac observable of $\t{\delta E}{^a_i}$ up to second order in ${\cal H}_I$ when computing Poisson brackets with the constraints. Hence, as far as the set of original constraints ${\cal C}_I$ are considered $\delta {\bf H}$ is a weak Dirac observable.

Explicitly we obtain, see appendix \ref{app:CanHamNewVariables}:
\begin{align}\label{eq:linHamredps}
\delta\mathbf{H} = &\int_{\mathbb{R}^3} d^3x \;\epsilon(\vec{x},t) \nonumber\\
&+\kappa \int_{\mathbb{R}^3}d^3k\; \frac{1}{2\beta^2}\, \sum_{r\in\{\pm\}}\Big( \ft{\delta A}^r(\vec{k},t) \,\ft{\delta A}^r (-\vec{k},t)+  2r\,||\vec{k}|| \ft{\delta E}^r(\vec{k},t) \, \ft{\delta A}^r(-\vec{k},t) \nonumber\\ &\hspace{1.8in}+ (\beta^2+1) ||\vec{k}||^2  \ft{\delta E}^r(\vec{k},t)\,  \ft{\delta E}^r(-\vec{k},t)\Big) \nonumber\\
&+\kappa \int_{\mathbb{R}^3} d^3k \;\delta_{b}^i \ft{\varphi} {}_a^b(\vec{k},t) \, \ft{\delta \mathcal{E}}\t{}{^a_i}(-\vec{k},t)\nonumber\\
&{+\kappa \int d^3k  \frac{ik_b}{2} \kappa\ft{\delta \sigma}^{\prime a} \left[ \ft{\varphi}{}_a^b - \delta_a^b \left( \ft{\varphi}{}_c^c +m^2 \ft{\widetilde{V}}-\ft{\epsilon} \right)\right]+\kappa \int d^3k \left[ 2i \kappa ||\vec k|| \ft{\delta \tau} \hat{k}^b \ft{p}'{}_b \right]}\nonumber\\
&-\kappa\int_{\mathbb{R}^3} \frac{d^3k}{||\vec{k}||^2} \Bigg\{- \ft{p}\t{}{_c}(\vec{k},t) \,\ft{p}\t{}{_d}(-\vec{k},t) \left[ 4\delta^{cd}- \hat{k}^c \hat{k}^d \right] + \ft{\epsilon}(\vec{k},t)\,\ft{\epsilon}(-\vec{k},t) \left[\frac{1}{4}+\frac{3}{2}\right] \nonumber\\
&\hspace{1in}-\frac{3}{2} m^2\,\ft{\widetilde{V}}(\vec{k},t) \,\ft{\epsilon}(-\vec{k},t)-\ft{\varphi} {}_a^a(\vec{k},t) \,\ft{\epsilon}(-\vec{k},t) \Bigg\}\,,
\end{align}
with
\begin{align}
    \epsilon(\vec{x},t)&:=\epsilon(\phi^{GI}(\vec{x},t),\pi^{GI}(\vec{x},t)),  & p_a(\vec{x},t)&:=p_a(\phi^{GI}(\vec{x},t),\pi^{GI}(\vec{x},t)), \\
    \widetilde{V}(\vec{x},t) &:= \phi^{GI}(\vec{x},t)^2, & \varphi_a^b(\vec{x},t) &:= (\partial_a \phi^{GI}(\vec{x},t))(\partial^b \phi^{GI}(\vec{x},t))\,,
\end{align}
where as before we considered terms up to linear order in $\kappa$ only. Note that an underbow denotes the spatial Fourier transform of the corresponding quantity, i.e.
\begin{equation*}
    \ft{f}(\vec{k},t):=\frac{1}{(2\pi)^{\frac{3}{2}}}\int_{\mathbb{R}^3}d^3x\; e^{-i\vec{k}\vec{x}} f(\vec{x},t), 
    \qquad \hat{k}^a := \frac{1}{||\vec{k}||} k^a.
\end{equation*}
Note that for the $\kappa^0$ order we also use the operator valued distributions of the physical matter variables because the difference to the gauge variant variables is of higher order than $\kappa^0$ and is neglected to obtain a model that is consistent with the Post-Minkowski approximation order by order in $\kappa$. 
As can be seen from the explicit form of the physical Hamiltonian in \eqref{eq:linHamredps} for our choice of $\delta\tau$ and $\delta\sigma^c$ in \eqref{eq:Choicetausigma}, $\delta\tau$ being linear in $t$ does not contribute to the physical Hamiltonian and $\delta\sigma^c$ is also absent if we chose it to be constant and enters via Kronecker delta in case we choose it to be linearly in $x^c$ in exact agreement to what was found in \cite{arnowitt1962dynamics} for the ADM variables. This additional Kronecker delta contribution for the second choice combines with terms including the scalar field and its derivatives respectively and has thus a suitable fall-off behaviour.

The different contributions of the Hamiltonian in the individual lines can be interpreted as follows: The first line corresponds to the Hamiltonian of a scalar field on a Minkowski background. The next two lines encode the Hamiltonian of a gravitational field in vacuum, the third integral denotes the interaction between the gravitational degrees of freedom and the matter field and finally the remaining contributions encode the gravitational self-interaction of the scalar field. Even though the latter does not contain any gravitational degrees of freedom, it only appears due to the coupling between matter and gravity and results from the coupling of the gauge degrees of freedom in the gravitational sector to the scalar field. Once these gauge degrees of freedom are expressed in terms of the independent physical degrees of freedom we end up with this result. The latter contribution was neglected in \cite{Dittrich:2006ee} by using the argument that only the transverse-traceless components need to be considered. While in the vacuum case one can use a gauge fixing that sets all but the transverse-traceless gravitational degrees of freedom to zero, this is no longer a valid choice for a gauge-fixing if we couple the scalar field, because the constraints involve additional contributions from the scalar field and suitable gauge fixing will result in such a contribution of gravitational self-interaction of the scalar field. Hence, to our understanding, it is not justified to neglect this contributions in the order of perturbation theory considered in \cite{Dittrich:2006ee}.
 This finalises the discussion on the classical model that will be the starting point for the Fock quantisation in the next section. Having identified the physical degrees of freedom in the model, now we are in the situation to separate the total system into a system and an environmental part, as usually done in the framework of decoherence models. We choose the following separation: 
\begin{equation*}
{\rm system:}\,\, \phi^{GI},\, \pi^{GI}  \quad\quad {\rm environment:}\,\, \t{\delta{\cal A}}{_a^i},\, \t{\delta {\cal E}}{^a_i}
\end{equation*}
and hence the gauge invariant matter sector becomes the system and the physical degrees freedom in the gravitational sector provide the environment. 
~\\
~\\
Finally let us comment on the relation of the fully gauge invariant framework to the gauge fixing discussed in the appendix \ref{app:GaugeFixing}. As discussed for instance in \cite{Giesel:2007wi} for certain choices of the ${\cal G}^I$ one can relate a model formulated in terms of fully gauge invariant quantities to a corresponding gauge-fixed model. The observables constructed here also fall into this class of models, which means that the gauge-fixed Hamiltonian and the gauge invariant physical Hamiltonian formally agree if we replace all gauge invariant quantities by their gauge-fixed counter parts. Practically, this can be achieved by setting ${\cal G}={\cal G}^a={\cal G}^j=0$, that is strongly equal to zero and then we can formally identify the Dirac observables with the gauge-fixed quantities involved in the gauge fixing discussed in the appendix \ref{app:GaugeFixing}. For this specific gauge fixing, the constructed observables have a standard interpretation. The Hamiltonian of the gauge fixed theory can be obtained by inserting all constraints, gauge fixing conditions as well as the Lagrange multipliers that ensure the stability of the gauge fixing into $\delta \mathbf{H}_{\rm can}$. As can be seen from the explicit form of $\delta \mathbf{H}_{can}$ in \eqref{eq:deltaHcanNewVar} the Hamiltonian of the gauge fixed theory agrees formally with $\delta\mathbf{H}$ under the identification of the Dirac observables with their corresponding gauge-fixed quantities. In order to compare our results with the existing literature we discuss in appendix \ref{app:GFADMclocks} the geometrical clocks used in \cite{Dittrich:2006ee} that are denoted as ADM clocks there because they were introduced in the seminal paper \cite{arnowitt1962dynamics}. Also this set of geometrical clocks encodes the physical gravitational degrees of freedom in the symmetric transverse-traceless linearised connection and triad variables. The main difference we see among the two sets of geometrical clocks is how the reference field associated with the linearised Gau\ss{} constraint is chosen. In our work we choose a Lorentz-like condition if we set the corresponding gauge fixing condition $\delta{\cal G}^j=0$, whereas in \cite{Dittrich:2006ee} the resulting $\delta {\cal G}^j=0$ is only equivalent to a Lorentz-like condition in the vacuum case but not if we couple a scalar field. Then the resulting reference fields for the linearised spatial diffeomorphism and Hamiltonian constraint also differ because for both sets of clocks we require that the all geometrical clocks mutually commute. The physical Hamiltonian for the ADM clocks is also presented in \eqref{app:GFADMclocks} and it turns out that compared to the physical Hamiltonian in \ref{eq:linHamredps} it only differs in the term describing the self-interaction of the scalar field. There we find that it differs only by numerical factors in all but one term and it involves one additional term where in our case only a trace enters, but for the ADM clocks an additional contraction with the Fourier basis is present, see \eqref{eq:gfhamditadm} in appendix \ref{app:GFADMclocks}.
~\\
~\\
However, working at the gauge invariant level and not choosing one specific gauge fixing allows us also to choose a different gauge fixing than the one discussed in the appendix \ref{app:GaugeFixing} and then the observables and their dynamics discussed here are still valid because everything was formulated in a gauge invariant manner. The only difference for other gauge fixing choices is that the interpretation of the gauge invariant Dirac observables and their relation to the gauge-fixed quantities is modified because then, the gauge fixing does not necessarily correspond to setting all ${\cal G}^I$ equal to zero. Hence in this sense, if we consider a class of models where a relation to one gauge fixing is possible, then there is always a convenient choice for the reference fields in order to make such a relation as simple as possible. This was exactly our motivation for choosing the reference fields for this model the way we did, apart from the additional requirements (i) to (iii) listed below \eqref{eq:ObsFormulaDual}.

\section{Fock quantisation of the model}
\label{sec:QuantumModel}
In this section we present the results of a Fock quantisation of the model under consideration. We work with units where $\hbar = c = 1$. For convenience we choose to quantise the following scalar and tensor fields:
\begin{equation}
\phi^{GI}(\vec{x},t) \hspace{0.2in} \pi^{GI}(\vec{x},t) \hspace{0.2in} \delta \mathcal{E}\t{}{^a_i} (\vec{x},t) \hspace{0.2in} \delta \mathcal{C}\t{}{_a^i}(\vec{x},t)\,, 
\end{equation}
where the latter is the Fourier transform of $\ft{\delta \mathcal{C}}\t{}{_a^i}(\vec{k},t) = \ft{\delta \mathcal{C}}^+(\vec{k},t) m_a(\vec{k})m^i(\vec{k}) + \ft{\delta \mathcal{C}}^-(\vec{k},t) \overline{m}_a(\vec{k})\overline{m}^i(\vec{k})$ with $\ft{\delta \mathcal{C}}^\pm(\vec{k},t)=-\frac{1}{\beta}\left(\ft{\delta A}^\pm(\vec{k},t) \pm ||\vec{k}||\, \ft{\delta E}^\pm(\vec{k},t) \right)$; see appendix \ref{app:Basis} for the definition of the basis $(\hat{k}_a, m_a(\vec k), \overline{m}_a(\vec k))$ in Fourier space. The reason for this choice is that by using $\delta \mathcal{C}$ instead of $\delta \mathcal{A}$ the terms in the linearised Hamiltonian containing gravitational degrees of freedom only in \eqref{eq:linHamredps} can be rewritten in the following way:
\begin{align}
\kappa \int_{\mathbb{R}^3} d^3x \frac{1}{2} \left[\delta^{ab}\delta_{ij} \delta \mathcal{C}\t{}{_a^i}(\vec{x},t) \, \delta \mathcal{C}\t{}{_b^j}(\vec{x},t) +\delta_{bc}\delta^{ij} (\partial_a \delta \mathcal{E}\t{}{^b_i}(\vec{x},t)) (\partial^a \delta \mathcal{E}\t{}{^c_j} (\vec{x},t)) \right]\nonumber\\
=\kappa \int_{\mathbb{R}^3} d^3x \sum\limits_{r\in\{+,-\}} \frac{1}{2} \left[ \delta \mathcal{C}^r(\vec{x},t) \, \delta \mathcal{C}^r(\vec{x},t) + (\partial_a \delta E^r(\vec{x},t)) (\partial^a \delta E^r (\vec{x},t)) \right]\,,
\end{align}
which has the same form as the energy density of two massless scalar fields $\delta E^\pm(\vec{x},t)$ since 
\begin{equation}
    \{\delta E^\pm(\vec{x},t), \delta \mathcal{C}^\pm(\vec{y},t)\} = -\frac{1}{\beta}\{\delta E^\pm(\vec{x},t), \delta A^\pm(\vec{y},t)\} = \frac{1}{\kappa} \delta^3(\vec{x}-\vec{y})\,.
\end{equation}
A mode expansion of the fields and the Fock quantisation yield the following operator-valued distributions for the physical degrees of freedom:
\begin{align}
\phi^{GI}(\vec{x},t) &=\int_{\mathbb{R}^3} \frac{d^3k}{(2\pi)^{\frac{3}{2}}}\,\frac{1}{\sqrt{2\omega_k}} \left[a_{k}\, e^{-i\omega_k t +i\vec{k}\vec{x}} + {a}_k^\dagger\, e^{i\omega_k t - i\vec{k}\vec{x}}\right]\\
\pi^{GI}(\vec{x},t) &= \int_{\mathbb{R}^3} \frac{d^3k}{(2\pi)^{\frac{3}{2}}}\, (-i)\sqrt{\frac{\omega_k}{2}} \left[a_{k}\, e^{-i\omega_k t +i\vec{k}\vec{x}} - {a}_k^\dagger\, e^{i\omega_k t - i\vec{k}\vec{x}}\right]\\
\delta \mathcal{E}\t{}{^a_i}(\vec{x},t) &= \int_{\mathbb{R}^3} \frac{d^3k}{(2\pi)^\frac{3}{2}}\,\sqrt{\frac{1}{2\kappa\Omega_k}} \sum\limits_{r\in\{\pm\}} \left[[P^r(-\vec{k})]^a_i \: b_{k}^r\, e^{-i\Omega_k t+i\vec{k}\vec{x}} +[P^r(\vec{k})]^a_i \: (b_k^r)^\dagger \, e^{i\Omega_kt-i\vec{k}\vec{x}} \right]\\
\delta \mathcal{C}\t{}{_a^i}(\vec{x},t) &= \int_{\mathbb{R}^3} \frac{d^3k}{(2\pi)^\frac{3}{2}}\,(-i) \sqrt{\frac{\Omega_k}{2\kappa}} \sum\limits_{r\in\{\pm\}} \left[[P^r(-\vec{k})]_a^i \:b_{k}^\pm\, e^{-i\Omega_k t+i\vec{k}\vec{x}} -[P^r(\vec{k})]_a^i \:(b_k^\pm)^\dagger\, e^{i\Omega_kt-i\vec{k}\vec{x}}\right]\,,
\end{align}
where the $a_k^{(\dagger)}$ denote annihilation (creation) operator-valued distributions for scalar particles while the $(b_k^\pm)^{(\dagger)}$ denote annihilation (creation) operator-valued distributions for gravitons with polarisation label $\pm$. We have also introduced $\omega_k := \sqrt{\vec{k}^2+m^2}$ and $\Omega_k := \sqrt{\vec{k}^2}$ as well as the transverse-traceless  projectors $[P^r(\vec{k})]_a^i$ defined in \eqref{transverseprojector}, which play the role of the polarisation tensors of the quantised fields. From their definitions follows that $\overline{[P^r(\vec{k})]^a_i} = [P^r(-\vec{k})]^a_i$ and hence $\delta \mathcal{E}\t{}{^a_i}(\vec{x},t)^\dagger = \delta \mathcal{E}\t{}{^a_i}(\vec{x},t)$ and $\delta \mathcal{C}\t{}{_a^i}(\vec{x},t)^\dagger = \delta \mathcal{C}\t{}{_a^i}(\vec{x},t)$. Note that in contrast to \cite{anastopoulos2013master} the polarisation tensors are different for the positive and negative frequency modes. The creation and annihilation operator-valued distributions satisfy the following commutation relations:
\begin{align}
[a_k, a_l^\dagger] &= \delta^3(\vec{k}-\vec{l}) & [a_k, a_l] = [a_k^\dagger, a_l^\dagger] &=0 \\
[b_k^\pm, (b_l^\pm)^{(\dagger)}]& = \delta^3(\vec{k}-\vec{l}) &[b_k^\pm,b_l^\pm] = [(b_k^\pm)^{(\dagger)},(b_l^\pm)^{(\dagger)}] &=0\,,
\end{align}
here we omit the vector arrow on the mode labels in all index positions in order to keep our notation more compact. The total Fock space consists of a tensor product of three individual bosonic Fock spaces, one for the scalar particles, one for the $+$ polarised gravitons and one for the $-$ polarised ones. As usual the annihilation and creation operators for different fields or polarisations mutually commute. The Hamiltonian operator corresponding to \eqref{eq:linHamredps} can then be implemented in the Schrödinger picture as
\begin{align}\label{eq:HamiltonFockQU}
H = &\int_{\mathbb{R}^3} d^3k \left\{  \omega_k\,a_k^\dagger\, a_k + \Omega_k\, \left[ (b_k^+)^\dagger\, b_k^+ + (b_k^-)^\dagger\, b_k^-\right]\right\}\nonumber \\
+ &\sqrt{\frac{\kappa}{2}} \int d^3k \frac{1}{\sqrt{\Omega_k}} \sum\limits_{r\in\{\pm\}} \left[ b_k^r \: J_r^\dagger(\vec{k}) + (b_k^r)^\dagger \: J_r(\vec{k})\right]  \nonumber\\
+&\kappa U \otimes \mathds{1}_{\mathcal{E}}\,,
\end{align}
where we have introduced some kind of normal-ordered current operator for the scalar field that couples to the gravitational environment and which is quadratic in the scalar field and its derivatives:
\begin{multline}\label{eq:defJschroed}
J_r(\vec{k}) := \hat{\ft{\varphi}}{}_a^b(\vec{k},0) [P^{r}(\vec{k})]^a_b \\=  \int \frac{d^3p}{(2\pi)^\frac{3}{2}} \frac{1}{2\sqrt{\omega_p \omega_{k+p}}} \left[ p_a p^b [P^{r}(\vec{k})]^a_b\right] \left( 2 a_p^\dagger a_{k+p} +  a_{-p} a_{k+p} + a_{p}^\dagger a_{-p-k}^\dagger\right)\,.
\end{multline}
Furthermore, $U$ denotes a self-interaction operator that is present only due to coupling of the scalar field to linearised gravity, as its contribution to the Hamiltonian operator will vanish if the coupling constant $\kappa$ is set to zero. It involves fourth powers of the annihilation and creation operators of the scalar field and its momentum and its contributions can be understood as additional self-interaction vertices of the scalar field that are not present in the corresponding free theory. We chose to implement this operator in a completely normal ordered form, that is  $:U:$, where $:\cdot:$ denotes normal ordering. This is in contrast to the quantisation procedure of similar systems in \cite{anastopoulos2013master,oniga2016quantum,lagouvardos2021gravitational}, where either no specific ordering is mentioned \cite{oniga2016quantum}, or the individual operators corresponding to the operators $\epsilon, p_a, \widetilde{V}, \varphi$ and $J_r$ introduced above
are normal ordered \cite{anastopoulos2013master, lagouvardos2021gravitational}, but no normal ordering is applied  to the entire self-interaction operator $U$. For the normal ordered self-interaction operator we obtain
\begin{align}\label{eq:defU}
    U:=&{ \int d^3k  \frac{ik_b}{2} \kappa\ft{\delta \sigma}^{a}(-\vec k,0) \left[ \ft{\varphi}{}_a^b(\vec k,0) - \delta_a^b \left( \ft{\varphi}{}_c^c(\vec k,0) +m^2 \ft{\widetilde{V}}(\vec k,0)-\ft{\epsilon}(\vec k,0) \right)\right]}\nonumber\\
&- \int d^3k \left[2i \kappa ||\vec k|| \ft{\delta \tau}(-\vec k,0) \hat{k}^b \ft{p}{}_b(\vec k,0)\right]\nonumber\\
   &-\int_{\mathbb{R}^3} \frac{d^3k}{||\vec{k}||^2} \Bigg\{- :\ft{\hat{p}}\t{}{_c}(\vec{k},0) \,\ft{\hat{p}}\t{}{_d}(-\vec{k},0): \left[4\delta^{cd} - \hat{k}^c \hat{k}^d \right] + :\ft{\hat{\epsilon}}(\vec{k},0)\,\ft{\hat{\epsilon}}(-\vec{k},0): \left[\frac{1}{4} +\frac{3}{2}\right] \nonumber\\
&\hspace{1in}-\frac{3}{2} m^2\,:\ft{\hat{\widetilde{V}}}(\vec{k},0) \,\ft{\hat{\epsilon}}(-\vec{k},0):-:\ft{\hat{\varphi}} {}_a^a(\vec{k},0) \,\ft{\hat{\epsilon}}(-\vec{k},0): \Bigg\} \,.
\end{align}
We realise that the parameters $\delta\tau$ and $\delta\sigma^c$ enter into the self-interaction operator $U$ of the scalar field only. For our choice of $\delta\tau$ in \eqref{eq:Choicetausigma} where $\delta\tau$ is linear in $t$ the term including $\delta\tau$ drops out completely. The same happens for all contributions involving $\delta\sigma^c$ if we choose it to be constant. In case we choose it linearly in $x^c$ then we have $\delta\sigma^c_{,a}=\delta^a_c$ and hence the corresponding contributions in Fourier space will not vanish but $x^c$ will not enter explicitly into  the self-interaction operator $U$. 

Here we introduced several new operators involved in the physical Hamiltonian operator of the model. First of all a (normal-ordered) operator corresponding to the scalar field's momentum density:
\begin{align}
    \hat{\ft{p}}{}_a(\vec{k},t)&:=\frac{1}{2} \int_{\mathbb{R}^3} \frac{d^3q}{(2\pi)^{\frac{3}{2}}} q_a \sqrt{\frac{\omega_{k-q}}{\omega_q}} \bigg[ a_q a_{k-q} e^{-it(\omega_q + \omega_{k-q})} + a_{-q}^\dagger a_{k-q} e^{-it(\omega_{k-q} - \omega_{q})} \nonumber\\ &\hspace{1.8in} - a_{q-k}^\dagger a_q e^{it(\omega_{k-q} - \omega_{q})} - a_{-q}^\dagger a_{q-k}^\dagger e^{it(\omega_q + \omega_{k-q})} \bigg] \,.
    \end{align}
Additionally, an operator corresponding to the scalar field's energy density:
    \begin{align}
\hat{\ft{\epsilon}}(\vec{k},t)&:=\frac{1}{4} \int_{\mathbb{R}^3} \frac{d^3q}{(2\pi)^{\frac{3}{2}}} \Bigg\{ -\sqrt{\omega_q \omega_{k-q}} \bigg[ a_q a_{k-q} e^{-it(\omega_q + \omega_{k-q})} - a_{-q}^\dagger a_{k-q} e^{-it(\omega_{k-q} - \omega_{q})} \nonumber\\ &\hspace{2in} - a_{q-k}^\dagger a_q e^{it(\omega_{k-q} - \omega_{q})} + a_{-q}^\dagger a_{q-k}^\dagger e^{it(\omega_q + \omega_{k-q})} \bigg] \nonumber\\
& \hspace{1.15in}-\frac{q_a(k^a-q^a)-m^2}{\sqrt{\omega_q \omega_{k-q}}} \bigg[ a_q a_{k-q} e^{-it(\omega_q + \omega_{k-q})} + a_{-q}^\dagger a_{k-q} e^{-it(\omega_{k-q} - \omega_{q})} \nonumber\\ &\hspace{2.5in} + a_{q-k}^\dagger a_q e^{it(\omega_{k-q} - \omega_{q})} + a_{-q}^\dagger a_{q-k}^\dagger e^{it(\omega_q + \omega_{k-q})} \bigg] \Bigg\} \,.
    \end{align}
Finally, two more operators that correspond to certain different terms of the scalar field's energy momentum tensor that appear in the classical Hamiltonian, namely to $\phi^2$ and $\partial_a\phi \: \partial^b \phi$:
\begin{align}
    \hat{\widetilde{\ft{V}}}(\vec{k},t)&:=\frac{1}{2} \int_{\mathbb{R}^3} \frac{d^3q}{(2\pi)^{\frac{3}{2}}} \frac{1}{\sqrt{\omega_{q}\omega_{k-q}}} \bigg[ a_q a_{k-q} e^{-it(\omega_q + \omega_{k-q})} + a_{-q}^\dagger a_{k-q} e^{-it(\omega_{k-q} - \omega_{q})} \nonumber\\ &\hspace{1.8in} + a_{q-k}^\dagger a_q e^{it(\omega_{k-q} - \omega_{q})} + a_{-q}^\dagger a_{q-k}^\dagger e^{it(\omega_q + \omega_{k-q})} \bigg]
    \end{align}
    \begin{align}
\hat{\ft{\varphi}}{}_a^b(\vec{k},t)&:=-\frac{1}{2} \int_{\mathbb{R}^3} \frac{d^3q}{(2\pi)^{\frac{3}{2}}} \frac{q_a(k-q)^b}{\sqrt{\omega_{q}\omega_{k-q}}} \bigg[ a_q a_{k-q} e^{-it(\omega_q + \omega_{k-q})} + a_{-q}^\dagger a_{k-q} e^{-it(\omega_{k-q} - \omega_{q})} \nonumber\\ &\hspace{1.8in} + a_{q-k}^\dagger a_q e^{it(\omega_{k-q} - \omega_{q})} + a_{-q}^\dagger a_{q-k}^\dagger e^{it(\omega_q + \omega_{k-q})} \bigg]\,.\label{defvarphi}
\end{align}
Note that all these constituents are symmetric in the sense that $\ft{\epsilon}^\dagger(-\vec k,t) = \ft{\epsilon}(\vec k,t)$ which implies $\epsilon^\dagger(\vec x,t) = \epsilon(\vec x,t)$, and hence also $U^\dagger=U$. In the interaction picture, which we denote by a tilde, the Hamiltonian  operator that involves the gravity-matter interaction is given by
\begin{align}\label{eq:HaminIntp}
\widetilde{H}(t) = & \sqrt{\kappa}\;\underbrace{\frac{1}{\sqrt{2}} \int d^3k \frac{1}{\sqrt{\Omega_k}} \sum\limits_r \left[ b_k^r \: e^{-i\Omega_k t} \: J_r^\dagger(\vec{k},t) + (b_k^r)^\dagger\: e^{i\Omega_k t} \: J_r(\vec{k},t)\right]}_{=: \widetilde{H}_{TI}(t)}  +\kappa \widetilde{U}(t) \otimes \mathds{1}_{\mathcal{E}}
\end{align}
with an appropriate current operator obtained directly from the one in the Schrödinger picture by using the time-dependent constituents,
\begin{align}\label{eq:defJt}
J_r(\vec{k},t):=& \hat{\ft{\varphi}}{}_a^b(\vec{k},t) [P^{r}(\vec{k})]^a_b \nonumber\\=&\int \frac{d^3p}{(2\pi)^\frac{3}{2}} \frac{1}{2\sqrt{\omega_p \omega_{k+p}}} \left[ p_a p^b [P^{r}(\vec{k})]^a_b\right] \Big( a_p^\dagger a_{k+p} e^{it(\omega_p -\omega_{k+p})} +  a_{-p} a_{k+p} e^{-it(\omega_p +\omega_{k+p})}\nonumber\\&\hspace{2in}+a_{-p-k}^\dagger a_{-p} e^{it(\omega_{k+p} -\omega_{p})}+ a_{p}^\dagger a_{-p-k}^\dagger e^{it(\omega_p +\omega_{k+p})} \Big)\,,
\end{align}
and the total normal-ordered self-interaction operator
\begin{align}
    \widetilde{U}(t):=&{ \int d^3k  \frac{ik_b}{2} \kappa\ft{\delta \sigma}^{a}(-\vec k,t) \left[ \ft{\varphi}{}_a^b(\vec k,t) - \delta_a^b \left( \ft{\varphi}{}_c^c(\vec k,t) +m^2 \ft{\widetilde{V}}(\vec k,t)-\ft{\epsilon}(\vec k,t) \right)\right]}\nonumber\\
&- \int d^3k \left[ 2i \kappa ||\vec k|| \ft{\delta \tau}(-\vec k,t) \hat{k}^b \ft{p}{}_b(\vec k,t)\right]\nonumber\\
    &-\int_{\mathbb{R}^3} \frac{d^3k}{||\vec{k}||^2} \Bigg\{ -:\ft{\hat{p}}\t{}{_c}(\vec{k},t) \,\ft{\hat{p}}\t{}{_d}(-\vec{k},t): \left[4\delta^{cd}-\hat{k}^c \hat{k}^d \right] + :\ft{\hat{\epsilon}}(\vec{k},t)\,\ft{\hat{\epsilon}}(-\vec{k},t): \left[\frac{1}{4} + \frac{3}{2}\right] \nonumber\\
&\hspace{1in}-\frac{3}{2} m^2\,:\ft{\hat{\widetilde{V}}}(\vec{k},t) \,\ft{\hat{\epsilon}}(-\vec{k},t):- :\ft{\hat{\varphi}} {}_a^a(\vec{k},t) \,\ft{\hat{\epsilon}}(-\vec{k},t): \Bigg\} \,.
\end{align}
With this, the classical model has been carried over to the quantum field theoretic framework. In the next section, we will derive the so called master equation that governs the effective time evolution of the matter system without the need of tracking all the details of the gravitational degrees of freedom. 

\section{Deriving the master equation for the gravitationally induced decoherence model}
\label{sec:DerivMasterEqn}
In this section we derive the master equation for the gravitationally induced decoherence model considered in this article from first principle. For this purpose we need to trace out the environmental and thus the gravitational degrees of freedom. Before we apply this to the model in subsection \ref{sec:derivMeq} we briefly review two methods to derive master equations in the context of decoherence models. The first method, namely the projection operator technique, will be discussed in subsection \ref{sec:ReviewPOT} and a second method based on an influence phase functional is presented in subsection \ref{sec:ReviewIPFA}.
\subsection{Review of the projection operator technique}
\label{sec:ReviewPOT}
The starting point in this top-down approach of the derivation of every quantum master equation is the microscopic dynamics in the form of the Liouville-von Neumann equation, regardless of the exact approach that is taken thereafter. With an interaction Hamiltonian operator $\tilde{H}(t)$ and the density operator of the total system in the interaction picture given by $\tilde{\rho}(t)$, where we denote operators in the interaction picture with a tilde, the evolution equation initially amounts to:

\begin{equation}
    \frac{\partial}{\partial t} \tilde{\rho}(t) = -i \alpha [\tilde{H}(t), \tilde{\rho}(t)] =: \alpha \mathcal{L}(t) \tilde{\rho}(t), \label{eq:vonNeumann}
\end{equation}

where $\alpha$ denotes the dimensionless\footnote{due to the involvement of gravity, the coupling will not be dimensionless in the application of this formalism, see chapter (\ref{sec:PostMinkowski}) for further details} coupling strength between the system and the environment and $\mathcal{L}(t)$ is the so-called Liouville superoperator. The goal is to eliminate the explicit evolution of the many environmental degrees of freedom in order to arrive at an effective equation for the relevant system's degrees of freedom, generally of the following form:

\begin{align}
    \frac{\partial}{\partial t} \tilde{\rho}_S(t) &= -i \alpha [\tilde{H}_{LS}(t), \tilde{\rho}_S(t)] + \tilde{\mathcal{D}}\big( \tilde{\rho}_S(t) \big) \qquad \qquad \qquad \textrm{(interaction picture)} \\[0.5em]
    \frac{\partial}{\partial t} \rho_S(t) &= -i \alpha [H_S(t) + H_{LS}(t), \rho_S(t)] + \mathcal{D}\big( \rho_S(t) \big) \qquad \; \textrm{(Schrödinger picture)}
\end{align}

In general, there can be a Lamb shift-type contribution $H_{LS}(t)$ in addition to the unitary dynamics generated by $H_S(t)$ in the Schrödinger picture equation, alongside the dissipator $\mathcal{D}\big( \tilde{\rho}_S(t) \big)$,  the latter encodes both decoherence and dissipation. Note that the dissipator can in general retain some time dependence even after the transition to the Schrödinger picture, this critically depends on the model and the applied approximations. At this point there are various possibilities to proceed, the easiest and most frequently used path is towards the so-called Lindblad equation. We would like to mention that the Lindblad equation is a very specific form of a master equation in the sense that it originates from a bounded generator of a quantum-dynamical semigroup \cite{gorini1976completely,lindblad1976generators}. Naturally, the Lindblad equation is completely positive and trace-preserving, a feature not all master equations exhibit \cite{breuer2003concepts} without further adjustments as we will see an the end of this section. We will start with a fully analytical and in principle exact approach towards the Nakajima-Zwanzig equation, which will provide the basis for the time-convolutionless (TCL) master equation we will later use for the model considered in this work. This relies fundamentally on the results in \cite{Nakajima1958OnQT, Zwanzig1960, Shibata1977AGS, chaturvedi1979time}. Hereby we will closely follow the pedagogical outline given in \cite{breuer2003concepts} and \cite{breuer2002theory}. The density operator $\tilde{\rho}(t)$ in \eqref{eq:vonNeumann} is trace class by construction, hence is contained  in the set $\mathcal{B}(\mathcal{H})$ of linear, bounded operators. We proceed to define a pair of projection \textit{super}operators called $\mathcal{P}$ and $\mathcal{Q}$:

\begin{align}
    \mathcal{P}: \mathcal{B}(\mathcal{H}) \to \mathcal{B}(\mathcal{H}), \qquad \mathcal{P}(\rho) &:= tr_{\mathcal{E}}(\rho) \otimes \rho_{\mathcal{E}}, \label{eq:rel_proj}\\[0.3em]
    \mathcal{Q}: \mathcal{B}(\mathcal{H}) \to \mathcal{B}(\mathcal{H}), \qquad \mathcal{Q}(\rho) &:= \rho - \mathcal{P}(\rho), \label{eq:irrel_proj}
\end{align}

where $\rho_{\mathcal{E}}$ is a stationary environmental state that we will specify later and $tr_{\mathcal{E}}$ defines the environmental trace, i.e. the trace over all degrees of freedom that do not belong to the system of interest. The fact that these operators satisfy $\mathcal{P}^2 = \mathcal{P}$, $\mathcal{Q}^2 = \mathcal{Q}$, $\mathcal{P}+\mathcal{Q} = \mathds{1}$ and $\mathcal{P}\mathcal{Q} = \mathcal{Q}\mathcal{P} = 0$ and thus are indeed projectors is directly evident from the definitions above. The operators $\mathcal{P}$ and $\mathcal{Q}$ signify the relevant and irrelevant part of the combined dynamics, respectively, in the sense that everything projected with $\mathcal{P}$ exclusively contains the dynamics of the system degrees of freedom, with only an effective influence coming from the traced-out environment. These operators do furthermore commute with the partial time-derivative and hence can be directly applied to the Liouville-von Neumann equation (\ref{eq:vonNeumann}) to obtain:

\begin{align}
    \mathcal{P} \frac{\partial}{\partial t} \tilde{\rho}(t) &= \frac{\partial}{\partial t} \mathcal{P} \tilde{\rho}(t) = \alpha \mathcal{P} \mathcal{L}(t) \tilde{\rho}(t) = \alpha \big( \mathcal{P} \mathcal{L}(t)  \mathcal{P} \tilde{\rho}(t) +  \mathcal{P} \mathcal{L}(t)  \mathcal{Q} \tilde{\rho}(t) \big), \label{eq:rel_projection} \\[0.3em]
    \mathcal{Q} \frac{\partial}{\partial t} \tilde{\rho}(t) &= \frac{\partial}{\partial t} \mathcal{Q} \tilde{\rho}(t) = \alpha \mathcal{Q} \mathcal{L}(t) \tilde{\rho}(t) = \alpha \big( \mathcal{Q} \mathcal{L}(t)  \mathcal{P} \tilde{\rho}(t) +  \mathcal{Q} \mathcal{L}(t)  \mathcal{Q} \tilde{\rho}(t) \big) \label{eq:nonrel_projection}.
\end{align}

Hereby we used that the projection operators are linear. In the last step of each line we used that the two projectors combine to unity. In the end we would like to find a solution for the relevant part $\mathcal{P} \tilde{\rho}(t)$ of the density operator, the projectors do however not commute with the Liouvillian $\mathcal{L}(t)$. As a first step to eliminate the appearance of $\mathcal{Q} \tilde{\rho}(t)$ in the first and hence for us interesting and relevant equation (\ref{eq:rel_projection}), we write down a formal solution of equation (\ref{eq:nonrel_projection}):

\begin{align}
    \mathcal{Q} \tilde{\rho}(t) = \mathcal{G}(t,t_0) \mathcal{Q} \tilde{\rho}(t_0) + \alpha \int^t_{t_0} ds \, \mathcal{G}(t,s) \mathcal{Q} \mathcal{L}(s) \mathcal{P} \tilde{\rho}(s), \quad \mathcal{G}(t,s) := \mathcal{T}_{\leftarrow} \exp \Big[ \int^t_s d\tau \, \mathcal{Q} \mathcal{L}(\tau) \Big], \label{eq:irrel_Dyson}
\end{align}

where $\mathcal{T}_{\leftarrow}$ stands for time-ordering with the largest argument to the very left. Applying a partial derivative with respect to $t$ yields equation (\ref{eq:nonrel_projection}) while the initial condition for $t=t_0$ is satisfied due to $\mathcal{G}(t_0,t_0) = \mathds{1}$ because of a vanishing integral, so the stated solution is unique. Inserting (\ref{eq:irrel_Dyson}) into the relevant equation (\ref{eq:rel_projection}) gives:

\begin{align}
    \frac{\partial}{\partial t} \mathcal{P} \tilde{\rho}(t) = \alpha \mathcal{P} \mathcal{L}(t)  \mathcal{P} \tilde{\rho}(t) +  \alpha \mathcal{P} \mathcal{L}(t)  \mathcal{G}(t,t_0) \mathcal{Q} \tilde{\rho}(t_0) + \alpha^2 \int^t_{t_0} ds \, \mathcal{P} \mathcal{L}(t) \mathcal{G}(t,s) \mathcal{Q} \mathcal{L}(s) \mathcal{P} \tilde{\rho}(s) \label{eq:NZ}.
\end{align}

This is the \textit{Nakajima-Zwanzig} equation, albeit admittedly not less complicated than the attempt of solving the initial problem, mostly due to the appearance of the integral kernel

\begin{equation}
    \mathcal{K}(t,s) := \mathcal{P} \mathcal{L}(t) \mathcal{G}(t,s) \mathcal{Q} \mathcal{L}(s) \mathcal{P}.
\end{equation}

It is worth noting that this manifestly non-Markovian equation (\ref{eq:NZ}) is in principle exact, there was no need for approximations whatsoever until this point. If we assume that the interaction Hamiltonian density is of the general form $\sum_\alpha S_\alpha \otimes E_\alpha$ with a discrete summation index $\alpha$ (omitting model-specific time- or mode dependencies for the sake of brevity) and $E_\alpha$ consisting of linear contributions of annihilation- and creation-type variables, the first contribution in equation (\ref{eq:NZ}) vanishes. This is simply due to the fact that odd numbers of the environmental interaction term do not contribute in the environmental trace, given a suitable initial state, specifically a Gibbs state. This observation is contained in the very definition (\ref{eq:rel_proj}) of the relevant projector $\mathcal{P}$, which turns out to be  useful even beyond the mere simplification of equation (\ref{eq:NZ}) as we will see later in this section. For this broad class of models, a method called \textit{cumulant expansion} turns out to be a powerful tool for a step-by-step approximation of a time-convolutionless approach to non-Markovian dynamics, first developed in \cite{VANKAMPEN1974215} and \cite{VANKAMPEN1974239}, respectively. Up to now the main complexity of the Nakajima-Zwanzig equation remains, the convolution integral makes it nearly impossible to be solved analytically. To this extent, we introduce the two propagators $F(t,s)$ and $G(t,s)$ such that $G(t,s)\tilde{\rho}(s) = \tilde{\rho}(t)$ and $F(t,s) \circ G(t,s) = \mathds{1}$ with $s \leq t \in \mathbb{R}$. Note the distinction $\mathcal{G}(t,s) \neq G(t,s)$ in the previously used notation, the latter being the propagator of the \textit{total} system and not merely the one of the irrelevant projection. Hence:

\begin{align}
    \frac{\partial}{\partial t} \Big( F(t,s) \circ G(t,s) \tilde{\rho}(s) \Big) = \Big( \frac{\partial}{\partial t} F(t,s) \Big) \tilde{\rho}(t) + \alpha F(t,s) \mathcal{L}(t) \tilde{\rho}(t) = 0.
\end{align}

This can be easily solved formally for the inverse propagator:

\begin{align}
    \frac{\partial}{\partial t} F(t,s) = - \alpha F(t,s) \mathcal{L}(t) \quad \Longleftrightarrow \quad F(t,s) = \mathcal{T}_{\rightarrow} \exp \Big[- \alpha \int^t_s d\tau \, \mathcal{L}(\tau) \Big],
\end{align}

where $\mathcal{T}_{\rightarrow}$ denotes anti-time-ordering, that is with the largest argument to the very right. As the next step, we insert $\tilde{\rho}(s) = F(t,s) \big( \mathcal{P} + \mathcal{Q} \big) \tilde{\rho}(t)$ into the solution of equation (\ref{eq:nonrel_projection}) and obtain:

\begin{align}\label{deriv_beq}
    \mathcal{Q} \tilde{\rho}(t) &= \mathcal{G}(t,t_0) \mathcal{Q} \tilde{\rho}(t_0) + \alpha \int_{t_0}^t ds \, \mathcal{G}(t,s) \mathcal{Q} \mathcal{L}(s) \mathcal{P} F(t,s)(\mathcal{P} + \mathcal{Q}) \tilde{\rho}(t)\nonumber \\
    &=: \mathcal{G}(t,t_0) \mathcal{Q} \tilde{\rho}(t_0) + \Sigma(t,t_0)(\mathcal{P} + \mathcal{Q}) \tilde{\rho}(t),
\end{align}

where we collected the convolution part into a separate, now central entity $\Sigma(t,t_0)$:

\begin{align}
    \Sigma(t,t_0) := \alpha \int_{t_0}^t ds \, \mathcal{G}(t,s) \mathcal{Q} \mathcal{L}(s) \mathcal{P} F(t,s). \label{eq:Sigmat}
\end{align}

Now our aim is to solve \eqref{deriv_beq} for $\mathcal{Q} \tilde{\rho}(t)$ by collecting all terms on one side:

\begin{align}
    \big( \mathds{1} - \Sigma(t,t_0) \big) \mathcal{Q} \tilde{\rho}(t) = \mathcal{G}(t,t_0) \mathcal{Q} \tilde{\rho}(t_0) + \Sigma(t,t_0) \mathcal{P} \tilde{\rho}(t).
\end{align}

The key point is the treatment of $\big( \mathds{1} - \Sigma(t,t_0) \big)$ or rather the inverse of it. Evidently, the initial conditions $\Sigma(t_0, t_0) = 0$ and $\Sigma(t,t_0)|_{\alpha = 0} = 0$ are easy to check, our approach involves the representation of $\big( \mathds{1} - \Sigma(t,t_0) \big)^{-1}$ as the limit of a geometric series of operators, that is:

\begin{align}
    \big( \mathds{1} - \Sigma(t,t_0) \big)^{-1} = \sum\limits_{n=0}^\infty \big( \Sigma(t,t_0)\big)^n \label{eq:sigma_inv}
\end{align}

It is indeed the case that there exists some neighbourhood around $t_0$ where the expression in (\ref{eq:sigma_inv}) is invertible. These steps require some additional mathematical justification, for a more detailed discussion on this point we refer the reader to section (\ref{sec:sigma_inv}) in the appendix. Hence, under the assumption that we restrict to an interval $[t_0,t]$ where invertibility is guaranteed:

\begin{align}
    \mathcal{Q} \tilde{\rho}(t) = \big( \mathds{1} - \Sigma(t,t_0) \big)^{-1} \mathcal{G}(t,t_0) \mathcal{Q} \tilde{\rho}(t_0) + \big( \mathds{1} - \Sigma(t,t_0) \big)^{-1} \Sigma(t,t_0) \mathcal{P} \tilde{\rho}(t).
\end{align}

Finally inserting this into the Nakajima-Zwanzig equation (\ref{eq:NZ}) for $\mathcal{P}\tilde{\rho}(t)$ yields:

\begin{align}
    \frac{\partial}{\partial t} \mathcal{P} \tilde{\rho}(t) &= \alpha \mathcal{P} \mathcal{L}(t) \Big( \mathds{1} + \big( \mathds{1} - \Sigma(t,t_0) \big)^{-1} \Sigma(t,t_0) \Big) \mathcal{P} \tilde{\rho}(t) + \alpha \mathcal{P} \mathcal{L}(t) \big( \mathds{1} - \Sigma(t,t_0) \big)^{-1} \mathcal{G}(t,t_0) \mathcal{Q} \tilde{\rho}(t_0) \nonumber \\[0.4em]
    &=: K(t,t_0) \mathcal{P} \tilde{\rho}(t) + \mathcal{I}(t,t_0) \mathcal{Q} \tilde{\rho}(t_0).
\end{align}

This is the so-called \textit{time convolution-less} (TCL) master equation. Note that this master equation, although time-local, still conveys non-Markovian processes in contributions with higher than second order in $\alpha$. In principle it is even exact in the neighbourhood where $\big( \mathds{1} - \Sigma(t,t_0) \big)^{-1}$ is invertible, considering terms of all orders. We will not further investigate the inhomogeneity $\mathcal{I}(t,t_0) \mathcal{Q} \tilde{\rho}(t_0)$ at this point since we are focused on separable initial conditions for now, for which this expression vanishes. To this extent, consider the explicit form of the TCL generator $K(t,t_0)$ in terms of the geometric series expansion of the inverse of $\big( \mathds{1} - \Sigma(t,t_0) \big)$, neglecting $\mathcal{P} \mathcal{L}(t) \mathcal{P}$ based on the assumption that the environmental part of the interaction Hamiltonian density is linearly comprised of creation- and annihilation-type operators:

\begin{align}
    K(t,t_0) = \alpha \mathcal{P} \mathcal{L}(t) \big( \mathds{1} - \Sigma(t,t_0) \big)^{-1} \Sigma(t,t_0) \mathcal{P} &= \alpha \mathcal{P} \mathcal{L}(t) \sum\limits_{n=1}^\infty \big( \Sigma(t,t_0)\big)^n \mathcal{P} \nonumber \\
    &= \alpha \mathcal{P} \mathcal{L}(t) \sum\limits_{n=1}^\infty \Big( \sum\limits_{m=1}^\infty \alpha^m \Sigma_m(t,t_0) \Big)^n \mathcal{P}, \label{eq:TCL_gen}
\end{align}

where we noticed that it is apparent that $\Sigma(t,t_0)$, given its explicit form in equation (\ref{eq:Sigmat}), can be expanded in powers of $\alpha$ as well. Since we are interested in a weak coupling scenario, it is sufficient to consider contributions up to second order in $\alpha$, that is only the lowest-order terms that occur in the series expansion of $\Sigma(t,t_0)$. In our particular case this entails \textit{only} the second-order contributions since $\mathcal{P} \mathcal{L}(t) \mathcal{P} = 0$. This fact amounts to the Born approximation on the one hand, by virtue of the assumption of separable initial conditions with a stationary thermal environment and to the (first) Markov approximation on the other hand. The latter fact becomes evident once we realise that by eliminating the convolution integral with the density operator altogether without correcting for it with higher-order terms, we end up with an equation that has simply replaced $\tilde{\rho}(s)$ with $\tilde{\rho}(t)$. The second Markov approximation is commonly depicted as the limit $t_0 \to -\infty$, based on the assumption that the characteristic time scales of system and environment are vastly different, inducing a peakedness of the correlation functions. The latter arise naturally as functions of the form $\mathrm{tr}_\varepsilon \big(E_\alpha(t) E_\beta(s) \rho_\varepsilon \big)$ in the explicit computation of the environmental trace with $E_\alpha(t)$ meaning the environmental part of the interaction Hamiltonian. This peakedness property means that an extension of the integration interval has a negligible effect on the effective dynamics of the system, which however has to be checked explicitly for any given model. In the specific decoherence model we consider, we do not take this limit yet, all of the previous assumptions are nevertheless justified based on the weak coupling of matter to gravity. The only relevant contribution is the one where $\mathcal{G}(t,s)$ and $F(t,s)$ do not contribute, since everything beyond the identity contains the coupling constant.

\begin{equation}
    K(t,t_0) = \alpha^2 \mathcal{P} \mathcal{L}(t) \Sigma_1(t,t_0) \mathcal{P} =\alpha^2 \mathcal{P} \mathcal{L}(t) \int_{t_0}^t ds \, \mathcal{Q} \mathcal{L}(s) \mathcal{P} = \alpha^2 \mathcal{P} \int_{t_0}^t ds \, \mathcal{L}(t) \mathcal{L}(s) \mathcal{P}.
\end{equation}

Note that higher-order contributions in the coupling constant get increasingly difficult to evaluate explicitly, based on the complexity of (\ref{eq:TCL_gen}), see for example \cite{breuer2002timelocal} for the fourth-order TCL generator. The cumulant expansion procedure \cite{VANKAMPEN1974215, VANKAMPEN1974239} provides a systematic framework in which the individual contributions in $K(t,t_0)$ can be evaluated, especially for models in which all terms involving $\mathcal{P} \mathcal{L}(t) \mathcal{P}$ vanish due to the explicit form of the interaction. Now we can explicitly write down the desired master equation we will use for the model considered in this article later:

\begin{align}
    \frac{\partial}{\partial t} \mathcal{P} \tilde{\rho}(t) &= \alpha^2 \mathcal{P} \int_{t_0}^t ds \, \mathcal{L}(t) \mathcal{L}(s)  \tilde{\rho}(t) \nonumber \\
    &= -\alpha^2  \int_{t_0}^t ds \, tr_{\mathcal{E}} \left\{ \Big[\tilde{H}(t), \Big[\tilde{H}(s), tr_\mathcal{E} \left\{\tilde{\rho}(t)\right\} \otimes \tilde{\rho}_\mathcal{E} \Big] \Big] \right\} \otimes \tilde{\rho}_\mathcal{E}.\label{proj_gentclmeq}
\end{align}

This second-order TCL master equation could have been obtained in a straightforward fashion: First integrate the von Neumann equation (\ref{eq:vonNeumann}) and subsequently insert the formal solution into the original equation, akin to the second-order truncation of the Dyson series. Secondly, assume separable initial conditions and replace $\tilde{\rho}_S(s) \to \tilde{\rho}_S(t)$, eliminating the convolution integral. It is however not clear from this perspective how non-Markovian corrections or initial conditions with nonvanishing entanglement entropy can be dealt with. The presented projection operator technique in conjunction with the TCL formalism answers these questions order by order and simultaneously provides a differential (as opposed to an integro-differential) equation for $\tilde{\rho}(t)$. An alternative but largely equivalent route towards a time-local master equation is presented in the next section. Lastly we would like to add a few sentences regarding the relation between the TCL master equation (\ref{proj_gentclmeq}) and the frequently used Lindblad equation \cite{lindblad1976generators, gorini1976completely}. Trace preservation and the notion of positivity are of paramount importance when it comes to solutions of the master equation. However, not every master equation is guaranteed to admit solutions that adhere to these principles. The Lindblad form ensures that these criteria are automatically met:

\begin{equation}
\label{eq:Lindblad}
    \frac{\partial}{\partial t} \rho_S(t) = -i \alpha [H_S(t) + H_{LS}(t), \rho_S(t)] + \sum\limits_{\alpha \beta} \lambda_{\alpha \beta} \Big( L_\alpha \rho_S(t) L^\dagger_\beta  - \frac{1}{2} \{ L^\dagger_\beta L_\alpha, \rho_S(t) \} \Big), 
\end{equation}

with a finite set of traceless Lindblad operators $L_\alpha$ and positive, semi-definite coefficients $\lambda_{\alpha \beta}$. This equation of so-called \textit{first} standard form can be diagonalised by a unitary transformation of the operators $L_\alpha$, after which it is commonly called the \textit{second} standard form. It is immediately clear that not every equation of the form (\ref{proj_gentclmeq}) exhibits Lindblad form, firstly because of the remaining time dependence that has to be removed by hand based on the properties of the environmental correlation functions. Secondly, the operator structure also differs in a way that positivity is not automatically ensured, even if the derivation has been performed in a top-down manner \cite{homa2019positivity}. There is no universal pathway that remedies these issues, for certain models it turns out to be feasible to expand the dissipator to obtain Lindblad form \cite{breuer2002theory}, whereas it can be sufficient to neglect certain contributions in the course of a rotating-wave approximation \cite{fleming2010rotating}.

\subsection{Review of the influence phase functional approach}
\label{sec:ReviewIPFA}
A different albeit mostly equivalent approach of formulating a master equation is the influence phase functional approach, which has the same starting point as the previously introduced projection operator formalism. During the course of this brief review we will follow the idea first presented in \cite{FEYNMAN1963118} while we also follow the pedagogical presentation in \cite{breuer2002theory}. Regarding the notation in terms of propagators and thermal Wightman functions, we work with the same notation as for instance used in \cite{boyanovsky2015effective} or \cite{Hollowood:2017bil}. As before, we start at the von Neumann equation in the interaction picture:

\begin{equation}
    \frac{\partial}{\partial t} \tilde{\rho}(t) = -i \alpha [\tilde{H}(t), \tilde{\rho}(t)] =: \alpha \mathcal{L}(t) \tilde{\rho}(t), \quad
    \tilde{\rho}_S(t) = tr_{\mathcal{E}} \Big\{ \mathcal{T}_\leftarrow \exp \Big\{ \alpha \int_{t_0}^t \int_{\mathds{R}^3} d\tau \, d^3x \, \ell(x) \Big\} \tilde{\rho}(t_0) \Big\}, \label{eq:it_Dyson}
\end{equation}

where $\ell(x)= \ell(\tau, \vec{x})$ is the superoperator density associated to $\mathcal{L}(\tau)$ and $\mathcal{T}_\leftarrow := \mathcal{T}^{(S)}_\leftarrow \mathcal{T}^{(\varepsilon)}_\leftarrow$ corresponds to chronological time-ordering of system an environmental operators, respectively, with larger time arguments to the left of the expression. Contrary to the Nakajima-Zwanzig formalism, the strategy is to directly eliminate the environmental time-ordering at the level of the iterated Dyson series (\ref{eq:it_Dyson}). Note that system's operators can be considered as commuting under the time-ordering $\mathcal{T}^{(S)}_\leftarrow$. We assume an interaction Hamiltonian density of the general form $\mathcal{H}(x) = \sum\limits_\alpha S_\alpha(x) \otimes E_\alpha(x)$, the indices $\alpha, \beta$ can be understood as (a mixture of either co- or contravariant spacetime- or internal) multi-indices depending on the specific model. The goal of this section is to reformulate equation (\ref{eq:it_Dyson}) into a more tractable form similar to

\begin{equation}
    \tilde{\rho}_S(t) = \exp \{ i \Phi_t[S^+, S^-] \} \tilde{\rho}_S(t_0),
\end{equation}

where $\Phi_t[S^+, S^-]$ is called the influence phase functional and the $S^\pm$ are superoperators associated to the system of interest. A convenient decomposition of the time ordering in equation (\ref{eq:it_Dyson}) can be straightforwardly shown by induction for \textit{almost} commuting environmental operators, i.e. the case that $[E_\alpha(x), E_\beta(x')] \propto ic_{\alpha \beta}(x, x') \mathds{1}_\varepsilon$ with $c_{\alpha \beta}(x, x') \in \mathds{R}$. This is done by introducing an ansatz where the time-ordering amounts to a phase factor times the ordinary, non-time-ordered exponential of the Liouvillian, leading to a differential equation which can be directly integrated. A very similar procedure has been performed at the level of the time-evolution operator in \cite{Hornberger2009} for the spin-boson model. This can be carried over almost immediately by recalling the adjoint action of these operators on the density operator and the fact that the $S_\alpha(t)$ can be ordered arbitrarily \textit{inside} the time-ordering $\mathcal{T}^{(S)}_\leftarrow$. This is the setup this work will rely on, however in the more general case of strictly non-commuting environmental operators, the proof still holds but relies on the application of the finite-temperature Wick theorem \cite{breuer2002theory,evans1996wick}. The result is:

\begin{align}
    \tilde{\rho}_S(t) &= \mathcal{T}^{(S)}_\leftarrow tr_{\mathcal{E}} \Big\{ \mathcal{T}^{(\varepsilon)}_\leftarrow \exp \Big\{ \alpha \int_{t_0}^t \, \int_{\mathds{R}^3} d\tau \, d^3x \, \ell(x) \Big\} \tilde{\rho}(t_0) \Big\} \nonumber \\[0.5em] \nonumber
    &= \mathcal{T}^{(S)}_\leftarrow \exp \Big\{ \frac{\alpha^2}{2} \int_{t_0}^t \int_{\mathds{R}^3} d\tau d^3x \int_{t_0}^\tau \int_{\mathds{R}^3} d\tau' d^3x' \, \big[ \ell(x), \ell(x') \big] \Big\} tr_{\mathcal{E}} \Big\{ \exp \Big\{ \alpha \int_{t_0}^t \, \int_{\mathds{R}^3} d\tau \, d^3x \, \ell(x) \Big\} \tilde{\rho}(t_0) \Big\} \\[0.5em] \nonumber
    &= \mathcal{T}^{(S)}_\leftarrow \exp \Big\{ \frac{\alpha^2}{2} \int_{t_0}^t \int_{\mathds{R}^3} d\tau d^3x \int_{t_0}^\tau \int_{\mathds{R}^3} d\tau' d^3x' \, \big[ \ell(x), \ell(x') \big] \Big\} \\[0.5em]
    & \; \qquad \times \exp \Big\{ \frac{\alpha^2}{2} \int_{t_0}^t \int_{\mathds{R}^3} d\tau d^3x \int_{t_0}^t \int_{\mathds{R}^3} d\tau' d^3x' \, \big\langle \ell(x) \ell(x') \big\rangle_\varepsilon \Big\} \tilde{\rho}_S(t_0), \label{eq:influence_phase_dec}
\end{align}

with the environmental expectation value $\big\langle \ell(x) \ell(x') \big\rangle_\varepsilon := tr_\varepsilon \big\{ \ell(x) \ell(x') \rho_\varepsilon \big\}$ under the assumption that the initial state $\tilde{\rho}(t_0) = \tilde{\rho}_S(t_0) \otimes \tilde{\rho}_\varepsilon(t_0)$ is uncorrelated and hence separable. Note that the integrals in the first exponent are nested, i.e. the commutator of the Liouvillians is explicitly time-ordered whereas the second exponent holds integrals with identical limits. Furthermore we would like to point out that the last line of (\ref{eq:influence_phase_dec}) only acts upon $\tilde{\rho}_S(t_0)$. This is due to the fact that for an uncorrelated thermal initial state, the action of the exponentiated Liouvillian can be directly computed in terms of a cumulant expansion with a vanishing first order, terminating after the second order, assuming that $\tilde{\rho}_\varepsilon$ describes a Gaussian state with respect to the environmental interaction variables. Hence it is possible to absorb the expectation value $\big\langle \ell(x) \ell(x') \big\rangle_\varepsilon$ into the exponential. Use of the Jacobi identity and the above definitions lets us further simplify (\ref{eq:influence_phase_dec}):

\begin{align*}
    \big[ \ell(x), \ell(x') \big] \tilde{\rho} = - \big[ \big[\mathcal{H}(x), \mathcal{H}(x') \big], \tilde{\rho} \big] &= - \sum\limits_\alpha \sum\limits_\beta \big[ \big[S_\alpha(x) \otimes E_\alpha(x), S_\beta(x') \otimes E_\beta(x') \big], \tilde{\rho}_S \otimes \tilde{\rho}_\varepsilon \big] \\
    &\cong - \sum\limits_\alpha \sum\limits_\beta \big[ E_\alpha(x), E_\beta(x') \big] \big[ S_\alpha(x) S_\beta(x') \otimes \mathds{1}_\varepsilon,  \tilde{\rho}_S \otimes \tilde{\rho}_\varepsilon \big].
\end{align*}

Note that the last equality holds under consideration of the overall system time-ordering to the very left of expression (\ref{eq:influence_phase_dec}) and under the assumption that the commutator $\big[ E_\alpha(x), E_\beta(x') \big]$ is purely a phase factor, as it is the case for the model in this work. We have indicated that a certain equality only holds under additional time-ordering $\mathcal{T}^{(S)}_\leftarrow$ with $\cong$. This justifies acting upon $\tilde{\rho}_S$ instead of $\tilde{\rho}_S \otimes \tilde{\rho}_\varepsilon$ in (\ref{eq:influence_phase_dec}), where the environmental part has been completely reduced to the expectation value in the second exponential. The action of these operator densities on the density operator is to be understood in a formal sense, all quantities are spatially integrated once we insert these identities back into the phase functional. In order to represent the action of the exponentiated commutators, let us introduce the following notation:

\begin{align*}
    S_\alpha^+(x) \tilde{\rho}_S := S_\alpha(x) \tilde{\rho}_S, \qquad S_\alpha^-(x) \tilde{\rho}_S := \tilde{\rho}_S S_\alpha(x),
\end{align*}

where the commutator of Liouvillians $\big[ \ell(x), \ell(x') \big]$ immediately assumes the form

\begin{align*}
    \big[ \ell(x), \ell(x') \big] = - \sum\limits_\alpha \sum\limits_\beta \Big( \big[ E_\alpha(x), E_\beta(x') \big] S_\alpha^+(x) S_\beta^+(x') + \big[ E_\alpha(x), E_\beta(x') \big] S_\alpha^-(x) S_\beta^-(x') \Big).
\end{align*}

This concludes the first part of the derivation of the influence phase functional based on (\ref{eq:influence_phase_dec}):
\begin{align}
    \tilde{\rho}_S(t) &= \mathcal{T}^{(S)}_\leftarrow \exp \Big\{ - \frac{\alpha^2}{2} \int_{t_0}^t \int_{\mathds{R}^3} d\tau d^3x \int_{t_0}^t \int_{\mathds{R}^3} d\tau' d^3x' \, \theta(\tau - \tau') \sum\limits_\alpha \sum\limits_\beta \big[ E_\alpha(x), E_\beta(x') \big] S_\alpha^+(x) S_\beta^+(x') \Big\} \nonumber \\[0.5em]
    & \; \qquad \times \exp \Big\{\frac{\alpha^2}{2} \int_{t_0}^t \int_{\mathds{R}^3} d\tau d^3x \int_{t_0}^t \int_{\mathds{R}^3} d\tau' d^3x' \, \theta(\tau - \tau') \sum\limits_\alpha \sum\limits_\beta \big[ E_\alpha(x), E_\beta(x') \big] S_\alpha^-(x) S_\beta^-(x') \Big\} \nonumber \\[0.5em]
    & \; \qquad \times \exp \Big\{ \frac{\alpha^2}{2} \int_{t_0}^t \int_{\mathds{R}^3} d\tau d^3x \int_{t_0}^t \int_{\mathds{R}^3} d\tau' d^3x' \, \big\langle \ell(x) \ell(x') \big\rangle_\varepsilon \Big\} \tilde{\rho}_S(t_0). \label{eq:influence_phase_dec2}
\end{align}

Next we have a closer look at the thermal expectation value $\big\langle \ell(x) \ell(x') \big\rangle_\varepsilon$ and apply its definition:

\begin{align*}
\hspace{-1.5cm} \int_{t_0}^t \int_{\mathds{R}^3} d\tau d^3x \int_{t_0}^t \int_{\mathds{R}^3} d\tau' d^3x' \, \big\langle \ell(x) \ell(x') \big\rangle_\varepsilon \, \tilde{\rho}_S  &
   =  \int_{t_0}^t \int_{\mathds{R}^3} d\tau d^3x \int_{t_0}^t \int_{\mathds{R}^3} d\tau' d^3x' \, tr_\varepsilon\Big(- \big[\mathcal{H}(x), \big[\mathcal{H}(x), \tilde{\rho}_S \otimes \tilde{\rho}_\varepsilon \big] \big] \Big)  \\
    \cong  \int_{t_0}^t \int_{\mathds{R}^3} d\tau d^3x \int_{t_0}^t \int_{\mathds{R}^3} d\tau' d^3x' \, \sum\limits_\alpha \sum\limits_\beta \Big( - &\big\langle E_\beta(x') E_\alpha(x) \big\rangle_\varepsilon S_\alpha^+(x) S_\beta^+(x') - \big\langle E_\alpha(x) E_\beta(x') \big\rangle_\varepsilon S_\alpha^-(x) S_\beta^-(x')  \\
   & \hspace{-0.5cm} + \big\langle E_\beta(x') E_\alpha(x) \big\rangle_\varepsilon S_\alpha^+(x) S_\beta^-(x') + \big\langle E_\alpha(x) E_\beta(x') \big\rangle_\varepsilon S_\alpha^-(x) S_\beta^+(x') \Big),
\end{align*}

where it is important that the last equality again only holds under the time-ordering still present in (\ref{eq:influence_phase_dec2}). This is because we exchanged the operators $S_\alpha^+(x), S_\beta^+(x')$ and switched the integration variables and summation indices respectively. Afterwards we ended up with what seem to be switched arguments in the correlation function $\big\langle E_\beta(x') E_\alpha(x) \big\rangle_\varepsilon$, likewise for the contribution from $S_\alpha^-(x) S_\beta^+(x')$. This gives us the possibility to group the individual terms in the influence functional (\ref{eq:influence_phase_dec}) such that we can express it in terms of thermal Wightman functions.

\begin{align}
    \tilde{\rho}_S(t) &= \mathcal{T}^{(S)}_\leftarrow \exp \bigg\{ \frac{\alpha^2}{2} \int_{t_0}^t \int_{\mathds{R}^3} d\tau d^3x \int_{t_0}^t \int_{\mathds{R}^3} d\tau' d^3x' \sum\limits_\alpha \sum\limits_\beta \nonumber \\[0.5em]
    \times \bigg[ - &\Big( \theta(\tau - \tau') \big[ E_\alpha(x), E_\beta(x') \big] + \big\langle E_\beta(x') E_\alpha(x) \big\rangle_\varepsilon \Big) S_\alpha^+(x) S_\beta^+(x') \nonumber \\[0.5em]
    + &\Big( \theta(\tau - \tau') \big[ E_\alpha(x), E_\beta(x') \big] - \big\langle E_\alpha(x) E_\beta(x') \big\rangle_\varepsilon \Big) S_\alpha^-(x) S_\beta^-(x') \nonumber \\[0.5em]
    + &\big\langle E_\beta(x') E_\alpha(x) \big\rangle_\varepsilon S_\alpha^+(x) S_\beta^-(x') + \big\langle E_\alpha(x) E_\beta(x') \big\rangle_\varepsilon S_\alpha^-(x) S_\beta^+(x')\bigg] \bigg\} \tilde{\rho}_S(t_0). \label{eq:influence_phase_dec3}
\end{align}

Upon closer inspection, these combinations of correlation functions amount to the propagators:

\begin{align}
    G^{++}_{\alpha \beta}(x-x') &:= \big\langle \mathcal{T}^{(\varepsilon)}_\leftarrow \big( E_\alpha(x) E_\beta(x') \big) \big\rangle_\varepsilon = \theta(\tau - \tau') \big[ E_\alpha(x), E_\beta(x') \big] + \big\langle E_\beta(x') E_\alpha(x) \big\rangle_\varepsilon \nonumber \\[0.3em]
    &= \big\langle E_\alpha(x) E_\beta(x') \big\rangle_\varepsilon \theta(\tau - \tau')  + \big\langle E_\beta(x') E_\alpha(x) \big\rangle_\varepsilon \theta(\tau' - \tau) \nonumber \\[0.3em]
    &=: G_{\alpha \beta}^{>}(x-x') \theta(\tau - \tau') + G_{\alpha \beta}^{<}(x-x') \theta(\tau' - \tau), \label{eq:G++_prop}
\end{align}

and similarly although anti-time-ordered for the other contribution:

\begin{align}
    -G^{--}_{\alpha \beta}(x-x') &:= - \big\langle \mathcal{T}^{(\varepsilon)}_\rightarrow \big( E_\alpha(x) E_\beta(x') \big) \big\rangle_\varepsilon = \theta(\tau - \tau') \big[ E_\alpha(x), E_\beta(x') \big] - \big\langle E_\alpha(x) E_\beta(x') \big\rangle_\varepsilon \nonumber \\[0.3em]
    &= - \big\langle E_\alpha(x) E_\beta(x') \big\rangle_\varepsilon \theta(\tau' - \tau) - \big\langle E_\beta(x') E_\alpha(x) \big\rangle_\varepsilon \theta(\tau - \tau') \nonumber \\[0.3em]
    &=: - G_{\alpha \beta}^{>}(x-x') \theta(\tau' - \tau) - G_{\alpha \beta}^{<}(x-x') \theta(\tau - \tau'). \label{eq:G--_prop}
\end{align}

Inserting $G^{++}_{\alpha \beta}(x-x')$ and $G^{--}_{\alpha \beta}(x-x')$ into equation (\ref{eq:influence_phase_dec3}) yields

\begin{align*}
    \tilde{\rho}_S(t) &= \mathcal{T}^{(S)}_\leftarrow \exp \bigg\{ \frac{\alpha^2}{2} \int_{t_0}^t \int_{\mathds{R}^3} d\tau d^3x \int_{t_0}^t \int_{\mathds{R}^3} d\tau' d^3x' \sum\limits_\alpha \sum\limits_\beta \\[0.5em]
    \times \bigg[ - &\Big( G_{\alpha \beta}^{>}(x-x') \theta(\tau - \tau') + G_{\alpha \beta}^{<}(x-x') \theta(\tau' - \tau) \Big) S_\alpha^+(x) S_\beta^+(x') \\[0.5em]
    - &\Big( G_{\alpha \beta}^{>}(x-x') \theta(\tau' - \tau) + G_{\alpha \beta}^{<}(x-x') \theta(\tau - \tau') \Big) S_\alpha^-(x) S_\beta^-(x') \\[0.5em]
    + &G_{\alpha \beta}^{<}(x-x') S_\alpha^+(x) S_\beta^-(x') + G_{\alpha \beta}^{>}(x-x') S_\alpha^-(x) S_\beta^+(x')\bigg] \bigg\} \tilde{\rho}_S(t_0). \\[0.5em]
\end{align*}

Now we would like to eliminate the Heaviside functions $\theta(\tau-\tau')$ from this expression in order to be left purely with the thermal Wightman functions $G_{\alpha \beta}^{\lessgtr}(x-x')$. This is achieved by expanding the mixed contributions with $1 = \theta(\tau - \tau') + \theta(\tau' - \tau)$ and renaming the integration variables in the individual terms while changing the integration domain into a nested integral \cite{boyanovsky2015effective}:

\begin{align}
   \tilde{\rho}_S(t) &= \mathcal{T}^{(S)}_\leftarrow \exp \bigg\{ \alpha^2 \int_{t_0}^t \int_{\mathds{R}^3} d\tau d^3x \int_{t_0}^\tau \int_{\mathds{R}^3} d\tau' d^3x' \sum\limits_\alpha \sum\limits_\beta \nonumber \\[0.5em]
    \times \bigg[ - & G_{\alpha \beta}^{>}(x-x') S_\alpha^+(x) S_\beta^+(x') 
    - G_{\alpha \beta}^{<}(x-x') S_\alpha^-(x) S_\beta^-(x') \nonumber \\[0.5em]
    + &G_{\alpha \beta}^{<}(x-x') S_\alpha^+(x) S_\beta^-(x') + G_{\alpha \beta}^{>}(x-x') S_\alpha^-(x) S_\beta^+(x')\bigg] \bigg\} \tilde{\rho}_S(t_0).
\end{align}

This concludes the derivation of the influence phase functional, while strictly speaking $\Phi_t[S^+, S^-]$ is both a functional in the $S^{\pm}$ operators as well as a superoperator that acts on the space of density operators associated to the system. Explicitly, we obtain

\begin{align}
    i\Phi_t[&S^+, S^-] := \alpha^2 \int_{t_0}^t \int_{\mathds{R}^3} d\tau d^3x \int_{t_0}^\tau \int_{\mathds{R}^3} d\tau' d^3x' \sum\limits_\alpha \sum\limits_\beta \nonumber \\[0.5em]
    \times \bigg[ -&G_{\alpha \beta}^{>}(x-x') S_\alpha^+(x) S_\beta^+(x') 
    - G_{\alpha \beta}^{<}(x-x') S_\alpha^-(x) S_\beta^-(x') \nonumber \\[0.5em]
    + &G_{\alpha \beta}^{<}(x-x') S_\alpha^+(x) S_\beta^-(x') + G_{\alpha \beta}^{>}(x-x') S_\alpha^-(x) S_\beta^+(x')\bigg]. \label{eq:inf_func}
\end{align}

The influence phase functional is often depicted in a slightly different form, including functions called the dissipation- and noise kernel based on their interpretation in well-established decoherence models. The transition from the expression in (\ref{eq:inf_func}) is rather straightforward. To this extent, let us first introduce $D_{\alpha \beta}(x-x')$ and $D_{\alpha \beta}^1(x-x')$ based on the thermal Wightman functions:

\begin{align}
    D_{\alpha \beta}(x-x') &:= \big\langle \big[E_\alpha(x), E_\beta(x') \big] \big\rangle_\varepsilon = i \big(G_{\alpha \beta}^{>}(x-x') - G_{\alpha \beta}^{<}(x-x')\big), \label{eq:diss_kernel} \\[0.3em]
    D_{\alpha \beta}^1(x-x') &:= \big\langle \big\{ E_\alpha(x), E_\beta(x') \big\} \big\rangle_\varepsilon = G_{\alpha \beta}^{>}(x-x') + G_{\alpha \beta}^{<}(x-x'). \label{eq:noise_kernel}
\end{align}

These relations can be readily inverted:

\begin{align*}
     G_{\alpha \beta}^{>}(x-x') = \frac{1}{2} \big( D_{\alpha \beta}^1(x-x') - iD_{\alpha \beta}(x-x') \big), \quad G_{\alpha \beta}^{<}(x-x') = \frac{1}{2} \big( D_{\alpha \beta}^1(x-x') + iD_{\alpha \beta}(x-x') \big).
\end{align*}

In terms of the noise- and dissipation kernel, the influence phase functional is given by:

\begin{align}
    i\Phi_t&[S^+, S^-] := \alpha^2 \int_{t_0}^t \int_{\mathds{R}^3} d\tau d^3x \int_{t_0}^\tau \int_{\mathds{R}^3} d\tau' d^3x' \sum\limits_\alpha \sum\limits_\beta \nonumber \\[0.5em]
    \times \bigg[ \frac{i}{2}&D_{\alpha \beta}(x-x') \Big( S_\alpha^+(x) S_\beta^+(x') - S_\alpha^-(x) S_\beta^-(x') + S_\alpha^+(x) S_\beta^-(x') - S_\alpha^-(x) S_\beta^+(x') \Big) \nonumber \\
    - \frac{1}{2}&D_{\alpha \beta}^1(x-x') \Big( S_\alpha^+(x) S_\beta^+(x') + S_\alpha^-(x) S_\beta^-(x') - S_\alpha^+(x) S_\beta^-(x') - S_\alpha^-(x) S_\beta^+(x') \Big) \bigg]. 
\end{align}

We realise that these particular combinations of $S^{\pm}$-superoperators can be cast into a familiar form in terms of combinations of commutators and anti-commutators

\begin{align*}
    S_\alpha^c(x)(\cdot) := [S_\alpha(x), \cdot] = \big( S_\alpha^+(x) - S_\alpha^-(x) \big) (\cdot), \quad S_\alpha^a(x)(\cdot) := \{ S_\alpha(x), \cdot \} = \big( S_\alpha^+(x) + S_\alpha^-(x) \big) (\cdot).
\end{align*}

Finally, the most convenient (and most abundant) version of the influence phase functional reads:

\begin{align}
    i\Phi_t&[S^+, S^-] := \alpha^2 \int_{t_0}^t \int_{\mathds{R}^3} d\tau d^3x \int_{t_0}^\tau \int_{\mathds{R}^3} d\tau' d^3x' \sum\limits_\alpha \sum\limits_\beta \nonumber \\[0.5em]
    \times \bigg[ \frac{i}{2}&D_{\alpha \beta}(x-x') S_\alpha^c(x) S_\beta^a(x')
    - \frac{1}{2}D_{\alpha \beta}^1(x-x') S_\alpha^c(x) S_\beta^c(x') \bigg].
\end{align}

At this stage it is evident that the projection operator formalism and the influence phase functional approach are equivalent formulations. Strictly speaking, this is only the case for Gaussian environmental initial states \cite{breuer2002theory}. Up to second order this equality can be easily verified by deriving equation (\ref{eq:inf_func}) with respect to time and consequently applying all the commutators and anti-commutators to the density operator to the right. Once we recall the definition of the kernels, this precisely amounts to the Born-Markov master equation prior to the second Markov approximations regarding the upper integral limit. A similar procedure can be implemented order by order \cite{breuer2002timelocal}.

\subsection{Derivation of the master equation}
\label{sec:derivMeq}
In this section we apply the projection operator method reviewed in subsection \ref{sec:ReviewPOT} to the model in this work.  Assuming factorising initial conditions and a Gibbs state for the environment, we can take the TCL master equation \eqref{proj_gentclmeq} to second order in the coupling strength $\alpha=\sqrt{\kappa}$ and evaluate it for the model under our consideration described by the interaction Hamiltonian operator shown in \eqref{eq:HaminIntp}:

\begin{align}\label{meqStart}
    \frac{\partial}{\partial t} \widetilde{\rho}_S(t) = -i\kappa \left[ \widetilde{U}(t),\widetilde{\rho}_S(t) \right]- \kappa \int_{0}^t ds\; tr_{\mathcal{E}} \left\{ \left[ \widetilde{H}_{TI}(t), \left[ \widetilde{H}_{TI}(s), \widetilde{\rho}_S(t) \otimes \rho_{\mathcal{E}} \right] \right] \right\}\,.
\end{align}

Now to obtain the final master equations we have to evaluate the trace over the environmental degrees of freedom in this expression and thus to obtain the correlation functions. As a first step we take into account that the second term on the right hand side of \eqref{meqStart} can be written in terms of thermal Wightman functions, using the expression of $\widetilde{H}_{TI}(t)$ in position space given by
\begin{equation}
    \widetilde{H}_{TI} = \sqrt{\kappa}\int_{\mathbb{R}^3} d^3x \; \delta^i_b\; \varphi^b_a(\vec{x},t)\; \delta \mathcal{E}\t{}{^a_i}(\vec{x},t) \,.
\end{equation}
where $\varphi^b_a(\vec{x},t)$ denotes the three dimensional Fourier transforms of $\ft{\varphi}{}^b_a(\vec{k},t)$ which was defined in \eqref{defvarphi}. 
Following the procedure outlined in \cite{boyanovsky2015effective}, where in our case the environmental part of the interaction is linear in the environmental fields and we use a Gibbs state for the environment, hence all one-point correlation functions vanish, we can define the thermal Wightman functions:
\begin{align}
    G^{>}\t{}{^a_i^b_j}(\vec{x}-\vec{y},t-s) &:= \big\langle \delta \mathcal{E}\t{}{^a_i}(\vec{x},t) \: \delta \mathcal{E}\t{}{^b_j}(\vec{y},s) \big\rangle_\varepsilon = \t{P}{^a_i^b_j} G^{>}(\vec{x}-\vec{y},t-s)\label{eq:Ggr} \\
    G^{<}\t{}{^a_i^b_j}(\vec{x}-\vec{y},t-s) &:= \big\langle \delta \mathcal{E}\t{}{^b_j}(\vec{y},s) \: \delta \mathcal{E}\t{}{^a_i}(\vec{x},t) \big\rangle_\varepsilon = \t{P}{^a_i^b_j} G^{<}(\vec{x}-\vec{y},t-s) =G^{>}\t{}{^a_i^b_j}(\vec{y}-\vec{x},s-t)\,,\label{eq:Gkl}
\end{align}
where $\big\langle \cdot \big\rangle_\varepsilon$ denotes the expectation value with respect to a thermal Gibbs state and $\t{P}{^a_i^b_j}$ is the transverse-traceless projector given in \eqref{ttprojectorPositionSpace}. The detailed derivations of the identities in \eqref{eq:Ggr} and \eqref{eq:Gkl} as well as the derivations for the results up to \eqref{eq:GFeynman} can be found in appendix \ref{AppendixCorrelationFunctions}. Note that this definition coincides with the one in \eqref{eq:G++_prop} adapted to the model we consider here, hence the $\alpha$ and $\beta$ indices from \eqref{eq:G++_prop} here correspond each to the indices carried by $\delta \mathcal{E}$, i.e. to a spatial index $a$ as well as an internal index $i$. The spectral representations of the thermal Wightman functions in terms of a spectral density denoted by $\rho(k^0,||\vec{k}||)$ read:
\begin{align}
    G^{>}(\vec{x}-\vec{y},t-s) &= \int_{\mathbb{R}^4} \frac{d^4k}{(2\pi)^{\frac{4}{2}}} \rho^{>}(k^0,||\vec{k}||) e^{-ik^0 (t-s) +i\vec{k}(\vec{x}-\vec{y})} \\
     G^{<}(\vec{x}-\vec{y},t-s) &= \int_{\mathbb{R}^4} \frac{d^4k}{(2\pi)^{\frac{4}{2}}} \rho^{<}(k^0,||\vec{k}||) e^{-ik^0 (t-s) +i\vec{k}(\vec{x}-\vec{y})}
\end{align}
with
\begin{align}
    \rho^{>}(k^0,||\vec{k}||) := (1+ N(k^0)) \rho(k^0,||\vec{k}||) &&
    \rho^{<}(k^0,||\vec{k}||) := N(k^0) \rho(k^0,||\vec{k}||) 
\end{align}
and the spectral density
\begin{equation}
    \rho(k^0,||\vec{k}||) := \sqrt{\frac{\pi}{2}} \frac{1}{\Omega_k} \left[ \delta(k^0-\Omega_k) - \delta(k^0+\Omega_k) \right]\,.
\end{equation}
From this decomposition, it is immediately possible to see the additional effect caused by the finite temperature in the environment, evident by the presence of the Bose-Einstein distribution
\begin{equation}
    N(k^0) := \frac{1}{e^{\beta k^0}-1} = \frac{\coth\left(\frac{\beta k^0}{2}\right)}{2}-\frac{1}{2}\,,
\end{equation}
which vanishes for zero temperature parameter $\Theta = \beta^{-1}=0$ of the Gibbs state.  As usual the two thermal Wightman functions can be combined to build the thermal Feynman propagator:
\begin{align}
G^{(F)}\t{}{^a_i^b_j}(\vec{x}-\vec{y},t-s) &=  \big\langle \mathcal{T}_{\leftarrow} \; \delta \mathcal{E}\t{}{^a_i}(\vec{x},t) \: \delta \mathcal{E}\t{}{^b_j}(\vec{y},s) \big\rangle_\varepsilon \nonumber\\ &= \t{P}{^a_i^b_j}G^{>}(\vec{x}-\vec{y},t-s) \theta(t-s) + \t{P}{^a_i^b_j}G^{<}(\vec{x}-\vec{y},t-s) \theta(s-t) \,,
\end{align}
where $\theta(t)$ denotes the Heaviside step function which is one for  non-negative arguments and zero otherwise. A four-dimensional Fourier transform yields
\begin{align}\label{eq:GFeynman}
    \ft{G}^{(F)}\t{}{^a_i^b_j}(p) = \ft{P}\t{}{^a_i^b_j}(\vec{p}) \left[ \frac{i}{p^\mu p_\mu+i\epsilon} + 2\pi N(\Omega_p) \delta(p^\mu p_\mu)\right]\,,
\end{align}
where the first part is the ordinary Feynman propagator one obtains at a vanishing temperature parameter $\Theta$, that is when the Gibbs state merges into a vacuum state, and a second part the thermal contribution that obviously vanishes for $\Theta=0$. Note that we obtain a decomposition into the ordinary Feynman propagator and a thermal correction, this is caused by the fact that we use a mode expansion that involves a splitting into positive and negative frequency modes. As a consequence, normal ordered expectation values with respect to thermal states are in general non-vanishing as for instance $<b_k^\dagger b_k>_\epsilon = N(\Omega_k) \neq 0$. Vanishing expectation values of normal ordered products are usually an important property in the proof of Wick's theorem.  Given that we are working with a Gaussian state for the  environment and an interaction Hamiltonian being linear in the environmental fields, we can calculate the relevant expectation values in the model we consider directly without the use of Wick's theorem. However, for different and more complicated models the application of Wick's theorem might be of advantage and for this purpose it is often convenient to consider a different splitting than in the zero temperature case such that the expectation values of normal ordered products with respect to thermal states also vanish. A detailed discussion on possible splittings in this context as well as the relation between different choices of splittings is presented in \cite{evans1996wick}. Now we can write down the second term on the right side of in the master equation \eqref{meqStart} in terms of the thermal Wightman functions:
\begin{align}
    &\kappa \int_{0}^t ds\; tr_{\mathcal{E}} \left\{ \left[ \widetilde{H}_{TI}(t), \left[ \widetilde{H}_{TI}(s), \widetilde{\rho}_S(t) \otimes \rho_{\mathcal{E}} \right] \right] \right\} = \nonumber\\
   & = -\kappa \int_0^t ds\; \int_{\mathbb{R}^3} d^3x \int_{\mathbb{R}^3} d^3y \sum\limits_{r\in\{\pm\}} \Bigg\{ \left[ J_r(\vec{x},t) J_r(\vec{y},s) \widetilde{\rho}_S(t) -  J_r(\vec{y},s)\widetilde{\rho}_S(t)  J_r(\vec{x},t) \right] G^>(\vec{x}-\vec{y},t-s) \nonumber\\
   &\hspace{2.35in}  +\left[ \widetilde{\rho}_S(t)  J_r(\vec{y},s) J_r(\vec{x},t)-  J_r(\vec{x},t)\widetilde{\rho}_S(t)  J_r(\vec{y},s) \right] G^<(\vec{x}-\vec{y},t-s) \Bigg\}\,.
\end{align}

Since we consider a Gibbs state on Fock space which is not well defined, we would need to work with KMS states \cite{Kubo:1957mj, Schwinger:1959many} or alternatively regularise the system within a finite volume, where we will follow the latter in this work. For this purpose we put the system into a box of volume $V=L^3$ allowing us to explicitly show the above identities and evaluate thermal two-point functions. This kind of regularisation leads to the discreteness of modes that belong to the set $\mathds{K}$ now and to a replacement of the operator-valued distributions by operators:
\begin{equation}\label{BoxIntro}
    \int_{\mathbb{R}^3} \frac{d^3k}{(2\pi)^3} \longrightarrow \frac{1}{V} \sum\limits_{\vec{k}\in\mathds{K}} \hspace{1in} b_k^r \longrightarrow \sqrt{\frac{V}{(2\pi)^3}}\; b_{\vec{k}}^r\,,
\end{equation}
where a rescaling of the operators was introduced in order to keep the exponential $e^{-\beta H_{\mathcal{E}}}$ dimensionless in the regularised model:
\begin{equation}
    e^{-\beta H_{\mathcal{E}}} = \exp\left\{ -\beta \sum\limits_{r\in\{+,-\}} \sum\limits_{\vec{k}\in\mathds{K}}  \Omega_{\vec{k}} \left[(b_{\vec{k}}^r)^\dagger\; b_{\vec{k}}^r\right] \right\}\,.
\end{equation}
The Gibbs state for the environment is then of the form
\begin{equation}
    \rho_{\mathcal{E}} = Z_{\mathcal{E}}^{-1} e^{-\beta H_{\mathcal{E}}}
\end{equation}
with the partition sum given by

\begin{equation}
    Z_{\mathcal{E}} := tr_{\mathcal{E}} \left\{ e^{-\beta H_{\mathcal{E}}} \right\} = \left[ \prod_{j\in\mathbb{N}} \frac{1}{1-e^{-\beta \Omega_{\vec{k}_j}}} \right]^2\,,
\end{equation}

where the $j\in\mathbb{N}$ label the $\vec{k}_j\in\mathds{K}$.  Moreover, for the calculation we had to assume $\vec{k}=0 \not\in \mathds{K}$, which corresponds to $\Omega_{\vec{k}}=0$ and is the usual infrared divergence. More details on computing this partition sum can be found in appendix \ref{AppendixCorrelationFunctions}. Removing the regulator by taking the limit $L\to\infty$ in the Schrödinger picture we obtain the following TCL master equation for the system density operator $\rho_S(t)$:
\begin{align}\label{eq:mastereq1}
\frac{\partial}{\partial t} \rho_S(t) =& -i \left[H_S+\kappa \: {U}, \rho_S(t)\right]\nonumber\\ &-\frac{\kappa}{2}\int_{0}^t ds \sum\limits_{r} \int_{\mathbb{R}^3} d^3k \; \bigg\{ i D(\vec{k},t-s) \left[{J}_{r}(\vec k),\left\{ J_{r}(-\vec k,s-t), {\rho}_S(t)\right\}\right] \nonumber\\&\hspace{2.0in} + D_1(\vec{k},t-s) \left[{J}_{r}(\vec k),\left[ J_{r}(-\vec k,s-t),{\rho}_S(t)\right]\right] \bigg\}
\end{align}
with the matter system's Hamiltonian operator
\begin{equation}
    H_S := \int_{\mathbb{R}^3} d^3k \; \omega_k \, a_k^\dagger \, a_k
\end{equation}
as well as the operator $U$  for the gravitational self-interaction of the scalar field that was defined in \eqref{eq:defU}, the operator-valued distribution $J_r(\vec{k},t)$ defined in \eqref{eq:defJt} and $J_r(\vec{k}) = J_r(\vec{k},0)$.  The quantities $D(\vec{k},t-s)$ and $D_1(\vec{k},t-s)$ are the three-dimensional Fourier transforms of the dissipation and noise kernel. They can be related to the commutator's and anticommutator's correlation functions and also to the thermal Wightman functions $G^>(\vec{x}-\vec{y},t-s)$ and $G^<(\vec{x}-\vec{y},t-s)$, see also \eqref{eq:diss_kernel} and \eqref{eq:noise_kernel}:
\begin{align}
    \left[\delta \mathcal{E}\t{}{^a_i}(\vec{x},t), \delta \mathcal{E}\t{}{^b_j}(\vec{y},s)\right] &= \t{P}{^a_i^b_j} [G^>(\vec{x}-\vec{y},t-s) - G^<(\vec{x}-\vec{y},t-s)]\nonumber\\ &=i \t{P}{^a_i^b_j} \int_{\mathbb{R}^3} \frac{d^3k}{(2\pi)^{\frac{3}{2}}} e^{i\vec{k}(\vec{x}-\vec{y})}\; D(\vec{k},t-s) \\
    tr_{\mathcal{E}}\left(\left\{\delta \mathcal{E}\t{}{^a_i}(\vec{x},t), \delta \mathcal{E}\t{}{^b_j}(\vec{y},s)\right\}\right) &= \t{P}{^a_i^b_j} [G^>(\vec{x}-\vec{y},t-s) + G^<(\vec{x}-\vec{y},t-s)]\nonumber\\ &=\t{P}{^a_i^b_j} \int_{\mathbb{R}^3} \frac{d^3k}{(2\pi)^{\frac{3}{2}}} e^{i\vec{k}(\vec{x}-\vec{y})}\; D_1(\vec{k},t-s)\,,
\end{align}
where we used that for the model considered here the commutator in the first equation is proportional to the identity operator and thus can be pulled out of the trace. Hence, we get
\begin{align}\label{eq:diskern}
    D(\vec{k},t-s) &:=  -\frac{\sin(\Omega_k(t-s))}{\Omega_k}\\
\label{eq:noikern} D_1(\vec{k},t-s) &:= [2 N(\Omega_k)+1]\,\frac{\cos(\Omega_k(t-s))}{\Omega_k}\,.
\end{align}
To compare our results better with those already existing in the literature like for instance the ones in  \cite{anastopoulos2013master,oniga2016quantum,lagouvardos2021gravitational} it is of advantage to rewrite the master equation in \eqref{eq:mastereq1} in two equivalent ways. The first alternative and equivalent form is given by
\begin{align}\label{eq:mastereq2}
\frac{\partial}{\partial t} \rho_S(t) =& -i \left[H_S+\kappa\: {U}, \rho_S(t)\right]\nonumber\\ &+\frac{\kappa}{4} \int d^3k \sum\limits_r \Bigg\{ \frac{\coth\left(\frac{\beta\Omega_k}{2}\right)}{\Omega_k} \left[ J_r(\vec{k}), \left[ \rho_S(t), \widetilde{J}_r^\dagger(\vec{k},t) \right] \right] +\textrm{h.c.}  \nonumber \\
&\hspace{1.7in}+ \frac{1}{\Omega_k} \left[ J_r(\vec{k}), \left\{ \rho_S(t),\widetilde{J}_r^\dagger(\vec{k},t) \right\} \right] +\textrm{h.c.}\Bigg\}\,,
\end{align}
with
\begin{equation}
    \widetilde{J}_r(\vec{k},t) := \int_0^t ds \; e^{-i\Omega_k(t-s)} J_r(\vec{k},s-t)
\end{equation}
and where h.c. denotes the hermitian conjugate. This form of master equation is similar to the master equations derived  in \cite{anastopoulos2013master} and \cite{lagouvardos2021gravitational}. While the first reference investigates also a model where a scalar field is coupled to linearised gravity, the second one replaces the scalar field by photons. Let us briefly discuss the similarities and difference of these models and the one considered here: At the classical level one of the main difference is that we formulate the model in terms of Ashtekar variables whereas the models in \cite{anastopoulos2013master,lagouvardos2021gravitational} are based on ADM variables. Thus, we have to deal with an additional Gau\ss{} constraint in the model that also needs to be taken into account when gauge invariance is considered. While we presented a way to work with a gauge invariant formulation of the classical theory by means of constructing suitable Dirac observables and also showed that one can construct a canonical transformation on the original phase space that provides a natural separation of the physical and gauge sector of the phase space by the transformation in appendix \ref{app:Trafo}, in the papers \cite{anastopoulos2013master,lagouvardos2021gravitational} a specific gauge fixing is used to eliminate the gauge degrees of freedom. Furthermore, we consider a standard boundary term  \cite{thiemann1995generalized} that ensures that the action in terms of Ashtekar variables  as well as its variation is well defined and we expect that the corresponding ADM boundary term should be included in the formulation of the models in \cite{anastopoulos2013master,lagouvardos2021gravitational} as well. Note that in our case the boundary terms cancels the term in $\mathbf{\delta H}_{can}$ that is linear in the linearised geometric contribution to the Hamiltonian constraint and the remaining matter part combines with the $\frac{1}{\kappa}$ in front of the action to the energy density of the scalar field in $\kappa^0$ order. In contrast in \cite{anastopoulos2013master,lagouvardos2021gravitational} no perturbations of linear order have been considered in the canonical Hamiltonian but the $\kappa^0$ order is present. Therefore, effects of the missing boundary term cannot be seen in the final master equation when comparing the two results. 

For the quantised model let us comment on the work in  \cite{lagouvardos2021gravitational} and \cite{anastopoulos2013master} separately, starting with the latter.  One difference in the Hamiltonian operators in our result and the one in \cite{anastopoulos2013master} is the form of the self-interaction part because we choose a different normal ordering compared to \cite{anastopoulos2013master}. We chose to normal order the entire self-interaction part of the Hamiltonian operator, whereas in \cite{anastopoulos2013master} only the individual contributions $\hat{\ft{p}}{}_a$, $\hat{\ft{\epsilon}}$, $\hat{\ft{\widetilde{V}}}$ and $\ft{\hat{\varphi}}{}^a_b$ were normal ordered. Albeit the current operator valued distribution $J_r$ defined in \eqref{eq:defJschroed} is the same as in \citep{anastopoulos2013master}, it appears in their Hamiltonian \eqref{eq:HamiltonFockQU} with a different factor of $\sqrt{2}$ compared to our coupling. This difference has no physical effect because it is absorbed at the level of the master equation for the following reason: The different factor of $\sqrt{2}$ arises because the Hamiltonian for the pure gravitational part contains an additional factor of $\frac{1}{2}$ when expressed in Ashtekar variables and working with $\delta \mathcal{E}$ and $\delta \mathcal{C}$ compared to the one in ADM variables in \citep{anastopoulos2013master}. 
This leads to an additional factor $\frac{1}{\sqrt{2}}$ in our Fock quantised gravitational variables compared to the ADM variables. In addition in terms of ADM variables the interaction term in the Hamiltonian density reads $-\frac{\kappa}{2} \delta h_{ab}^{TT}\, T^{ab}$ while in Ashtekar variables it is given by $+\kappa \delta \mathcal{E}\t{}{^a_i} \delta^{bi}\: T_{ab}$.
This difference can be explained by the relation between ADM and Ashtekar variables. In case of ADM variables the gravitational physical degrees of freedom are given by  the transverse-traceless components $\delta h^{TT}_{ab} := P_{ab}^{cd} \delta h_{cd}$ and the interaction reads $\delta h^{TT}_{ab} T^{ab}$ which is then quantised in \cite{anastopoulos2013master} and where the coupling to the gravitational gauge degrees of freedom to the energy momentum tensor is contained in the self-interaction part. In contrast in our case, writing $\delta h_{ab}$ in terms of the perturbed cotriad, that  is $\delta h_{ab}(E) = -\delta_b^i\delta_{ac} \delta\t{E}{^c_i}-\delta_a^i\delta_{bc} \delta\t{E}{^c_i}+\delta_{ab} \delta_c^i \t{\delta E}{^c_i}$ and then using symmetry of the energy-momentum tensor, we obtain $\delta h_{ab}(E) T^{ab} \rightarrow [-2\delta_a^i \delta_{bc} +\delta_{ab}\delta_c^i]\t{\delta E}{^c_i} T^{ab}$, where the gravitational physical degrees of freedom only enter in the first term and we obtain in the linearised theory just $-2\delta \t{\mathcal{E}}{^a_i} \delta_b^i T^b_a$. This difference already present in the classical Hamiltonian has no effect on the final equations of motion because with our convention the Poisson bracket between $\delta \t{\mathcal{A}}{_a^i},\delta \t{\mathcal{E}}{^a_i}$ involves an additional factor of $\frac{1}{2}$ compared to the Poisson brackets used in \cite{anastopoulos2013master,lagouvardos2021gravitational}, cancelling the additional factor of 2. The minus sign is also cancelled because in our case the momentum variables couple to the energy momentum tensor whereas in \cite{anastopoulos2013master,lagouvardos2021gravitational} the coupling is via configuration variables. This kind of cancellation carries of course over to the commutator at the quantum level. 

A further difference is that  the final master equation presented in \cite{anastopoulos2013master} is of Lindblad form \eqref{eq:Lindblad}. This is a significant difference because as we will show in the following, the master equation in \eqref{eq:mastereq1} we obtain is not of Lindblad type yet and it requires further assumptions and approximations respectively to be of this form. For the purpose of discussing this we rewrite the master equation in \eqref{eq:mastereq1} in another equivalent form
\begin{align}\label{eq:mastereq3}
    \frac{\partial}{\partial t} \rho_S(t) =& -i [H_S+\kappa\: U, \rho_S(t)]\nonumber\\ &-\kappa \sum\limits_{r} \int_{\mathbb{R}^3} \frac{d^3k}{2 \Omega_k} \;\Big\{ [J_r^\dagger(\vec{k}),\widetilde{J}_r(\vec{k},t)\rho_S(t)] + N(\Omega_k) [J_r^\dagger(\vec{k}), [\widetilde{J}_r(\vec{k},t),\rho_S(t)]]+ \textrm{h.c.} \Big\}\,.
\end{align}
In this form, one can clearly separate the dissipator into a part associated with zero temperature (terms without $N(\Omega_k)$) and a second part corresponding to finite temperature (terms with $N(\Omega_k)$). All three forms of the master equation derived in this paper are time-convolutionless. However, they cannot (yet) be brought into Lindblad form, as there are still time-dependent quantities apart from the reduced density operator itself present in the dissipator. Nevertheless, we can bring the master equation into a form which is similar to the first standard form introduced in \eqref{eq:Lindblad} and which admits the splitting of the dissipator into a Lamb-shift Hamiltonian and a remaining contribution. To this extent, we start from the formulation in \eqref{eq:mastereq3}. First, we note that the $J_r$-operators can be split in the following way into pairs of creation and / or annihilation operator-valued distributions:
\begin{align}
J_r(\vec{k}) &= -\int \frac{d^3p}{(2\pi)^\frac{3}{2}} \sum\limits_{a=1}^4 j_r^a(\vec{k},\vec{p}) \\
\widetilde{J}_r(\vec{k},t) &= -\int \frac{d^3p}{(2\pi)^\frac{3}{2}} \sum\limits_{a=1}^4 j_r^a(\vec{k},\vec{p}) \; f(\Omega_k + \omega_a(\vec{k},\vec{p});t)
\end{align}
with
\begin{align}
j_r^1(\vec{k},\vec{p})&:=  a_p^\dagger a_{k+p} \frac{1}{2\sqrt{\omega_p \omega_{k+p}}}  \left[p_ap^b [P^{-r}(\vec{k})]^a_b \right] & \omega_1(\vec{k},\vec{p}) &:= \omega_p - \omega_{k+p}\\
j_r^2(\vec{k},\vec{p})&:=  a_{-p-k}^\dagger a_{-p} \frac{1}{2\sqrt{\omega_p \omega_{k+p}}} \left[p_ap^b [P^{-r}(\vec{k})]^a_b \right] & \omega_2(\vec{k},\vec{p}) &:= \omega_{k+p} - \omega_{p}\\
j_r^3(\vec{k},\vec{p})&:= a_{-p} a_{k+p} \frac{1}{2\sqrt{\omega_p \omega_{k+p}}} \left[p_ap^b [P^{-r}(\vec{k})]^a_b \right] & \omega_3(\vec{k},\vec{p}) &:= -\omega_p - \omega_{k+p}\\
j_r^4(\vec{k},\vec{p})&:= a_p^\dagger a_{-k-p}^\dagger \frac{1}{2\sqrt{\omega_p \omega_{k+p}}} \left[p_ap^b [P^{-r}(\vec{k})]^a_b \right] & \omega_4(\vec{k},\vec{p}) &:= \omega_p + \omega_{k+p}
\end{align}
and
\begin{equation}
f(\omega;t) := \int_0^t ds\; e^{-i\omega(t-s)} = \frac{i}{\omega} (e^{-i\omega t} -1)\,.
\end{equation}
Using this expansion in \eqref{eq:mastereq3}, the dissipator can be rewritten as
\begin{equation}
\mathcal{D}[\rho_S]=  \frac{\kappa}{2}\sum\limits_{r;a,b} \int_{\mathbb{R}^3} \frac{d^3k\: d^3p\: d^3l}{(2\pi)^\frac{6}{2}} \Bigg\{ \Delta_{ab}(\vec{p},\vec{l};\vec{k},t) \left[j_r^b(\vec{k},\vec{l})\rho_S(t), j^a_r(\vec{k},\vec{p})^\dagger\right]+h.c.\Bigg\}\,.
\end{equation}
with
\begin{align}
\label{eq:Deltaab}
\Delta_{ab}(\vec{p},\vec{l};\vec{k},t) :=& \frac{1}{\Omega_k}\left[(N(\Omega_k)+1) \:f(\Omega_k+\omega_b(\vec{k},\vec{l});t)  +N(\Omega_k)\;f(-\Omega_k+\omega_b(\vec{k},\vec{l});t) \right]\nonumber\\
=&2 \int_0^t ds\; G^>(\vec{k},t-s) e^{-i\omega_b(\vec{k},\vec{l})(t-s)} \,,
\end{align}
where $G^>(\vec{k},t-s)$ denotes the three dimensional Fourier transform of the Wightman function $G^>(\vec{x}-\vec{y},t-s)$.
Combining the new coefficient functions in the following form
\begin{align}
S_{ab}(\vec{p},\vec{l};\vec{k},t) &:= \frac{1}{2i} \left( \Delta_{ab}(\vec{p},\vec{l};\vec{k},t) - \Delta^*_{ba}(\vec{l},\vec{p};\vec{k},t) \right)\nonumber\\
&= \frac{1}{2\Omega_k} \Bigg[ [N(\Omega_k)+1] \Bigg\{ \frac{e^{-i(\Omega_k + \omega_b(\vec k,\vec l))t}-1}{\Omega_k + \omega_b(\vec k,\vec l)} +  \frac{e^{i(\Omega_k + \omega_a(\vec k,\vec p))t}-1}{\Omega_k + \omega_a(\vec k,\vec p)}\Bigg\}\nonumber\\
&\hspace{1in} - N(\Omega_k) \Bigg\{ \frac{e^{i(\Omega_k - \omega_b(\vec k,\vec l))t}-1}{\Omega_k - \omega_b(\vec k,\vec l)} +  \frac{e^{-i(\Omega_k - \omega_a(\vec k,\vec p))t}-1}{\Omega_k - \omega_a(\vec k,\vec p)}\Bigg\}\Bigg]\label{eq:SabMatrix}
\end{align}
\begin{align}
R_{ab}(\vec{p},\vec{l};\vec{k},t) &:=  \Delta_{ab}(\vec{p},\vec{l};\vec{k},t) + \Delta^*_{ba}(\vec{l},\vec{p};\vec{k},t)\nonumber\\
&= \frac{i}{\Omega_k} \Bigg[ [N(\Omega_k)+1] \Bigg\{ \frac{e^{-i(\Omega_k + \omega_b(\vec k,\vec l))t}-1}{\Omega_k + \omega_b(\vec k,\vec l)} -  \frac{e^{i(\Omega_k + \omega_a(\vec k,\vec p))t}-1}{\Omega_k + \omega_a(\vec k,\vec p)}\Bigg\}\nonumber\\
&\hspace{1in} - N(\Omega_k) \Bigg\{ \frac{e^{i(\Omega_k - \omega_b(\vec k,\vec l))t}-1}{\Omega_k - \omega_b(\vec k,\vec l)} -  \frac{e^{-i(\Omega_k - \omega_a(\vec k,\vec p))t}-1}{\Omega_k - \omega_a(\vec k,\vec p)}\Bigg\}\Bigg]\label{eq:RabMatrix}\,,
\end{align}
we can split the dissipator into two parts:
\begin{equation}
\mathcal{D}[\rho_S] = -i\kappa[H_{LS},\rho_S(t)] + \mathcal{D}_{\rm first}[\rho_S]
\end{equation}
with the Lamb-shift Hamiltonian operator
\begin{equation}\label{lsham}
H_{LS} := \frac{1}{2} \int \frac{d^3k\, d^3p\, d^3l}{(2\pi)^\frac{6}{2}} \sum\limits_{r;a,b} \; S_{ab}(\vec{p},\vec{l};\vec{k},t) \; j_r^a(\vec{k},\vec{p})^\dagger\, j_r^b(\vec{k},\vec{l})
\end{equation}
and a new dissipator term being in a form similar to the first standard form:
\begin{equation}\label{Dfsf}
\mathcal{D}_{\rm first}[\rho_S] := \frac{\kappa}{2} \int \frac{d^3k\, d^3p\, d^3l}{(2\pi)^\frac{6}{2}} \sum\limits_{r;a,b} \; {R}_{ab}(\vec{p},\vec{l};\vec{k},t) \left( j_r^b(\vec{k},\vec{l}) \rho_S(t)j^a_r(\vec{k},\vec{p})^\dagger-\frac{1}{2} \left\{ j^a_r(\vec{k},\vec{p})^\dagger j_r^b(\vec{k},\vec{l}), \rho_S(t)\right\} \right)\,,
\end{equation}
where the label 'first' refers to the dissipator in first standard form.
The difference to the first standard form shown in \eqref{eq:Lindblad} is that the coefficient function $R_{ab}$ is still time-dependent and thus the corresponding master equation is not of Lindblad type. Often a Lindblad form can be obtained for a given master equation by applying the second Markov approximation, that is by formally sending the upper limit of the integral involved in \eqref{eq:Deltaab} or directly $t$ in \eqref{eq:SabMatrix} and \eqref{eq:RabMatrix} to infinity leading to time-independent coefficients. Note that this is a priori problematic here as the complex exponentials in \eqref{eq:SabMatrix} and \eqref{eq:RabMatrix} do not have a well-defined limit for $t\rightarrow \infty$. This can be considered in the context of distributions along the lines of the Fourier transform of the step functions, see for instance \cite{Burrows:1990} and this will be analysed together with the remaining integration over the modes in detail in future work in \cite{fgkSoon}.
If we ignore such subtleties for the moment and simply send $t$ formally to infinity then the master equation from \cite{anastopoulos2013master} coincides structurally with the one in \eqref{eq:mastereq2} apart from the explicit form of their last term, which is is given by $\{J_r,[\rho_S,\widetilde{J}_r^\dagger]\} + h.c.$. In contrast in our derivation we actually end up with  $[J_r,\{\rho_S,\widetilde{J}_r^\dagger\}] +h.c.$, which yields different terms and from our calculation we do not see a way how their result can be reproduced. In \cite{anastopoulos2013master} no detailed derivation of this result is presented but the authors cite \cite{breuer2002theory}, where however also such terms are not involved in the master equations.

 As already mentioned, the master equation in \cite{lagouvardos2021gravitational} involves a photon field instead of a scalar field and thus the projectors involved are those that project on the physical degrees of freedom of the photon field. Most of the differences have already been pointed out above because the procedure in \cite{lagouvardos2021gravitational} is quite similar to the one followed in \cite{anastopoulos2013master}. Compared to \cite{anastopoulos2013master}, more details on the derivation of the Lindblad equation are presented in \cite{lagouvardos2021gravitational} and it is derived by means of a Born-Markov and a (weak) rotating wave approximation. The form of their master equations after the Born-Markov approximation is structurally the same as the one in \eqref{eq:mastereq2}, also the commutator-anticommutator structure agrees.

Note that the master equation in \eqref{eq:mastereq3} looks, apart from the expected differences due to their usage of ADM variables, similar to the one derived in \cite{oniga2016quantum}, where a Dirac quantisation was carried out and the physical degrees of freedom were identified by imposing the gauge conditions in the quantum theory and specialising to the transverse-traceless degrees of freedom. In a second step, they then derived a master equation using the influence functional approach for a general bosonic field. Their final master equation is a TCL master equation that, similar to the result derived in this paper, does also not exhibit Lindblad form. 

The physical investigation of the final master equation is complex, as the equation is in general not completely positive, in contrast to an equation of Lindblad-type. Apart from applying a second Markov approximation in order to arrive at second standard form, it is required to diagonalise the coefficient functions with respect to the labels $(a,\vec{p})$ and $(b,\vec{l})$. In this context an application of a rotating wave approximation is beneficial as has for instance been used in \cite{lagouvardos2021gravitational}. For the model considered in our work this needs a further detailed analysis because applying the above-proposed additional approximations to cast it into a completely positive form in general one looses some features of the dynamics of the system. Hence, we postpone this kind of detailed analysis into our companion paper \cite{fgkSoon} where we also will discuss the special case of a one-particle master equations. 

\section{Conclusions}
\label{sec:Conclusions}
In this article we considered a scalar field coupled to linearised gravity in a Post Minkowski approximation scheme as the classical starting point for a gravitationally induced decoherence model. To enable a later generalisation of the model to the case that  the gravitational sector is quantised using techniques from loop quantum gravity we used Ashtekar variables to construct the model. In contrast to \cite{anastopoulos2013master,Blencowe:2012mp,oniga2016quantum,lagouvardos2021gravitational} we include a suitable boundary term in our action ensuring that the action as well as its variation is well defined. As a preparation for applying a reduced phase space quantisation, the model is formulated at the gauge invariant level by means of suitable Dirac observables. The latter have been constructed using geometrical clocks in the relational formalism. Moreover, we constructed a canonical transformation on the full phase space that separates the gauge and physical degrees of freedom into two sets by introducing a dual observable map in which the role of constraints and clocks is interchanged. That the relevance of physical degrees of freedom and the implementation of gauge invariance plays a pivotal role in decoherence models has for instance been pointed out in \cite{Wilson-Gerow:2017masterthesis,Wilson-Gerow:2020jmv}, showing that if this is not taken care of it can lead to a misleading interpretation of results \cite{Blencowe:2012mp}. The results for the reduced phase space quantisation obtained here extend the results in the literature where so far ADM variables were used and the construction of Dirac observables \cite{anastopoulos2013master} was discussed only briefly and to the understanding of the authors not in a complete manner. We further considered a second choice of a set of geometrical clocks used in \cite{Dittrich:2006ee} based on the ADM clocks in \cite{arnowitt1962dynamics} and showed that the difference in the corresponding physical Hamiltonians only lies in the term encoding the self-interaction of the scalar field. That term, however, was not discussed in detail in \cite{Dittrich:2006ee} where only the coupling of the transverse-traceless variables to matter was considered which seems to our understanding problematic in the non-vacuum case.  Compared to former models in the literature we further analysed two different choices of geometrical clocks and discussed differences and similarities for these two choices. In particular for these two choices we obtain that the physical Hamiltonian differs in the self-interaction part of the scalar field only. In addition we also briefly investigated how different choices of the parameters for a given set of geometrical clocks associated with physical spatial and temporal coordinates affects the form of the physical Hamiltonian and its required fall-off behaviour in the linearised theory.  We showed  that for the choices taken in this work no issues arise as far as the fall-off behaviour of the physical Hamiltonian is concerned and we can confirm the results in \cite{arnowitt1962dynamics} using ADM variables that for the choices considered here the final physical Hamiltonian does not depend on the physical spatial and temporal coordinates.  The results obtained in \cite{oniga2016quantum} using a Dirac quantisation approach in the ADM framework are consistent with our results. The environment is chosen to be the physical sector of the quantum gravitational sector corresponding to the symmetric and transverse-traceless projections of the Dirac observables associated with the linearised Ashtekar variables. 

In this model we quantised the physical Hamiltonian using a Fock quantisation which should be understood as a first step towards the aim of formulating gravitationally induced decoherence models inspired by loop quantum gravity. In the quantisation we choose to normal order the entire physical Hamiltonian which differs from the operator orderings chosen in \cite{anastopoulos2013master,lagouvardos2021gravitational,oniga2016quantum}. However, what kind of possible effects these different orderings might have at the level of the master equations is complex to analyse at the current level of the model due to its complexity. We will postpone such an analysis to future work where we will apply further approximations and consider also the special case of one particle \cite{fgkSoon}. The final master equation was derived in the full relativistic framework by assuming a Gibbs state for the quantum gravitational environment and applying the the projection operator technique, no restriction to the non-relativistic sector was considered, as for instance done in \cite{Anastopoulos:1995ya,asprea2021gravitational,asprea2021gravitationalspin,Asprea:2021jag} during the derivation. We further discuss that for the assumptions the discussed model is based upon, the influence functional approach yields the same result.

The final master equation we end up with is not of Lindblad type and not of first standard form since it still carries time-dependent coefficient functions. Applying the the second Markov approximation for the model considered here is less trivial than in for instance some of the standard decoherence models in quantum mechanics because in the limit in which we send the upper limit of the temporal integral to infinity, we can no longer work with ordinary functions but need to consider distributions along the lines of \cite{Burrows:1990}. Combined with the fact that the result obtained from the remaining integration over all modes might not have the properties ordinary test functions have, it requires a careful analysis how this limit can be taken in general. Moreover, since properties like the positivity of the density matrix might only be valid for a restricted temporal interval, the second Markov approximation might be problematic from this aspect as well. An example of a model where the second Markov approximation cannot be applied can be found in \cite{Feller:2016zuk}. The reason here is that the chosen environmental operators are non-dynamical, yielding correlation functions with no dependence on the temporal coordinate. In the model in \cite{Feller:2016zuk} this results in a set of effective system operators in the final master equation that depend linearly on the temporal coordinate preventing them from applying the second Markov approximation. In contrast, the environmental operators in the model here depend linearly on the densitised triad operators present in the interaction Hamiltonian. This means that the correlation functions contain a combination of complex time dependent exponentials, where the application of the second Markov approximation is not excluded a priori. This aspect will be analysed in future work in \cite{fgkSoon}.

Let us ignore those subtleties for the moment and just formally apply the second Markov approximation and compare the resulting master equation to the results in \cite{anastopoulos2013master}. Here we obtain a slightly different result for the last term in their equation, where the roles of commutators and anticommutators are switched compared to our result and we are not able to reproduce the dissipator in the form presented in \cite{anastopoulos2013master}. The final master equation in \cite{lagouvardos2021gravitational} is based on a model in which  photons are coupled to linearised gravity. Compared to \cite{anastopoulos2013master} the derivation of the master equation is presented more in detail involving a Born-Markov as well as a rotating wave approximation. Our results structurally agree with the ones in \cite{lagouvardos2021gravitational} considering that the projector on the physical degrees of freedom are adapted to a coupling with photons. In this case commutators and anti-commutators enter in the same way as in our result.  Our results structurally agree with the master equation derived in \cite{oniga2016quantum} where the model is based on Dirac quantisation. In particular this model also ends up with time-dependent coefficients in the master equations that is, as in our case, not of Lindblad type unless further approximations are applied. Because the model presented in this work here uses Ashtekar variables compared to the models in \cite{anastopoulos2013master,Blencowe:2012mp,oniga2016quantum,lagouvardos2021gravitational,Anastopoulos:2021jdz}
it can be easier generalised to models where fermions are coupled whereas in the other cases one has to reformulate the model in terms of tetrads first. As discussed when reviewing the projection operator and influence functional technique, the formulation of the final master equation in terms of thermal Wightman functions can be done in a rather model-independent way and thus provides a good framework comparing and/or generalising a given decoherence model.

Note that in principle there are conceivable differences between the various top-down approaches depending on the nature of the environmental correlation functions. A stochastic environmental noise \cite{Kok:2003mc, Breuer:2008rh, Asprea:2021jag} does in general not capture the full complexity of the system-environment interaction due to it's fundamentally classical nature \cite{Anastopoulos:2021jdz}. Truncating the TCL master equation at a certain order does however also not guarantee an outcome with a sound physical interpretation for arbitrarily large time scales, positivity violations \cite{homa2019positivity} can warrant the introduction of additional terms in order to restore positivity \cite{breuer2002theory}. In some cases, additional approximations are able to ensure Lindblad form. Hence, it is indispensable to evaluate the features of the master equation for every model individually based on the assumptions and approximations performed in the course of its derivation.
~\\ 
~\\
The results obtained in this work build the basis for further investigations and generalisations of gravitationally induced decoherence models along the lines discussed in this article. The formulation of the model in a gauge invariant manner at the classical level using geometrical clocks can be easily carried over to couplings of matter other than just a scalar field because the linearised constraints will be of the form that they split into a purely geometric and matter contribution which simplified the construction of the canonical transformation and Dirac observables. Because with the analysis of the results obtained here using a Fock quantisation, we understand now in detail how our results compare to those where ADM variables are used. If we consider a loop quantisation for the gravitational sector in future work, we can therefore judge then clearly what kind of differences and effects stem from the fact that we do no longer apply Fock quantisation. Using a loop quantisation we expect that the derivation of a master equation will be more difficult in two aspects. First the transition to the interaction picture will be more complex because the underlying algebra is the one of holonomies and fluxes and secondly this also has the consequence that the commutator of two environmental operators in general will no longer just be almost commuting, that is proportional to the identity operator, which was a crucial property in the derivation of the master equation. Next to understanding the effects of different choices of quantisation, we also aim at investigating more in detail the physical properties of the model. Since this is a rather complicated task due to the still involved time-dependence in the final master equation, we want to consider the one-particle sector in future work where additional approximations can be handled more easily and such a specialisation further allows to make contact to the already existing bottom-up models used for instance in the context of decoherence effects in neutrino oscillations. Following this route we will be able to evaluate whether some of the free parameters involved in such bottom-up models are determined by the model derived in this work. For neutrino oscillations for instance we expect this to be the dependence on the power with which the neutrino energy enters into the contribution relevant for decoherence effects. Since we did not consider a coupling of fermions here we understand such a one-particle model as a first toy model to work with and postpone the derivation of such a model to future work \cite{fgkSoon}.
\begin{acknowledgments}
M.J.F and M.K both thank the Heinrich-B\"oll foundation for financial support. M.J.F. and K.G thank Thomas Eberl for illuminating discussions on decoherence models in the context of neutrino oscillations. M.J.F. thanks Thomas Thiemann for useful comments on his master's thesis on which part of the work here is based on.
\end{acknowledgments}

\begin{appendices}
\renewcommand{\thesection}{A.\Roman{section}}
\renewcommand{\thesubsection}{\thesection.\arabic{subsection}}
\renewcommand{\thesubsubsection}{\thesubsection.\arabic{subsubsection}}
\renewcommand{\theequation}{\thesection.\arabic{equation}}

\makeatletter
\renewcommand{\p@subsection}{}
\renewcommand{\p@subsubsection}{}
\makeatother
\section{More details on the construction of Dirac observables}
\label{app:ConstrDiracObs}
In this appendix we present more details on the construction of the Dirac observables that we use in the main text to formulate the decoherence model at fully gauge invariant level.

\subsection{More detailed discussion on the choice of reference fields for the linearised constraints}
\label{app:DetailsRefFields}
Defining the smeared linearised Gauß, Hamiltonian and spatial diffeomorphism constraints as
\begin{align}
\delta G[\delta \Lambda](t) &:= \int_\sigma d^3y\; \delta \Lambda^i(\vec{x},t)\, \delta G_i(\vec{x},t)\\
    \delta C[\delta N](t) &:= \int_\sigma d^3x \; \delta N(\vec{x},t) \,\delta C(\vec{x},t) \\
    \overrightarrow{\delta C}[\overrightarrow{\delta N}](t) &:= \int_\sigma d^3x\; \delta N^a(\vec{x},t)\, \delta C_a(\vec{x},t)
\end{align}
one can evaluate the gauge transformations they infer on the phase space variables. As the Gauß constraint does not contain matter degrees of freedom, it leaves the matter fields unchanged and just modifies the geometrical degrees of freedom:
\begin{align}
\{\phi(\vec{x},t),\delta G[\delta \Lambda](t)\} =\{ \pi(\vec{x},t),\delta G[\delta \Lambda](t)\} &=0 \\
\{\t{\delta E}{^a_i}(\vec{x},t) ,\delta G[\delta \Lambda](t)\} &= \frac{1}{2} \epsilon^{ijk} \t{\delta \Lambda}{_j}(\vec{x},t)\, \delta_k^a\\
\{\t{\delta A}{_a^i}(\vec{x},t) , \delta G[\delta \Lambda](t)\} &=- \frac{1}{2} \t{\partial}{_a} (\delta \Lambda^i(\vec{x},t)) \,.
\end{align}
For the Hamiltonian constraint we find the following gauge transformations:
\begin{align}
\{\phi(\vec{x},t),\delta C[ \delta N](t) \} &= \kappa \delta N(\vec{x},t) \, \pi(\vec{x},t)  \label{eq:PhiTrafoHam}
\\\{\pi(\vec{x},t),\delta C[ \delta N](t) \} &= \kappa \partial_a \left[\delta N(\vec{x},t)\, \partial^a\phi(\vec{x},t)\right] -\kappa \delta N(\vec{x},t)\, m^2 \,\phi(\vec{x},t) 
\\ 
\{ \t{\delta E}{^a_i}(\vec{x},t) ,\delta C[\delta N](t)\} &= -\beta\, \epsilon^{ijk}\, \delta_j^a\, \delta_k^b\, \t{\partial}{_b} (\delta N(\vec{x},t)) \\
\{ \t{\delta A}{_a^i}(\vec{x},t) ,\delta C[\delta N](t)\} &=0\,.
\end{align}
Finally, the gauge transformations induced by the spatial diffeomorphism constraint are
\begin{align}
\{\phi(\vec{x},t),\overrightarrow{\delta C}[\overrightarrow{\delta N}](t)\} &=\kappa \delta N^a(\vec{x},t)\, \partial_a \phi(\vec{x},t)\label{eq:PhiTrafoDiffeo}
\\ \{\pi(\vec{x},t),\overrightarrow{\delta C}[\overrightarrow{\delta N}](t) \} &= \kappa \partial_a [\delta N^a(\vec{x},t)\, \pi(\vec{x},t)]
\\ \label{eq:GTEveccon}\{\t{\delta E}{^a_i}(\vec{x},t),\overrightarrow{\delta C}[\overrightarrow{\delta N}](t) \} &= -\frac{1}{2} \left( \delta_i^b\, \t{\partial}{_b} (\delta N^a(\vec{x},t)) - \delta_i^a\, \t{\partial}{_b} (\delta N^b(\vec{x},t)) \right)\\
\{\t{\delta A}{_a^i}(\vec{x},t),\overrightarrow{\delta C}[\overrightarrow{\delta N}](t) \} &=0\,.
\end{align}

\subsubsection{Choice of the reference field for the linearised Gauß constraint}
As the choice of reference fields can be understood as gauge fixing constraints in a corresponding gauge fixed theory, we choose a reference field for the Gauß constraint that implements a Lorentz-like gauge condition analogous to \cite{ashtekar1991gravitons} for the connection perturbation, i.e.
\begin{equation}
    \partial^a (\delta \t{A}{_a^i}(\vec{x},t))=0\,.
\end{equation}
Evaluation of the Poisson bracket between this condition and the Gauß constraint \eqref{eq:linGauss} using the linearised Poisson bracket \eqref{eq:poissongravitylin} yields
\begin{equation}
    \{\partial^a (\delta \t{A}{_a^i}(\vec{x},t)), \delta G_j(\vec{y},t)\} =- \frac{1}{2} \delta_j^i \Delta_{\vec{x}} \delta^3(\vec{x}-\vec{y})\,,
\end{equation}
where $\Delta_{\vec{x}}$ denotes the Laplacian with respect to $\vec{x}$. We will drop the subscript if only one coordinate is involved.
In order to have the commutation relation
\begin{equation}
\{ \delta\Xi^i(\vec{x},t), \delta G_j(\vec{y},t)\} = \frac{1}{\kappa}\delta_j^i \delta^3(\vec{x}-\vec{y})\,,
\end{equation}
we define the reference fields for the Gauß constraint as
\begin{equation}
\delta\Xi^i(\vec{x},t) := \frac{2}{\kappa} \partial^a \left(\delta \t{A}{_a^i} * G^\Delta\right)(\vec{x},t) = \frac{2}{\kappa} \int d^3y \; \partial_{\vec{x}}^a G^\Delta(\vec x -\vec y) \, \t{\delta A}{_a^i}(\vec y)
\end{equation}
with $\partial_{\vec{x}}^a := \frac{\partial}{\partial x_a}$.
The abbreviation $G^\Delta(\vec x-\vec y)$ denotes the Green's function of the Laplacian,
\begin{equation}
    G^\Delta(\vec x -\vec y) =  \int \frac{d^3k}{(2\pi)^3} \frac{1}{||\vec k||^2} e^{i\vec{k}(\vec{x}-\vec{y})}\,.
\end{equation}
Due to the fact that the connection remains invariant under the Hamiltonian and spatial diffeomorphism constraint, this reference field also remains invariant under transformations induced by these two constraints.

\subsubsection{Choice of the reference field for the linearised Hamiltonian constraint}
The Hamiltonian constraint leaves the connection invariant and only transforms the part $\t{\epsilon}{_a^c^b} \delta_c^i \partial_b \delta E\t{}{^a_i}(\vec{x},t)$ of the densitised triad in the following manner:
\begin{equation}
    \{ \t{\epsilon}{_a^c^b}\delta_c^i \partial_b \delta E\t{}{^a_i}(\vec{x},t), \delta C(\vec{y},t)\} = 2\beta \Delta_{\vec{x}}  \delta^3(\vec{x}-\vec{y})\,.
\end{equation}
This suggests to define a reference field
\begin{equation}
\delta\widetilde{T}(\vec{x},t) := -\frac{1}{2\beta\kappa} \t{\epsilon}{_a^c^b}\delta_c^i \partial_b \left(\delta E\t{}{^a_i} * G^\Delta \right)(\vec{x},t)  \,.
\end{equation}
However, in this form the reference field is not invariant under gauge transformations induced by the Gauß constraint:
\begin{equation}
    \{ \delta\widetilde{T}(\vec{x},t), \delta G_j(\vec{y},t)\} =-\frac{1}{2\beta\kappa } \delta_j^a\partial_a^{\vec{x}} G^\Delta(\vec{x}-\vec{y})\,,
\end{equation}
where $\partial_a^{\vec{x}}$ denotes the partial derivative with respect to $x^a$. To cure this, we seek to subtract some combination of geometrical phase space variables that transforms precisely the same way as $\delta\widetilde{T}$ under the Gauß constraint and remains invariant under the Hamiltonian constraint. The latter is true for any combination of the connection variables, and it turns out that the trace of the connection solves the problem. Hence a good choice for a reference field corresponding to the Hamiltonian constraint is
\begin{equation}
  \delta  T(\vec{x},t) := -\frac{1}{\kappa\beta} \left[ \frac{1}{2} \t{\epsilon}{_a^c^b} \delta_c^i\partial_b \left(\t{\delta E}{^a_i} * G^\Delta\right)(\vec{x},t) + \delta^a_i \left(\t{\delta A}{_a^i} * G^\Delta\right)(\vec{x},t)\right]\,,
\end{equation}
which also commutes with the spatial diffeomorphism constraint. Additionally, it commutes with the reference field for the Gauß constraint:
\begin{equation}
    \{\delta T(\vec{x},t), \Xi^i(\vec{y},t)\} =0\,.
\end{equation}

\subsubsection{Choice of the reference field for the linearised spatial diffeomorphism constraint}
In a last step we have to find a suitable reference field for the spatial diffeomorphism constraint. The list of requirements for this field is motivated by the application in the main part of this work and we will construct it in several steps. In the end, the final reference field $\delta T^a(\vec{x},t)$ should
\begin{itemize}
    \item consists of linearised elementary gravitational degrees of freedom only
    \item fulfill $\{ \delta T^a(\vec{x},t), \delta C_b(\vec{y},t)\} = \frac{1}{\kappa} \delta_b^a \delta^3(\vec{x}-\vec{y})$, 
    \item commute with the remaining linearised constraints and
    \item commute with the remaining reference fields.
\end{itemize}
For the first step it is helpful to realise from \eqref{eq:GTEveccon} that only the trace and one of the longitudinal parts of the densitised triad will be modified by this constraint in the following manner:
\begin{align}
    \{\delta^i_c\partial^c \delta \t{E}{^a_i}(\vec{x},t) , \delta C_b(\vec{y},t) \}&= \frac{1}{2} (\partial^a_{\vec{x}} \partial_b^{\vec{x}} - \delta_b^a \Delta_{\vec{x}} ) \delta^3(\vec{x}-\vec{y}) \\
    \{\delta_a^i \delta \t{E}{^a_i}(\vec{x},t) , \delta C_b(\vec{y},t) \}&= \partial_b^{\vec{x}} \delta^3(\vec{x}-\vec{y})\,.
\end{align}
A suitable combination for a reference field that is canonically conjugated to the linearised spatial diffeomorphism constraint is therefore
\begin{equation}
    \delta \widetilde{\widetilde{T}} {}^a(\vec{x},t):=\frac{2}{\kappa} \left(  \delta^a_b \delta_c^i \partial^c -\frac{1}{2} \delta^i_b \partial^a \right) \left(\delta \t{E}{^b_i} * G^\Delta \right)(\vec{x},t)\,.
\end{equation}
Unfortunately, this combination is not invariant under the linearised Gauß constraint:
\begin{equation}
    \{\delta \widetilde{\widetilde{T}} {}^a(\vec{x},t), \delta G_j(\vec{y},t)\} = -\frac{1}{\kappa} \t{\epsilon}{_c_b^a}\delta^c_j \partial^b_{\vec{x}} G^\Delta(\vec{x}-\vec{y})\,.
\end{equation}
To cure this, we can add a suitable form of 
\begin{equation}
    \partial_a \delta \t{E}{^a_i}(\vec{x},t)\,,
\end{equation}
which is just a scalar density and therefore invariant under the linearised spatial diffeomorphism constraint. It turns out that
\begin{equation}\label{eq:qFirstVectorClock}
   \delta  \widetilde{T}^a(\vec{x},t):= \frac{2}{\kappa} \left(  \delta^a_b \delta^i_c \partial^c -\frac{1}{2} \delta^i_b \partial^a + \delta^{ac}\delta_c^i \partial_b \right) \left(\delta \t{E}{^b_i} * G^\Delta\right)(\vec{x},t)
\end{equation} 
is invariant under the linearised Gauß constraint and still canonically conjugated to the linearised spatial diffeomorphism constraint. A quick calculation also shows that it is invariant under the gauge transformations generated by the linearised Hamiltonian constraint. The last bullet point which we required, namely vanishing Poisson bracket with the other reference fields, is not yet fulfilled. To construct an extension of \eqref{eq:qFirstVectorClock} that commutes with the remaining reference fields in the lowest order and maintains the additional requirements listed in the beginning, we use the linearised dual observable map introduced in \eqref{eq:LinObsdual}. The advantage is that we already know that the constraints are abelian up to first order in $\kappa$, so this extension will have exactly the same behaviour as $\delta\widetilde{T}^a(\vec{x},t)$ in the  Poisson brackets with the linearised constraints. As the clock in the dual observable map we choose the geometric contribution of the linearised Hamiltonian constraint given by $\delta C-\kappa\epsilon$ because we want the final clock to depend on the geometrical degrees of freedom only. Since we have $\{\delta T(x),(\delta C-\kappa\epsilon)(y)\}=\frac{1}{\kappa}\delta(x,y)$ this is a suitable choice. 
The result of the application of the dual linearised observable map on $\delta\widetilde{T}{}^a$ yields a suitable reference field corresponding to the spatial diffeomorphism constraint:
\begin{align}
\delta T^a(\vec{x},t) :=& \frac{2}{\kappa} \left(  \delta^a_b \delta^i_c \partial^c -\frac{1}{2} \delta^i_b \partial^a + \delta^{ac}\delta_c^i \partial_b \right) \left(\delta \t{E}{^b_i} * G^\Delta\right)(\vec{x},t) \nonumber\\ &+\frac{4\beta}{\kappa^2}\left[ \frac{1}{2} \delta^i_b \partial^a \partial^b \left( \delta G_i * G^{\Delta\Delta}\right)(\vec x,t) - \delta^{ab}\delta_b^i \left( \delta G_i * G^\Delta\right)(\vec x,t) \right]  + \frac{1}{\kappa^2} \partial^a \left[\left(\delta C-\kappa\epsilon\right) * G^{\Delta\Delta} \right](\vec{x},t)\,,
\end{align}
where $G^{\Delta\Delta}$ denotes the Green's function of the squared Laplacian, that is
\begin{equation}
    G^{\Delta\Delta}(\vec x-\vec y) = \int \frac{d^3k}{(2\pi)^3} \frac{1}{||\vec k||^4} e^{i\vec{k}(\vec x-\vec y)} = \int d^3z \; G^\Delta(\vec x -\vec z) \, G^\Delta(\vec y-\vec z)\,.
\end{equation}
This reference field is defined on the entire phase space. On the constraint hypersurface, the additional constraint terms drop and we are left again with $\delta\widetilde{T}^a(\vec{x},t)$, which still commutes with the constraints in the desired way.
This concludes the construction of the reference fields for the model under consideration.

\subsection{Choice of basis for the calculations in Fourier space}\label{app:Basis}
When discussing the physical degrees of freedom and also in the context of gauge fixing and the relation between the fully gauge invariant and gauge fixed model, it turns out to be helpful to work in Fourier space. For these calculations we choose a certain complex basis in Fourier space, which is the same as in \cite{ashtekar1991gravitons} consisting of 
\begin{equation}
    \hat{k}^a := \frac{k^a}{||\vec{k}||} \hspace{0.3in} m^a(\vec{k}) \hspace{0.3in} \overline{m}^a(\vec{k})\,,
\end{equation}
where $\vec{k}$ is the momentum and we understand the plane perpendicular to it as a complex plane spanned by the two unit vectors $m^a(\vec{k})$ and $\overline{m}^a(\vec{k})$, where the bar denotes complex conjugation. This basis is orthonormal in the following manner:
\begin{align}
\hat{k}^a \hat{k}_a = m^a(\vec{k}) \overline{m}_a(\vec{k}) = \overline{m}^a(\vec{k}) m_a(\vec{k}) &= 1\\
\hat{k}^a m_a(\vec{k}) = \hat{k}^a \overline{m}_a(\vec{k}) = m^a(\vec{k}) m_a(\vec{k}) = \overline{m}^a(\vec{k}) \overline{m}_a(\vec{k}) &= 0\,.
\end{align}
The expansion of the metric in this basis is
\begin{equation}
q_{ab}(\vec{k}) = \hat{k}_a \hat{k}_b + m_a(\vec{k})\, \overline{m}_b(\vec{k}) + \overline{m}_a(\vec{k})\, m_b(\vec{k})\,,
\end{equation}
the soldering form reads
\begin{equation}
\delta_a^i(\vec{k}) = \hat{k}_a \hat{k}^i + m_a(\vec{k})\, \overline{m}^i(\vec{k}) + \overline{m}_a(\vec{k})\, m^i(\vec{k})\,,
\end{equation}
and we fix the orientation of the basis vectors in the complex plane relative to each other by imposing
\begin{equation}
\epsilon_{abc}\, \hat{k}^a\, m^b(\vec{k})\, \overline{m}^c(\vec{k}) = -i\,.
\end{equation}
We additionally impose (see \cite{fahn2020masterthesis} section 5.3):
\begin{equation}
    m^a(-\vec{k})= \overline{m}^a(\vec{k})\,.
\end{equation}
We expand the Fourier transform of the gravitational fields in this basis, following the notation in \cite{ashtekar1991gravitons}:
\begin{align}\label{eq:EinMomSpace}
\ft{\delta E}\t{}{^a_i}(\vec{k}) =& \ft{\delta E}^{+}(\vec{k}) m^a(\vec{k})m_i(\vec{k}) + \ft{\delta E}^{-}(\vec{k}) \overline{m}^a(\vec{k}) \overline{m}_i(\vec{k}) + \ft{\delta E}^1(\vec{k}) \hat{k}^a m_i(\vec{k}) + \ft{\delta E}^{\overline{1}}(\vec{k}) \hat{k}^a \overline{m}_i(\vec{k}) \nonumber\\&+ \ft{\delta E}^2(\vec{k}) m^a(\vec{k}) \hat{k}_i + \ft{\delta E}^{\overline{2}}(\vec{k}) \overline{m}^a(\vec{k}) \hat{k}_i + \ft{\delta E}^3(\vec{k}) \hat{k}^a \hat{k}_i \nonumber\\ &+ \ft{\delta E}^4(\vec{k}) m^a(\vec{k}) \overline{m}_i(\vec{k}) + \ft{\delta E}^5(\vec{k}) \overline{m}^a(\vec{k}) m_i(\vec{k})
\end{align}
and
\begin{align}\label{eq:AinMomSpace}
\ft{\delta A}\t{}{_a^i}(\vec{k}) =& \ft{\delta A}^{+}(\vec{k}) m_a(\vec{k})m^i(\vec{k}) + \ft{\delta A}^{-}(\vec{k}) \overline{m}_a(\vec{k}) \overline{m}^i(\vec{k}) + \ft{\delta A}^1(\vec{k}) \hat{k}_a m^i(\vec{k}) + \ft{\delta A}^{\overline{1}}(\vec{k}) \hat{k}_a \overline{m}^i(\vec{k}) \nonumber\\&+ \ft{\delta A}^2(\vec{k}) m_a(\vec{k}) \hat{k}^i + \ft{\delta A}^{\overline{2}}(\vec{k}) \overline{m}_a(\vec{k}) \hat{k}^i + \ft{\delta A}^3(\vec{k}) \hat{k}_a \hat{k}^i\nonumber\\ &+ \ft{\delta A}^4(\vec{k}) m_a(\vec{k}) \overline{m}^i(\vec{k}) + \ft{\delta A}^5(\vec{k}) \overline{m}_a(\vec{k}) m^i(\vec{k})\,.
\end{align}
From this expansion it is evident that $\ft{\delta A}^\pm$ and $\ft{\delta E}^\pm$ encode the transverse traceless degrees of freedom with corresponding projectors
\begin{equation}\label{TTprojectors}
    \ft{P}\t{}{^a_i^j_b}(\vec{k}) := \overline{m}^a(\vec{k}) \overline{m}_i(\vec{k}) m^j(\vec{k}) m_b(\vec{k}) + m^a(\vec{k}) m_i(\vec{k}) \overline{m}^j(\vec{k}) \overline{m}_b(\vec{k})
\end{equation}
for $\ft{\delta E}$ and $\ft{P}\t{}{_a^i_j^b}(\vec{k}) := \delta_{ac} \delta^{il} \delta_{jm} \delta^{bd}\ft{P}\t{}{^c_l^m_d}(\vec{k})$ for $\ft{\delta A}$. Their position space version was given in \eqref{ttprojectorPositionSpace} and involves highly-nonlocal terms. Hence we set
\begin{align}
    \ft{\delta \cal E}\t{}{^a_i}(\vec k) &:=  \ft{\delta E}^{+}(\vec{k}) m^a(\vec{k})m_i(\vec{k}) + \ft{\delta E}^{-}(\vec{k}) \overline{m}^a(\vec{k}) \overline{m}_i(\vec{k})\\
    \ft{\delta \cal A}\t{}{_a^i}(\vec k) &:=  \ft{\delta A}^{+}(\vec{k}) m_a(\vec{k})m^i(\vec{k}) + \ft{\delta A}^{-}(\vec{k}) \overline{m}_a(\vec{k}) \overline{m}^i(\vec{k})\,.
\end{align}
\\
The Dirac observables constructed in \eqref{obsdelE} and \eqref{obsdelA} can be expressed in terms of this basis up to linear order and read, where we drop the $\vec k$ dependency for better readability and already set $\delta \xi^j=0$:
\begin{align}
    (\ft{\delta E}\t{}{^a_i})^{GI}=& \ft{\delta \cal E}\t{}{^a_i}+ \hat{k}^a m_i \left(\ft{\delta E}^1 +\frac{1}{||\vec k||} \ft{\delta A}^1\right) + \hat{k}^a \overline{m}_i \left(\ft{\delta E}^{\overline{1}} -\frac{1}{||\vec k||} \ft{\delta A}^{\overline{1}}\right) -m^a \hat{k}_i \frac{1}{||\vec k||} \ft{\delta A}^2\nonumber\\
    &\hspace{0.3in} +\overline{m}^a \hat{k}_i \frac{1}{||\vec k||} \ft{\delta A}^{\overline{2}} + \hat{k}^a\hat{k}_i \ft{\delta E}^3 - m^a \overline{m}_i \frac{2}{||\vec k||} \ft{\delta A}^4 + \overline{m}^a m_i \frac{2}{||\vec k||} \ft{\delta A}^5\nonumber\\ \label{eq:dirobsmombE}
    &\hspace{0.3in} +\frac{i\kappa}{2}\left(\delta_i^a k_c - \delta^a_c \delta_i^b k_b \right) \ft{\delta \sigma}^c- i\beta\kappa \t{\epsilon}{_c^a^b}\delta_i^c k_b \ft{\delta \tau}
    +O(\delta^2,\kappa)\\ \label{eq:dirobsmombA}
     (\ft{\delta A}\t{}{_a^i})^{GI}=& \ft{\delta \cal A}\t{}{_a^i}+\ft{\delta A}^2 m_a \hat{k}^i + \ft{\delta A}^{\overline{2}} \overline{m}_a \hat{k}^i + \ft{\delta A}^4 m_a \overline{m}^i + \ft{\delta A}^5 \overline{m}_a m^i+O(\delta^2,\kappa)\,.
\end{align}

\subsection{Further details on computing the algebra of Dirac observables}
\label{app:AlgebraObs}
In this appendix we present further details on the algebra of Dirac observables used in the main text. 

\subsubsection{Original matter and geometrical variables commute with clocks}
\label{app:PhysDofCommClocks}
In this appendix we will discuss that the original phase space variables in the matter sector, these are $ (\phi,\pi)$ as well as the transverse-traceless components of the gravitational sector, these are  $(\t{P}{_a^i_j^b}(\t{\delta{A}}{_b^j}),\t{P}{^a_i^j_b}(\t{\delta{E}}{^b_j}))$ will commute with all clocks $\delta T, \delta T^a$ and $\delta\Xi^j$ chosen for the Hamiltonian, spatial diffeomorphism and Gau\ss{} constraint respectively. \\ 
First we note that the clocks being geometrical ones do not contain any matter variables and also the additional contributions form the dual observable map involved in the diffeomorphism clock contains the geometric degrees of freedom only. Hence, all clocks trivially commute with the matter fields $(\phi,\pi)$. 
To see that they also commute with the variables $(\t{P}{_a^i_j^b}(\t{\delta{A}}{_b^j}),\t{P}{^a_i^j_b}(\t{\delta{E}}{^b_j}))$, it is convenient to express them in Fourier space. This can be found in section \ref{app:Trafo} and one can see that all clocks are independent of $\ft{\delta A}^\pm$ and $\ft{\delta E}^\pm$. As the latter are the transverse-traceless degrees of freedom and these are elementary phase space variables they have vanishing Poisson brackets with all remaining degrees of freedom and thus also commute with all clocks.

\subsubsection{Computing the algebra of the Dirac observables by means of using the  Dirac observable of the corresponding Dirac bracket}
\label{app:AlgRelDiracBracket}
In this appendix we will show how the algebra of the Dirac observables is related to the Dirac observable of the Dirac bracket. Here we will discuss the special case and order in perturbation theory that is relevant for the model under consideration in this work and we refer the reader for the proof in the general case to \cite{Thiemann:2004wk,Dittrich:2005kc}. In the application where we need this relation we will assume that we consider linearised geometrical clocks $\delta T^I(x)$ with $\delta^2T^I(x)=0$ that only depend on the gravitational degrees of freedom. We then introduce $\delta{\cal G}^I(x):=\delta T^I-\tau^I(x)$ that mutually commute with a set of constraints ${\cal C}^{\prime}_J(x)$ up to corrections of order $\delta^2$ that is
\begin{equation*}
M^I_J(x,y):=\{\delta{\cal G}^I(x),\delta{\cal C}^{\prime}_J(y)\}=\frac{1}{\kappa}\delta^I_J\delta^{(3)}(x,y)+O(\delta^2).
\end{equation*}
Given the definition of $M^I_J(x,y)$ above its inverse is simply given by
\begin{equation*}
(M^{-1})^I_J(x,y):=\kappa\delta^I_J\delta^{(3)}(x,y)+O(\delta^2,\kappa^2)
\end{equation*}
with 
\begin{equation*}
\int d^3z  M^I_J(x,z)(M^{-1})^J_K(z,y) = \delta^I_K\delta^{(3)}(x,y). 
\end{equation*}
Note that we chose to denote the constraints with a prime here because this is the notation used in the main text. 
This is exactly the setup needed to analyse the algebra of Dirac observables for the model in this work.  Since we only perturb the gravitational degrees of freedom but not the matter ones, it is convenient to discuss these two types separately. We will start with the geometric sector and afterwards discuss in detail how the results can be carried over to the algebra of the matter variables. In the framework of field theory the Dirac bracket $\{\cdot,\cdot\}^*$ reads
\begin{eqnarray}
\{ f(x), g(y)\}^* &=& \{f(x), g(y)\}\\
&&-\int d^3z^\prime\int d^3z^{\prime\prime}\{ f(x), {\cal C}^{\prime}_L(z^\prime)\}(M^{-1})^L_M(z^\prime, z^{\prime\prime})\{{\cal G}^M(z^{\prime\prime}),g(y)\} \nonumber \\
&&+\int d^3z^\prime\int d^3z^{\prime\prime}\{ g(y), {\cal C}^{\prime}_L(z^\prime)\}(M^{-1})^L_M(z^\prime, z^{\prime\prime})\{{\cal G}^M(z^{\prime\prime}),f(x)\} \nonumber \\
&=&\{f(x), g(y)\}-\int d^3z\{ f(x), {\cal C}^{\prime}_L(z)\}\delta^L_M\{T^M(z),g(y)\} \nonumber \\
&&+\int d^3z\{ g(y), {\cal C}^{\prime}_L(z)\}\delta^L_M\{T^M(z),f(x)\} \nonumber\, ,
\end{eqnarray}
where we used the explicit form of $(M^{-1})^J_K(z,y)$ together with the fact that inside the Poisson bracket we can replace ${\cal G}^M$ by $T^M$ because they differ only by a phase space independent function. Quantities involving gravitational variables are perturbed in the model in this article. Let us denote their linear and second-order perturbations by $\delta f$ and $\delta^2 f$ respectively. Then we consider the perturbation of the Dirac bracket up to linear order that is given by
\begin{eqnarray}
\{ f(x), g(y)\}^* &=& \{\delta f(x),\delta g(y)\}
-\kappa\int d^3z\{ \delta f(x), \delta {\cal C}^{\prime}_L(z)\}\delta^L_M\{\delta T^M(z),\delta g(y)\} \\
&&+\kappa\int d^3z\{ \delta g(y), \delta {\cal C}^{\prime}_L(z)\}\delta^L_M\{\delta T^M(z),\delta f(x)\} \nonumber \\
&&\{\delta f(x),\delta^2 g(y)\}+ \{\delta^2 f(x),\delta g(y)\} \nonumber \\
&&-\kappa\int d^3z\{ \delta^2 f(x), \delta {\cal C}^{\prime}_L(z)\}\delta^L_M\{\delta T^M(z),\delta g(y)\} \nonumber\\
&&-\kappa\int d^3z\{ \delta f(x), \delta^2{\cal C}^{\prime}_L(z)\}\delta^L_M\{\delta T^M(z),\delta g(y)\} \nonumber \\
&&-\kappa\int d^3z\{ \delta f(x), \delta{\cal C}^{\prime}_L(z)\}\delta^L_M\{\delta T^M(z),\delta^2 g(y)\} \nonumber \\
&&+\kappa\int d^3z\{ \delta^2 g(y), \delta {\cal C}^{\prime}_L(z)\}\delta^L_M\{\delta T^M(z),\delta f(x)\} \nonumber \\
&&+\kappa\int d^3z\{ \delta g(y), \delta^2 {\cal C}^{\prime}_L(z)\}\delta^L_M\{\delta T^M(z),\delta f(x)\} \nonumber \\
&&+\kappa\int d^3z\{ \delta g(y), \delta {\cal C}^{\prime}_L(z)\}\delta^L_M\{\delta T^M(z),\delta^2 f(x)\} +O(\delta^2,\kappa^2).\nonumber \end{eqnarray}
Now for elementary phase space variables  we have $\delta^2 f=0$ and $\delta^2 g=0$ then the Dirac bracket simplifies to
\begin{eqnarray}
\{ f(x), g(y)\}^* &=& \{\delta f(x),\delta g(y)\} \\
&&-\kappa\int d^3z\{ \delta f(x), \delta {\cal C}^{\prime}_L(z)\}\delta^L_M\{\delta T^M(z),\delta g(y)\}\nonumber \\
&&+\kappa\int d^3z\{ \delta g(y), \delta {\cal C}^{\prime}_L(z)\}\delta^L_M\{\delta T^M(z),\delta f(x)\} \nonumber \\
&&-\kappa\int d^3z\{ \delta f(x), \delta^2{\cal C}^{\prime}_L(z)\}\delta^L_M\{\delta T^M(z),\delta g(y)\} \nonumber \\
&&+\kappa\int d^3z\{ \delta g(y), \delta^2 {\cal C}^{\prime}_L(z)\}\delta^L_M\{\delta T^M(z),\delta f(x)\} +O(\delta^2,\kappa^2).\nonumber \\
 \end{eqnarray}
The Dirac observable associated with $\delta f$ is up to second order given by
\begin{eqnarray}
\label{eq:DiracObs2ndOrder}
O_{\delta f,\{T\}} &=& \delta f -\kappa\int d^3z \delta {\cal G}^{K}(z)\{\delta f,\delta{\cal C}^{\prime}_K(z)\} 
-\kappa\int d^3z \delta {\cal G}^{K}(z)\{\delta f,\delta^2{\cal C}^{\prime}_K(z)\} \\
&& 
-\frac{\kappa^2}{2}\int d^3z\int d^3z' \delta {\cal G}^{K}(z) \delta {\cal G}^{L}(z')\{\{\delta f,\delta^2{\cal C}^{\prime}(z)\},\delta{\cal C}^{\prime}_M(z')\}+O(\delta^3,\kappa^2)
\nonumber \\
&=:& O^{(1)}_{\delta f,\{T\}}+ O^{(2)}_{\delta f,\{T\}}+O(\delta^3,\kappa^2),
\end{eqnarray}
where, as in the main text, we include all terms of order $\delta$ and $\delta^2$ in $O^{(1)}_{\delta f,\{T\}}$ and $O^{(2)}_{\delta f,\{T\}}$ respectively.
Our aim is to show the following: For linearised quantities $\delta f,\delta g$ with $\delta^2f=0$ and $\delta^2g=0$ with a set of constraints ${\cal C}^{\prime}_I$ and reference fields ${\cal G}^I$ that satisfy $\{{\cal G}^I(x),{\cal C}^{\prime}_J(y)\}=\delta^I_J\delta^{(3)}(x,y)+O(\delta^2,\kappa^2)$ 
we have
\begin{equation}
\label{eq:RelObsAlgDB}
\{O_{\delta f,\{T\}}, O_{\delta g,\{T\}}\} = O_{\{\delta f,\delta g\}^*,\{T\}}  +O(\delta^2,\kappa^2),   
\end{equation}
where the crucial property for our later application is that we have a strong equality here and not a weak one that is present in the general case \cite{Thiemann:2004wk,Dittrich:2005kc}. For the purpose of showing the equality in \eqref{eq:RelObsAlgDB} we introduce the following notation:
\begin{eqnarray}
\label{eq:DBDeltaOrders}
\{f,g\}^*\big|_{\delta^0} &:=&
\{\delta f(x),\delta g(y)\} 
-\kappa\int d^3z\{ \delta f(x), \delta {\cal C}^{\prime}_L(z)\}\delta^L_M\{\delta T^M(z),\delta g(y)\} \\
&&+\kappa\int d^3z\{ \delta g(y), \delta {\cal C}^{\prime}_L(z)\}\delta^L_M\{\delta T^M(z),\delta f(x)\} \nonumber \\
\{f,g\}^*\big|_{\delta^1} &:=&-\kappa\int d^3z\{ \delta f(x), \delta^2{\cal C}^{\prime}_L(z)\}\delta^L_M\{\delta T^M(z),\delta g(y)\} \nonumber \\
&&+\kappa\int d^3z\{ \delta g(y), \delta^2 {\cal C}^{\prime}_L(z)\}\delta^L_M\{\delta T^M(z),\delta f(x)\}\, ,\nonumber
\end{eqnarray}
where we again used that $\delta^2f=\delta^2g=0$ and that we can replace $\delta{\cal G}^M$ by $\delta T^M$ inside the Poisson brackets. Given this notation we can write $O_{\{\delta f,\delta g\}^*,\{T\}}$ up to linear order as
\begin{eqnarray}
\label{eq:DiracObsDB}
O_{\{\delta f,\delta g\}^*,\{T\}} &=&
\{f,g\}^*\big|_{\delta^0}+\{f,g\}^*\big|_{\delta^1}
-\kappa\int d^3z \delta{\cal G}^K(z)\{ \{f,g\}^*\big|_{\delta^1}, \delta{\cal C}^{\prime}_K(z)\} + O(\delta^2,\kappa^2)\nonumber \\
&=&
\{f,g\}^*\big|_{\delta^0}+\{f,g\}^*\big|_{\delta^1} \\
&& -\kappa^2\int d^3z\int d^3z' \delta{\cal G}^L(z)\Big(-\delta^K_M\{\delta T^M(z'),\delta g\}\{\{\delta f, \delta^2{\cal C}^{\prime}_K(z')
\},\delta {\cal C}^{\prime}_L(z)\}\nonumber \\
&&\hspace{2cm}-\delta^K_M\{\delta T^M(z'),\delta f\}\{\{\delta g, \delta^2{\cal C}^{\prime}_K(z')
\},\delta {\cal C}^{\prime}_L(z)\}\nonumber
\Big)+O(\delta^2,\kappa^2)
\end{eqnarray}
Hence, we need to show that $\{O^{(1)}_{\delta f,\{T\}}+ O^{(2)}_{\delta f,\{T\}},O^{(1)}_{\delta g,\{T\}}+ O^{(2)}_{\delta g,\{T\}}\}$ agrees with the right hand side of \eqref{eq:DiracObsDB}. To confirm this we insert the Dirac observables of $\delta f$ and $\delta g$  up to second order in \eqref{eq:DiracObs2ndOrder} into the Poisson bracket and collect all terms in the individual orders. In order $\delta^0$ we obtain
\begin{eqnarray}
\delta^0 : && \{\delta f(x), \delta g(y)\} 
-\kappa\int d^3z \{\delta f(x),\delta T^L(z)\}\{ \delta g(y), \delta {\cal C}^{\prime}_L(z)\}\\
&&-\kappa\int d^3z \{\delta T^L(z),\delta g(y)\} \{ \delta f(x), \delta {\cal C}^{\prime}_L(z)\}
\nonumber \\
&=& \{f(x),g(y)\}^*\big|_{\delta^0}\nonumber,
\end{eqnarray}
where we used the antisymmetry of the Poisson bracket as well as again used that we can replace $\delta{\cal G}^M$ by $\delta T^M$ inside a given Poisson bracket.
For the linear order in $\delta$ the result is given by

\begin{eqnarray}
\delta^1 : && 
\underbrace{-\kappa\int d^3z{\cal G}^M(z)\{\delta f(x),\{\delta g(y), \delta^2{\cal C}^{\prime}_M(z)\}\}}_{=:A_1} 
\underbrace{-\kappa\int d^3z  \{\delta f(x),\delta T^M(z)\}\{\delta g(y),\delta^2{\cal C}^{\prime}_M(z)\}}_{=:B_1}\nonumber
\\
&&\underbrace{-\kappa\int d^3z {\cal G}^K(z)\{\{\delta f(x), \delta^2{\cal C}^{\prime}_K(z)\},\delta g(y)\}}_{=:A_2}
\underbrace{-\kappa\int d^3z \{\delta T^K(z),\delta g(y)\}\{\delta f(x),\delta^2{\cal C}^{\prime}_K(z)\}}_{=:B_2}\nonumber \\
&&
\underbrace{-\kappa^2\int d^3z \int d^3z^{\prime}\{\delta f(x), \delta T^{M}(z)\}\delta{\cal G}^N(z')\{\{\delta g(y) ,\delta^2{\cal C}^{\prime}_M(z)\}, \delta {\cal C}^{\prime}_N(z')\}}_{:=C_1}\nonumber \\
&&\underbrace{-\kappa^2\int d^3z \int d^3z^{\prime}\{\delta T^{K}(z),\delta g(y)\}\delta{\cal G}^L(z')\{\{\delta f(x) ,\delta^2{\cal C}^{\prime}_K(z)\}, \delta {\cal C}^{\prime}_L(z')\}}_{:=C_2},
\end{eqnarray}

where we again used that we can replace $\delta{\cal G}^M$ by $\delta T^M$ inside a given Poisson bracket.

Now let us discuss the individual contributions separately. First we show that $A_1+A_2=0$. Using the Jacobi identity for $A_1$ we obtain
\begin{eqnarray}
A_1 &=& -\kappa\int d^3z{\cal G}^M(z)\{\delta f(x),\{\delta g(y), \delta^2{\cal C}^{\prime}_M(z)\}\} \\
&=& \kappa\int d^3z {\cal G}^M(z)\Big(
\{\delta g(y),\{\delta^2{\cal C}^{\prime}_M(z), \delta f(x)\}\} 
+ \underbrace{\{\delta^2{\cal C}^{\prime}_M(z),\underbrace{\{\delta f(x),\delta g(y)\}}_{\sim \delta^0}\} }_{=0}\Big)\nonumber \\
&=&\kappa\int d^3z{\cal G}^M(z)
\{\delta g(y),\{\delta^2{\cal C}^{\prime}_M(z), \delta f(x)\}\} \nonumber \\
&=&
\kappa\int d^3z {\cal G}^M(z)
\{\{\delta f(x), \delta^2{\cal C}^{\prime}_M(z)\},\delta g(y)\}\}\nonumber \\
&=& -A_2.
\end{eqnarray}
Comparing $B_1+B_2$ with $\{f,g\}^*\big|_{\delta^1}$ in \eqref{eq:DBDeltaOrders} and using the antisymmetry of the Poisson bracket we realise that we have
\begin{equation}
B_1+B_2 = \{f,g\}^*\big|_{\delta^1}.    
\end{equation}
Comparing $C_1+C_2$ with the last two terms in \eqref{eq:DiracObsDB} and applying a suitable relabelling of the integration variables it turns out that these two expressions are exactly identical. Thus, collecting all intermediate results we have shown 
 \begin{equation*}
 O_{\{\delta f,\delta g\}^*,\{T\}} = A_1+A_2+B_1+B_2+C_1+C_2 +O(\delta^2, \kappa^2) 
 =\{O_{\delta f,\{T\}}, O_{\delta g,\{T\}}\} +O(\delta^2, \kappa^2),
 \end{equation*}
 which is exactly the equality in \eqref{eq:RelObsAlgDB} we wanted to prove.
 ~\\
 Now let us discuss the case of elementary variables from the matter sector. Because we use geometrical clocks if we consider for $\delta f$ and $\delta g$ both matter variables, then since these will commute with all geometrical clocks we immediately have $\{\delta f, \delta g\}^*=\{\delta f,\delta g\}$ so that we can work with the Poisson bracket instead of Dirac bracket in \eqref{eq:RelObsAlgDB} and as discussed in the main text the observable algebra drastically simplifies. The other case is if we consider the algebra of observables of one geometric and one matter quantity and without loss of the generality let us assume that $\delta f$ contains geometric and $\delta g$ matter variables. Then in general the Dirac bracket can differ from the Poisson bracket by those terms where $\delta f$ is involved in a Poisson bracket with the geometric clocks. In this case the proof just presented carries over if we just replace $\delta g$ by $g$ for the following reason: The Poisson bracket $\{g(y), \delta {\cal C}^\prime_L(z)\}$ for $g(y)$ a function of elementary matter variables is of order $\delta^0$ and hence also contributes in this case to the $\delta^0$-order of the Dirac bracket. The second relevant Poisson bracket $\{g(y), \delta^2{\cal C}^\prime_L(z)\}$ is as before of order $\delta^1$ because in the order of perturbation theory we consider, the contributions to this Poisson bracket come from terms that involve the gravitational perturbation linearly and the matter variables in quadratic order so that the final result will still be of order $\delta^1$.  Therefore, we can also apply the equality shown in \eqref{eq:RelObsAlgDB} to matter variables by simply replacing  $\delta f$ by $f$ and $\delta g$ by $g$ if we restrict $f,g$ to be elementary phase space variables from the set $\phi(x),\pi(y)$. Further because the Poisson brackets of the matter variables do not involve a factor $\frac{1}{\kappa}$ in contrast to the gravitational variables additional factors $\kappa$ can arise when the Dirac bracket with matter variables is considered. The application we need in the main text is even a more special case so we will not discuss these terms more in detail here since they will not be needed in any further computation. The main motivation for presenting this proof here in detail was to understand under which assumptions regarding perturbation theory we can ensure that the Dirac bracket agrees with the Poisson bracket and how, if this is not the case, the explicit form of the Dirac observable of the Dirac bracket looks like in perturbation theory.

\subsubsection{Poisson algebra of linearised Dirac observables}
\label{app:AlgebLinDiracObs}
In this appendix we will compute the Poisson algebra of the linearised Dirac observables explicitly and show that the zeroth order contributions are given by the standard canonical Poisson brackets, a result that we use in the main text of this work. In the notation of the main text for all observables $O_{f,\{T\}}=O^{(1)}_{f,\{T\}}+O^{(2)}_{f,\{T\}}+O(\delta^3,\kappa^3)$ we will neglect the contributions coming from  $O^{(2)}_{f,\{T\}}+O(\delta^3,\kappa^3)$. This means that the explicit computations presented in this appendix here do only consider the zeroth order result of the corresponding Poisson algebra. This causes no problem since we also know this as well as the linear contribution to the Poisson algebra already form the relation to the Dirac observable of the corresponding Dirac bracket as discussed in the main text. The computation presented here should be rather understood as double checking our results for the explicit form of the Dirac observables that would yield incorrect results for the algebra in zeroth order if they were not computed correctly. 
For the gauge invariant geometrical degrees of freedom, one can use the following way to obtain their algebra:
\begin{align}
    \{\t{\delta A}{_a^i}(\vec x,t),\t{\delta E}{^b_j}(\vec y,t)\} = \frac{\beta}{\kappa} \delta_j^i\delta_a^b \delta(\vec x-\vec y) \implies \{\ft{\delta A}\t{}{_a^i}(\vec k,t),\ft{\delta E}\t{}{^b_j}(\vec p,t)\} = \frac{\beta}{\kappa} \delta_j^i\delta_a^b \delta(\vec k+\vec p)
\end{align}
and hence for $r,u\in\{\pm\}$:
\begin{align}
    \{\ft{\delta A}^r(\vec k,t) ,\ft{\delta E}^u(\vec p,t) \} &= \overline{m}^a(r\vec{k})\overline{m}_i(r\vec{k})\overline{m}_b(u\vec{p})\overline{m}^j(u\vec{p}) \; \{ \ft{\delta A}\t{}{_a^i}(\vec k,t), \ft{\delta E}\t{}{^b_j}(\vec p,t) \}\nonumber\\
    &= \frac{\beta}{\kappa} \overline{m}^a(r\vec{k})\overline{m}_i(r\vec{k}){m}_a(u\vec{k}){m}^b(u\vec{k}) \delta(\vec{k}+\vec{p}) \nonumber\\
    &= \delta^{ru} \frac{\beta}{\kappa} \delta(\vec{k}+\vec{p})\,.
\end{align}
Analogously one can show that $\{\ft{\delta A}^r(\vec k,t) ,\ft{\delta A}^u(\vec p,t) \}=\{\ft{\delta E}^r(\vec k,t) ,\ft{\delta E}^u(\vec p,t) \}=0$. From this and the fact that $O^{(1)}_{\delta \mathcal{A}\t{}{_a^i},\{\delta T\}}(\vec x,t)= O^{(1)}_{P\delta A\t{}{_a^i},\{\delta T\}}(\vec x,t) = PO^{(1)}_{\delta A\t{}{_a^i},\{\delta T\}}(\vec x,t) = P\t{A}{_a^i}(\vec x,t)$ and similarly for $\delta E$, follows that
\begin{align}
    \{O^{(1)}_{\delta \mathcal{A}\t{}{_a^i},\{\delta T\}}(\vec x,t),O^{(1)}_{\delta \mathcal{E}\t{}{^b_j},\{\delta T\}}(\vec y,t)\} &= \int \frac{d^3k d^3p}{(2\pi)^3} e^{i\vec k\vec x + i\vec p \vec y} \ft{P}\t{}{_a^i_l^c}(\vec k) \ft{P}\t{}{^b_j^m_d}(\vec p) \{\ft{\delta A}\t{}{_c^l}(\vec k,t),\ft{\delta E}\t{}{^d_m}(\vec p,t)\}\nonumber\\
    &= \frac{\beta}{\kappa}\int \frac{d^3k}{(2\pi)^3} e^{i\vec k(\vec x -\vec y)} \ft{P}\t{}{_a^i_l^c}(\vec k) \ft{P}\t{}{^b_j^l_c}(-\vec k)\nonumber\\
    &=\frac{\beta}{\kappa}\int \frac{d^3k}{(2\pi)^3} e^{i\vec k(\vec x -\vec y)} \ft{P}\t{}{_a^i_j^b}(\vec k)\nonumber\\
    &= \frac{\beta}{\kappa} \t{P}{_a^i_j^b}\delta(\vec x-\vec y)\,,
\end{align}
where we used in the fourth line that $\ft{P}\t{}{^b_j^l_c}(-\vec k) = \ft{P}\t{}{^b_j^l_c}(\vec k) = \ft{P}\t{}{^l_c^b_j}(\vec k)$, as well as
\begin{align}
    \{O^{(1)}_{\delta \mathcal{A}\t{}{_a^i},\{\delta T\}}(\vec x,t),O^{(1)}_{\delta \mathcal{A}\t{}{_b^j},\{\delta T\}}(\vec y,t)\} &= 0\\
    \{O^{(1)}_{\delta \mathcal{E}\t{}{^a_i},\{\delta T\}}(\vec x,t),O^{(1)}_{\delta \mathcal{E}\t{}{^b_j},\{\delta T\}}(\vec y,t)\} &= 0\,.
\end{align}
For the matter variables we obtain, dropping terms in $O(\delta^2,\kappa^2)$ and higher:
\begin{align}
    \{ O^{(1)}_{\phi,\{\delta T\}}(\vec x,t),  O^{(1)}_{\phi,\{\delta T\}}(\vec y,t) \} &= -\kappa^2 (\delta \mathcal{G}^c(\vec x,t)) \delta^3(\vec x-\vec y) +\kappa^2 (\delta\mathcal{G}^c(\vec y,t)) \delta^3(\vec x-\vec y)=0\\
    \{O^{(1)}_{\pi,\{\delta T\}}(\vec x,t), O^{(1)}_{\pi,\{\delta T\}}(\vec y,t)\} &= \kappa^2 ((\partial_a^y \:\delta \mathcal{G}(\vec{y},t) + \partial_a^x \:\delta \mathcal{G}(\vec x,t)) \partial_y^a \delta^3(\vec x-\vec y) + (\delta \mathcal{G}(\vec y,t)-\delta \mathcal{G}(\vec x,t)) \Delta \delta^3(\vec x-\vec y)]\\
    \{O^{(1)}_{\phi,\{\delta T\}}(\vec x,t),O^{(1)}_{\pi,\{\delta T\}}(\vec y,t)\} &= \delta^3(\vec x-\vec y) -\kappa^2 \partial_a^y[\delta\mathcal{G}^a(\vec y,t) \delta^3(\vec x-\vec y)] -\kappa^2 \;\delta \mathcal{G}^a(\vec x,t) \partial_a^x\delta^3(\vec x-\vec y)\,.
\end{align}
Due to the presence of derivatives acting on the delta distributions, it is not immediately evident that all additional terms (which are of $O(\kappa)$ due to the presence of a factor $\kappa^{-1}$ in the reference fields) vanish. To see this, it is convenient to consider the smeared version of the linearised observables, then it turns out that:
\begin{align}
    \int d^3x \int d^3y \;f(\vec x) g(\vec y) \{O^{(1)}_{\pi,\{\delta T\}}(\vec x,t), O^{(1)}_{\pi,\{\delta T\}}(\vec y,t)\} &= 0\\
    \int d^3x \int d^3y \;f(\vec x) g(\vec y)  \{O^{(1)}_{\phi,\{\delta T\}}(\vec x,t),O^{(1)}_{\pi,\{\delta T\}}(\vec y,t)\} &= \int d^3x \; f(\vec x) g(\vec x)\,.
\end{align}
Hence we indeed end up with the desired algebra also for the matter observables:
\begin{align}
    &\{ O^{(1)}_{\phi,\{\delta T\}}(\vec x,t), O^{(1)}_{\phi,\{\delta T\}}(\vec y,t) \} =0 \\ 
    & \{ O^{(1)}_{\pi,\{\delta T\}}(\vec x,t), O^{(1)}_{\pi,\{\delta T\}}(\vec y,t) \} =0 \nonumber \\
    &\{ O^{(1)}_{\phi,\{\delta T\}}(\vec x,t), O^{(1)}_{\pi,\{\delta T\}}(\vec y,t) \} =\delta^3(\vec x-\vec y)\,.\nonumber
\end{align}
As $O^{(1)}_{\delta \mathcal{A}\t{}{_a^i},\{\delta T\}}$ and $O^{(1)}_{\delta \mathcal{E}\t{}{^a_i},\{\delta T\}}$ commute with the clocks and do not depend on the original matter fields, all Poisson brackets among the $(O^{(1)}_{\delta \mathcal{A},\{\delta T\}},O^{(1)}_{\delta \mathcal{E},\{\delta T\}})$  vanish up to $O(\delta^2,\kappa^2)$ and the $(O^{(1)}_{\phi,\{\delta T\}},O^{(1)}_{\pi,\{\delta T\}})$ vanish up to $O(\kappa^2)$.

\subsubsection{Comparison with the Dirac observables used by Anastopoulos and Hu}
\label{app:ComparisonDiracObs}
The model in \cite{anastopoulos2013master} also considers gauge invariant quantities for the matter variables in the ADM framework and we would like to compare their quantities to the constructed Dirac observables in this work. There, an extension of the basic gauge variant matter variables $\phi$ and $\pi$ is proposed that is required to commute with the linearised constraints. These observables take the form
\begin{align*}
    \widetilde{\phi}(\vec x,t) = \phi(\vec x-\vec q,t-\tau) && \widetilde{\pi}(\vec x,t) = \pi(\vec x-\vec q,t-\tau) -\partial_a q^a\,,
\end{align*}
where $\vec q$ and $\tau$, similarly to the geometrical clocks used in our work, consist of purely geometrical quantities in ADM variables. For these as well as the matter fields in \cite{anastopoulos2013master} the following transformation behaviour under the Hamiltonian and spatial diffeomorphism constraint respectively is used\footnote{Note that in \cite{anastopoulos2013master} the transformation for $\pi$ is given by $\pi \rightarrow \pi -\lambda \frac{\delta H_0}{\delta \pi}$ where we expect the $\frac{\delta}{\delta \pi}$ to be a typo and corrected this accordingly.}:
\begin{align*}
\text{Under the Hamilton constraint}: &&    \phi &\rightarrow \phi + \lambda \frac{\delta H_0}{\delta \pi} && \pi \rightarrow \pi -\lambda \frac{\delta H_0}{\delta \phi}\\ && \vec{q} &\rightarrow \vec{q} && \tau \rightarrow \tau + \lambda\\
\text{Under the spatial diffeomorphism constraint}: &&    \phi &\rightarrow \phi + \kappa \lambda^a \partial_a\phi && \pi \rightarrow \pi +\kappa \partial_a (\lambda^a \pi)\\ && q^i &\rightarrow q^i+\lambda^i && \tau \rightarrow \tau\,,
\end{align*}
where $H_0 = \int d^3x \; \epsilon(\vec x,t)$. Starting with the transformations under the Hamilton constraint, which is just $\epsilon(\vec x,t)$ in the matter part,
we obtain
\begin{equation*}
\{\phi(\vec x), \int d^3y \; \lambda(\vec y) \epsilon(\vec y)\} = \lambda(\vec x) \pi(\vec x) = \lambda(\vec x) \frac{\delta H_0}{\delta \pi}(\vec x)   
\end{equation*}
which hence indeed confirms their proposed transformation of the matter field, its momentum transforms differently reflecting the fact that it is a scalar density of weight one: 
\begin{align*}
    \left\{\pi(\vec x), \int d^3y \; \lambda(\vec y) \epsilon(\vec y)\right\} = \partial_a(\lambda(\vec x) \partial^a \phi(\vec x)) - \lambda(\vec x) V'(\phi(\vec x)) = -\lambda(\vec x) \frac{\delta H_0}{\delta\phi}(\vec x) + (\partial_a\lambda(\vec x)) \; \partial^a \phi(\vec x),
\end{align*}
where we assumed that the scalar field obeys a Klein-Gordon equation and that the energy density possesses an arbitrary potential $V(\phi)$. The transformation behaviour used in \cite{anastopoulos2013master} under the Hamiltonian constraint only agrees with the above expression if $\lambda$ is constant that in general cannot be assumed and  hence the second term coming from the Poisson bracket needs to be included in the transformation of $\pi(\vec x)$ and to our understanding has been omitted in \cite{anastopoulos2013master}.

Given the correct general transformation behaviours, we want to check whether the Dirac observables used in \cite{anastopoulos2013master} $\tilde{\phi}$ and $\widetilde{\pi}$ are indeed gauge invariant. First it is worth noticing that due to the fact that we are working in a linearised theory, to our understanding  the quantity $\widetilde{\phi}(\vec x,t)$ has to be understood as a Taylor expansion in the modifications truncated after linear order. In general this way of constructing observables looks like the canonical version of the strategy followed in \cite{Giddings:2019wmj}. To show this, we will add an additional $\kappa$ in front of the modifications and also include it into the constraints.  In \cite{anastopoulos2013master} this was done for the spatial diffeomorphism constraint only, as its action on the matter fields indeed yields a correction term of order $\kappa$. As shown in the discussion on the Post-Minkowski approximation scheme in \ref{sec:PostMinkowski}, for consistency such a factor is also required in the Hamiltonian constraint. Therefore, we obtain
\begin{align*}
    \widetilde{\phi}(\vec x,t)& = \phi(\vec x - \kappa \vec q(\vec x,t), t-\kappa\tau(\vec x,t)) = \phi(\vec x,t) - \kappa q^a(\vec x,t) \partial_a\phi(\vec x,t) -\kappa \tau(\vec x,t) \dot{\phi}(\vec x,t)\\
     \widetilde{\pi}(\vec x,t)& = \pi(\vec x - \kappa \vec q(\vec x,t), t-\kappa\tau(\vec x,t))-\partial_a q^a(\vec x,t) = \pi(\vec x,t) - \kappa q^a \partial_a\pi(\vec x,t) -\kappa \tau \dot{\pi}(\vec x,t)- \partial_a q^a(\vec x,t)\,.
\end{align*}
Now under the Hamilton constraint $\widetilde{\phi}$ transforms as:
\begin{align*}
    \widetilde{\phi}(\vec x,t) \rightarrow\widetilde{\phi}(\vec x,t) + \kappa(\lambda\pi)(\vec x,t) - \kappa (\lambda\dot{\phi})(\vec x,t) = \widetilde{\phi}(\vec x,t) \,,
\end{align*}
where we neglected terms of higher order in $\kappa$ and see that indeed $\widetilde{\phi}$ remains invariant under the transformations induced by the Hamilton constraint. For $\widetilde{\pi}$ one obtains:
\begin{align*}
    \widetilde{\pi}(\vec x,t) \rightarrow \widetilde{\pi}(\vec x,t)+ \kappa(\lambda \dot{\pi})(\vec x,t) - \kappa (\lambda \dot{\pi})(\vec x,t) =\widetilde{\pi}(\vec x,t)\,.
\end{align*}
Thus, working with the correct transformation behaviour of $\pi$ under the Hamiltonian constraint, also $\widetilde{\pi}$ remains invariant. For the spatial diffeomorphism constraint a quick calculation shows that its action on the matter fields is indeed given by the terms stated above and in \cite{anastopoulos2013master}. Under its action we obtain:
\begin{align*}
    \widetilde{\phi} &\rightarrow\widetilde{\phi} +\kappa \lambda^a \partial_a \phi -\kappa \lambda^a \partial_a \phi = \widetilde{\phi}\\
    \widetilde{\pi} &\rightarrow\widetilde{\pi} -\kappa \lambda^a \partial_a\pi -\partial_a\lambda^a \neq \widetilde{\pi}\,
\end{align*}
which demonstrates that the $\tilde{\pi}$ proposed in \cite{anastopoulos2013master} is no (linearised) Dirac observable with respect to the spatial diffeomorphism constraint.
Thus, actual linearised observables for $\phi(\vec x,t),\pi(\vec x,t)$ in \cite{anastopoulos2013master} would be given by
\begin{align*}
    \widetilde{\phi}(\vec x,t)& = \phi(\vec x,t) - \kappa q^a(\vec x,t) \partial_a\phi(\vec x,t) -\kappa \tau(\vec x,t) \dot{\phi}(\vec x,t)\\
     \widetilde{\pi}(\vec x,t)& = \pi(\vec x,t) - \kappa q^a \partial_a(q^a\pi)(\vec x,t) -\kappa \tau \dot{\pi}(\vec x,t)\,,
\end{align*}
where it is crucial that the partial derivative in the second term in $ \widetilde{\pi}$ acts on both $q^a(\vec x,t)$ and $\pi(\vec x,t)$.
Compared to our framework, the basic idea is similar to extend the matter fields with geometrical variables to obtain gauge invariant observables. Hence, the quantities $q^a$ and $\tau$ play the role of geometrical clocks in our setup. In our approach we in addition require that $q^a$ and $\tau$ mutually commute which requires to use an explicit choice of geometrical clocks that satisfies this additional property.

\subsection{Canonical transformation on the original phase space}\label{app:Trafo}
In this appendix the aim is to find an expression of the original phase space variables in terms of the new ones which consist of the physical degrees of freedom, the constraints and the clocks. For this, we first transform the clocks and constraints into momentum space using the basis introduced above in appendix \ref{app:Basis} and end up with the following expressions:
\begingroup
\allowdisplaybreaks
\begin{align}
\ft{\delta \mathcal{G}} &= \frac{1}{2\beta \kappa||\vec{k}||} (\ft{\delta E}^5 - \ft{\delta E}^4) - \frac{1}{\beta \kappa||\vec{k}||^2} (\ft{\delta A}^3+ \ft{\delta A}^4+\ft{\delta A}^5)-\ft{\delta \tau}\\
\ft{\delta C} &= \kappa ||\vec{k}|| (\ft{\delta A}^4 - \ft{\delta A}^5 ) +\kappa \ft{\epsilon}\\
m_a \ft{\delta \mathcal{G}}^a &= \frac{2i}{\kappa||\vec{k}||} (\ft{\delta E}^{\overline{1}} + \ft{\delta E}^{\overline{2}} ) - \frac{4\beta}{\kappa^2 ||\vec{k}||^2} m^i \ft{\delta G}{}_i - m_a  \ft{\delta \sigma}^a\\
\overline{m}_a \ft{\delta \mathcal{G}}^a &= \frac{2i}{\kappa||\vec{k}||} (\ft{\delta E}^{1} + \ft{\delta E}^{2} ) - \frac{4\beta}{\kappa^2||\vec{k}||^2} \overline{m}^i \ft{\delta G}{}_i- \overline{m}_a \ft{\delta \sigma}^a \\
\hat{k}_a \ft{\delta \mathcal{G}}^a &= \frac{i}{\kappa||\vec{k}||} (3\ft{\delta E}^{3} - \ft{\delta E}^{4} - \ft{\delta E}^5 ) - \frac{6\beta}{\kappa^2 ||\vec{k}||^2} \hat{k}^i \ft{\delta G}{}_i +\frac{i}{\kappa ||\vec{k}||^3} \left(\frac{1}{\kappa}\ft{\delta C}-\ft{\epsilon}\right)- \hat{k}_a \ft{\delta \sigma}^a\\
m^a \ft{\delta C}{}_a &= -\frac{i\kappa||\vec{k}||}{2\beta} \ft{\delta A}^{\overline{2}} + \kappa m^a \ft{p}{}_a\\
\overline{m}^a \ft{\delta C}{}_a &= -\frac{i\kappa||\vec{k}||}{2\beta} \ft{\delta A}^{2} + \kappa \overline{m}^a \ft{p}{}_a\\
\hat{k}^a \ft{\delta C}{}_a &= \frac{i\kappa||\vec{k}||}{2\beta} (\ft{\delta A}^{4} + \ft{\delta A}^5 ) + \kappa \hat{k}^a \ft{p}{}_a\\
m^i \ft{\delta G}{}_i &= \frac{\kappa i}{2\beta} (||\vec{k}|| \ft{\delta E}^{\overline{1}} - \ft{\delta A}^{\overline{1}} + \ft{\delta A}^{\overline{2}} ) \\
\overline{m}^i \ft{\delta G}{}_i &= \frac{\kappa i}{2\beta} (||\vec{k}|| \ft{\delta E}^{1} - \ft{\delta A}^{2} + \ft{\delta A}^{1} )\\
\hat{k}^i \ft{\delta G}{}_i &= \frac{\kappa i}{2\beta} (||\vec{k}|| \ft{\delta E}^{3} - \ft{\delta A}^{5} + \ft{\delta A}^{4} )\\
m_i \ft{\delta \mathcal{G}}^i &= \frac{2i}{\kappa||\vec{k}||} \ft{\delta A}^{\overline{1}} - m_i \ft{\delta \xi}^i \\
\overline{m}_i  \ft{\delta \mathcal{G}}^i &= \frac{2i}{\kappa||\vec{k}||} \ft{\delta A}^{1} -\overline{m}_i \ft{\delta \xi}^i \\
\hat{k}_i  \ft{\delta \mathcal{G}}^i &= \frac{2i}{\kappa||\vec{k}||} \ft{\delta A}^{3}-\hat{k}_i \ft{\delta \xi}^i\,.
\end{align}
\endgroup
In total, these are 14 equations that can be solved uniquely for the 14 old variables. The solution then is
\begingroup
\allowdisplaybreaks
\begin{align}
\ft{\delta A}^1 &= -\frac{i\kappa||\vec{k}||}{2} \overline{m}_i \left[\ft{\delta \mathcal{G}}^i+ \ft{\delta \xi}^i\right] \\
\ft{\delta A}^{\overline{1}} &= -\frac{i\kappa||\vec{k}||}{2} m_i \left[\ft{\delta\mathcal{G}}^i+ \ft{\delta \xi}^i\right] \\
\ft{\delta A}^2 &= \frac{2i\beta}{||\vec{k}||} \overline{m}^a \left(\frac{1}{\kappa}\ft{\delta C}{}_a - \ft{p}{}_a \right)\\
\ft{\delta A}^{\overline{2}} &= \frac{2i\beta}{||\vec{k}||} m^a \left(\frac{1}{\kappa}\ft{\delta C}{}_a - \ft{p}{}_a \right)\\
\ft{\delta A}^3 &= -\frac{i\kappa||\vec{k}||}{2} \hat{k}_i \left[\ft{\delta\mathcal{G}}^i+ \ft{\delta \xi}^i\right] \\
\ft{\delta A}^4 &=- \frac{i\beta}{||\vec{k}||}  \hat{k}^a \left( \frac{1}{\kappa} \ft{\delta C}{}_a - \ft{p}{}_a \right) + \frac{1}{2||\vec{k}||} \left(\frac{1}{\kappa} \ft{\delta C} - \ft{\epsilon} \right)\\
\ft{\delta A}^5 &=- \frac{i\beta}{||\vec{k}||}  \hat{k}^a \left( \frac{1}{\kappa} \ft{\delta C}{}_a - \ft{p}{}_a \right) - \frac{1}{2||\vec{k}||} \left(\frac{1}{\kappa} \ft{\delta C} - \ft{\epsilon} \right)\\
\ft{\delta E}^1 &= -\frac{2i\beta}{\kappa||\vec{k}||} \overline{m}^i \ft{\delta G}{}_i + \frac{2i\beta}{||\vec{k}||^2} \overline{m}^a\left(\frac{1}{\kappa} \ft{\delta C}{}_a- \ft{p}{}_a\right)  +\frac{i\kappa}{2} \overline{m}^i \left[\ft{\delta\mathcal{G}}{}_i + \ft{\delta \xi}{}_i\right]\\
\ft{\delta E}^{\overline{1}} &= -\frac{2i\beta}{\kappa||\vec{k}||} m^i \ft{\delta G}{}_i - \frac{2i\beta}{||\vec{k}||^2} m^a \left(\frac{1}{\kappa} \ft{\delta C}{}_a- \ft{p}{}_a\right)  -\frac{i\kappa}{2} m^i \left[\ft{\delta\mathcal{G}}{}_i+ \ft{\delta \xi}{}_i\right]  \\
\ft{\delta E}^2 &= -\frac{i\kappa||\vec{k}||}{2} \overline{m}_a \left[\ft{\delta \mathcal{G}}^a+ \ft{\delta \sigma}^a\right] -\frac{i\kappa}{2} \overline{m}_i \left[\ft{\delta\mathcal{G}}^i+ \ft{\delta \xi}^i\right] -\frac{2\beta i}{||\vec k||^2} \overline{m}_a \left(\frac{1}{\kappa} \ft{\delta C}^a - \ft{p}{}^a\right) \\
\ft{\delta E}^{\overline{2}} &= -\frac{i\kappa||\vec{k}||}{2} m_a\left[ \ft{\delta \mathcal{G}}^a+ \ft{\delta \sigma}^a\right] +\frac{i\kappa}{2} m_i \left[\ft{\delta\mathcal{G}}^i+ \ft{\delta \xi}^i\right] +\frac{2\beta i}{||\vec k||^2} m_a \left(\frac{1}{\kappa} \ft{\delta C}{}_a- \ft{p}{}^a\right) \\
\ft{\delta E}^3 &=-\frac{2i\beta}{\kappa ||\vec{k}||} \hat{k}^i \ft{\delta G}{}_i- \frac{1}{||\vec{k}||^2} \left( \frac{1}{\kappa} \ft{\delta C} - \ft{\epsilon}\right)\\
\ft{\delta E}^4 &= -\kappa\beta ||\vec{k}||\left[ \ft{\delta \mathcal{G}}+ \ft{\delta \tau}\right] + \frac{i\kappa ||\vec{k}||}{2} \hat{k}^a \left[\ft{\delta \mathcal{G}}{}_a+ \ft{\delta \sigma}{}_a\right] -\frac{1}{||\vec{k}||^2} \left(\frac{1}{\kappa}\ft{\delta C}-\ft{\epsilon}\right) +\frac{i\kappa}{2} \hat{k}^i\left[ \ft{\delta \mathcal{G}}{}_i+ \ft{\delta \xi}^i\right] \nonumber\\ &\hspace{1in}+ \frac{2i \beta}{||\vec{k}||^2} \hat{k}^a \left( \frac{1}{\kappa} \ft{\delta C}{}_a -  \ft{p}{}_a\right) \\
\ft{\delta E}^5 &= \kappa\beta ||\vec{k}|| \left[\ft{\delta \mathcal{G}}+ \ft{\delta \tau}\right] + \frac{i\kappa ||\vec{k}||}{2} \hat{k}^a\left[ \ft{\delta \mathcal{G}}{}_a+ \ft{\delta \sigma}{}_a\right] -\frac{1}{ ||\vec{k}||^2}\left(\frac{1}{\kappa} \ft{\delta C}-\ft{\epsilon}\right) -\frac{i\kappa}{2} \hat{k}^i\left[ \ft{\delta \mathcal{G}}{}_i+ \ft{\delta \xi}^i\right] \nonumber\\ &\hspace{1in}- \frac{2i \beta}{||\vec{k}||^2} \hat{k}^a \left( \frac{1}{\kappa} \ft{\delta C}{}_a -  \ft{p}{}_a\right) \,.
\end{align}
\endgroup
All the appearing matter degrees of freedom must be substituted by their gauge invariant extensions plus the corresponding correction terms, what can be read off from \eqref{phiGI} and \eqref{piGI}, from which we obtain up to first order in $\kappa$:
\begin{align}\label{phiGI2}
\phi(\vec{x},t) &= \phi^{GI}(\delta\sigma^c,\delta\tau,\delta\xi^j) + \kappa^2 (\delta T(\vec{x},t) - \delta\tau(\vec{x},t)) \pi(\vec{x},t) + \kappa^2 (\delta T^c(\vec{x},t) - \delta\sigma^c(\vec{x},t)) \partial_c \phi(\vec{x},t)\\
 \pi(\vec{x},t) &=\pi^{GI}(\delta\sigma^c,\delta\tau,\delta\xi^j) + \kappa^2 \partial_a [(\delta T(\vec{x},t) - \delta\tau(\vec{x},t)) \partial^a\phi(\vec{x},t)] + \kappa^2 \partial_c[(\delta T^c(\vec{x},t) - \delta\sigma^c(\vec{x},t)) \pi(\vec{x},t)] \nonumber\\&\hspace{0.62in}- \kappa^2 (\delta T(\vec{x},t) - \delta\tau(\vec{x},t))\, m^2 \phi(\vec{x},t)\,.\label{piGI2}
\end{align}
Note that when considering terms where any of the gravitational variables $\ft{\delta A}$ or $\ft{\delta E}$ appear with a prefactor $\kappa$, then we can drop all the correction terms of order $\kappa$ in \eqref{phiGI2} and \eqref{piGI2} as they would then lead to terms of order $\kappa^2$.

\subsection{Canonical Hamiltonian in terms of the transformed variables}
\label{app:CanHamNewVariables}
Using the transformation derived in the previous subsection, it is possible to express the canonical Hamiltonian $\delta{\mathbf{H}}_{\rm can}$ in \eqref{eq:totalhamlin} in terms of the new phase space variables consisting of the physical gauge invariant degrees of freedom, the constraints and the clocks. In a first step we rewrite the Hamiltonian in the following way:
\begin{align}\label{app:HCanStart}
\delta{\mathbf{H}}_{\rm can} =\int_\sigma d^3x \bigg[ &\epsilon(\phi,\pi) + \delta N^a\, \delta C_a + \delta N\, \delta C + \delta \Lambda^i\, \delta G_i \nonumber\\&+\kappa\, (\t{\partial}{_a}\phi) \,(\t{\partial}{^b}\phi)\, \t{\delta E}{^a_i}\, \delta^i_b +\frac{\kappa}{2} m^2 \phi^2\, \t{\delta E}{^a_i}\,\delta^i_a - \frac{\kappa}{2} \t{\delta E}{^a_i}\,\delta_a^i\, \epsilon(\phi,\pi)+  \nonumber\\&- \frac{\kappa}{2\beta^2} \epsilon^{jkl} \t{\epsilon}{_j_m_n}\, \delta^a_k\, \delta^b_l \left( \t{\delta A}{_a^m} \t{\delta A}{_b^n}+ (\beta^2+1) \t{\delta \Gamma}{_a^m} \t{\delta \Gamma}{_b^n} -2 \t{\delta \Gamma}{_a^m} \t{\delta A}{_b^n} \right) \bigg],
\end{align}
where in the first line there is the background matter Hamiltonian and the constraints, in the second line the interaction part and in the last line the second order of the Hamilton constraint. The first line contains the background Hamiltonian of the scalar field, which does not need to be transformed as there are no background constraints and hence it is already gauge invariant, see discussion above. In order to obtain the physical Hamiltonian as described in \eqref{eq:PhysHam} in terms of the new variables, one first has to determine the observables \eqref{eq:dirobsmombE} and \eqref{eq:dirobsmombA} in terms of the new variables, where we already set $\delta \xi^j=0$ as explained in the main text:
\begin{align}\label{eq:obsnvarE}
    (\ft{\delta E}\t{}{^a_i})^{GI}=& \ft{\delta \cal E}\t{}{^a_i}-\frac{2i\beta}{\kappa||\vec k||} \hat{k}^a \ft{\delta G}{}_i + \frac{2\beta}{\kappa||\vec k||^2} \t{\epsilon}{^b^a_i} \left[\ft{\delta C}{}_b-\kappa\ft{p}{}_b\right]-\frac{1}{\kappa||\vec k||^2} \delta^a_i \left[ \ft{\delta C}-\kappa\ft{\epsilon}\right] \nonumber\\
    &\hspace{0.4in}-\frac{i\kappa||\vec k||}{2} \ft{\delta \sigma}^b \left( \delta^a_b \hat{k}_i - \delta^a_i \hat{k}_b \right) -\kappa ||\vec k||\beta \ft{\delta \tau} i\hat{k}_c \t{\epsilon}{_i^a^c} +O(\delta^3,\kappa^2)\\\label{eq:obsnvarA}
     (\ft{\delta A}\t{}{_a^i})^{GI}=& \ft{\delta \cal A}\t{}{_a^i}+\frac{i\beta}{\kappa||\vec k||} (2\hat{k}_i\delta^{ab} -\hat{k}^b \delta^a_i -\hat{k}^i\hat{k}_a\hat{k}^b) \left[ \ft{\delta C}{}_b - \kappa\ft{p}{}_b\right] -\frac{1}{2\kappa||\vec k||} i\t{\epsilon}{_b_a^i} \hat{k}^b \left[ \ft{\delta C}-\kappa\ft{\epsilon}\right]+O(\delta^3,\kappa^2) \,.
\end{align}
The interaction part in \eqref{eq:PhysHam}, $O_{{\cal H}_I,\{T\}}$, then assumes the following form:
\begin{align}
    O_{{\cal H}_I,\{T\}} = &\kappa\int d^3k \; \ft{\varphi}{}_a^b \; \ft{\delta \mathcal{E}}'\t{}{^a_i}\delta_b^i   +\kappa \int d^3k  \frac{ik_b}{2} \ft{\delta \sigma}^{\prime a} \left[ \ft{\varphi}{}_a^b - \delta_a^b \left( \ft{\varphi}{}_c^c +m^2 \ft{\widetilde{V}}-\ft{\epsilon} \right)\right] \nonumber\\
    &+\int \frac{d^3k}{||\vec k||^2} \;\left(- \left[ \ft{\varphi}{}_a^a + \frac{3}{2} m^2\ft{\widetilde{V}} -\frac{3}{2}\ft{\epsilon}\right] \left(\ft{\delta C'}-\kappa\ft{\epsilon}'\right) +2i\beta k^a \ft{\delta G}'{}_i \left[ \ft{\varphi}{}_a^b \delta^i_b +\left(\frac{1}{2}m^2 \ft{\widetilde{V}}-\frac{1}{2}\ft{\epsilon}\right) \delta_a^i\right] \right)\nonumber\\
    &+ O(\delta^3,\kappa^2)\,,
\end{align}
where $\epsilon=\epsilon(\phi^{GI},\pi^{GI})$ and $p_b=p_b(\phi^{GI},\pi^{GI})$. 
We used here a notation where no argument means dependency on $(\vec k,t)$ and a prime after a quantity means dependency on $(-\vec k,t)$. 
The last line in \eqref{app:HCanStart} yields $O_{{\cal H}_{\rm geo},\{T\}}$:
\begin{align}
O_{{\cal H}_{\rm geo},\{T\}}=&\frac{\kappa}{2\beta^2}\int d^3k \sum_{r\in\{\pm\}}\left[\ft{\delta A}^r \ft{\delta A}^{\prime r} +||\vec{k}||^2 (\beta^2+1) \ft{\delta E}^r \ft{\delta E}^{\prime r}  + 2r||\vec{k}|| \ft{\delta E}^r \ft{\delta A}^{\prime r} \right] \nonumber\\
    &-\kappa\int \frac{d^3k}{||\vec{k}||^2} \bigg\{ -\left( \frac{1}{\kappa} \ft{\delta C}{}_a-\ft{p}{}_a\right)\left( \frac{1}{\kappa} \ft{\delta C}'{}_b-\ft{p}'{}_b\right) (4\delta^{ab}-\hat{k}{}^a\hat{k}{}^b)+\frac{1}{4} \left( \frac{1}{\kappa} \ft{\delta C} -\ft{\epsilon}\right) \left( \frac{1}{\kappa} \ft{\delta C}' -\ft{\epsilon}'\right)\Bigg\}\nonumber\\
    &-\kappa\int d^3k \; \ft{\delta G}{}_i \left[ \frac{\beta^2+1}{\kappa^2} \hat{k}^i\hat{k}^j\ \ft{\delta G}'{}_j +  \frac{i\beta}{\kappa||\vec{k}||} \hat{k}^i \left(\frac{1}{\kappa} \ft{\delta C}'-\ft{\epsilon}' \right) - \frac{4i}{\kappa||\vec{k}||} \epsilon_{abi} \hat{k}^a \left( \frac{1}{\kappa} \ft{\delta C}'{}_b-\ft{p}'{}_b\right)\right]\nonumber \\
    &-\kappa \int d^3k \left[ 2i \kappa ||\vec k|| \ft{\delta \tau} \hat{k}^b \left(\frac{1}{\kappa} \ft{\delta C}'{}_b-\ft{p}'{}_b \right) \right] \nonumber\\
    & +O(\delta^3,\kappa^2)\,.
\end{align}
 As discussed in the main text in section \ref{sec:ConstrObs} in \eqref{eq:PhysHam} the physical Hamiltonian $\delta \mathbf{H}$ consists in zeroth order of the Hamiltonian of the scalar field on Minkowski spacetime plus the Dirac observable associated with $\delta^2C^{\rm geo}$ as well as the Dirac observable corresponding to the interaction Hamiltonian ${\cal H}_I$. In the following equations these contributions have been marked in blue:
\begingroup
\allowdisplaybreaks
\begin{align}\label{eq:deltaHcanNewVar}
    \delta \mathbf{H}_{\rm can} =
    &{\color{blue}\int_{\mathbb{R}^3} d^3x \;\epsilon(\vec{x},t)} \nonumber\\
&{\color{blue}+\kappa \int_{\mathbb{R}^3}d^3k\; \frac{1}{2\beta^2}\sum_{r\in\{\pm\}}\Big( \ft{\delta A}^r(\vec{k},t) \,\ft{\delta A}^r (-\vec{k},t)+  2r\,||\vec{k}|| \ft{\delta E}^r(\vec{k},t) \, \ft{\delta A}^r(-\vec{k},t) }\nonumber\\ &\hspace{1.8in}{\color{blue}+ (\beta^2+1) ||\vec{k}||^2  \ft{\delta E}^r(\vec{k},t)\,  \ft{\delta E}^r(-\vec{k},t)\Big) }\nonumber\\
&{\color{blue}+\kappa \int_{\mathbb{R}^3} d^3k \;\delta_{b}^i \ft{\varphi} {}_a^b(\vec{k},t) \, \ft{\delta \mathcal{E}}\t{}{^a_i}(-\vec{k},t)}\nonumber\\
&{\color{blue}+\kappa \int d^3k  \frac{ik_b}{2} \kappa\ft{\delta \sigma}^{\prime a} \left[ \ft{\varphi}{}_a^b - \delta_a^b \left( \ft{\varphi}{}_c^c +m^2 \ft{\widetilde{V}}-\ft{\epsilon} \right)\right]+\kappa \int d^3k \left[2i \kappa ||\vec k|| \ft{\delta \tau} \hat{k}^b \ft{p}'{}_b \right]}\nonumber\\
&{\color{blue}-\kappa\int_{\mathbb{R}^3} \frac{d^3k}{||\vec{k}||^2} \Bigg( -\ft{p}\t{}{_c}(\vec{k},t) \,\ft{p}\t{}{_d}(-\vec{k},t) \left[ 4\delta^{cd}- \hat{k}^c \hat{k}^d \right] + \ft{\epsilon}(\vec{k},t)\,\ft{\epsilon}(-\vec{k},t) \left[\frac{1}{4}+\frac{3}{2}\right]} \nonumber\\
&\hspace{1in}{\color{blue}-\frac{3}{2} m^2\,\ft{\widetilde{V}}(\vec{k},t) \,\ft{\epsilon}(-\vec{k},t)-\ft{\varphi} {}_a^a(\vec{k},t) \,\ft{\epsilon}(-\vec{k},t) \Bigg)}\nonumber\\
    &+ \int {d^3k} \; \ft{\delta G}{}_i \Bigg( \ft{\delta \Lambda}{}'^i - 2i\beta k^a  \frac{1}{||\vec k||^2}\left[ \ft{\varphi}'{}_a^b \delta^i_b +\frac{1}{2}\left(m^2 \widetilde{V}'-\epsilon'\right) \delta_a^i\right] -\frac{\beta^2+1}{\kappa^2} \hat{k}^i\hat{k}^j\ \ft{\delta G}'{}_j \nonumber\\ &\hspace{1.2in}- \frac{i\beta}{\kappa||\vec{k}||} \hat{k}^i \left(\frac{1}{\kappa} \ft{\delta C}'-\ft{\epsilon}' \right) + \frac{4i}{||\vec{k}||} \epsilon_{bia} \hat{k}^a \left( \frac{1}{\kappa} \ft{\delta C}'{}_b-\ft{p}'{}_b\right) \Bigg)\nonumber\\ &+ \int d^3k \; \ft{\delta C}{}_a \Bigg(\ft{\delta N}^{\prime a} +\frac{1}{||\vec k||^2}  \left( \frac{1}{\kappa} \ft{\delta C}'{}_b -2 \ft{p}'{}_b \right) (4\delta^{ab}-\hat{k}{}^a\hat{k}{}^b)\Bigg)
    \nonumber\\
    &+ \int d^3k \; \ft{\delta C} \Bigg( \ft{\delta N}'- \frac{1}{||\vec k||^2}\left[ \ft{\varphi}'{}_a^a + \frac{3}{2} m^2\ft{\widetilde{V}}' -\frac{3}{2}\ft{\epsilon}'\right] -\frac{1}{4||\vec k||^2} \left( \frac{1}{\kappa} \ft{\delta C}' -2\ft{\epsilon}' \right) \Bigg)\nonumber\\ 
    &-\kappa \int d^3k \left[ +2i ||\vec k|| \ft{\delta \tau} \hat{k}^b \ft{\delta C}'{}_b\right] \nonumber\\
    &+ O(\delta^3,\kappa^2)\,
\end{align}
\endgroup
with
\begingroup
\allowdisplaybreaks
\begin{align}
\label{eq:deltaHNewVar}
   {\color{blue} \delta \mathbf{H} =}
    &{\color{blue}\int_{\mathbb{R}^3} d^3x \;\epsilon(\vec{x},t)} \nonumber\\
&{\color{blue}+\kappa \int_{\mathbb{R}^3}d^3k\; \frac{1}{2\beta^2}\, \sum_{r\in\{\pm\}}\Big( \ft{\delta A}^r(\vec{k},t) \,\ft{\delta A}^r (-\vec{k},t)+  2r\,||\vec{k}|| \ft{\delta E}^r(\vec{k},t) \, \ft{\delta A}^r(-\vec{k},t) }\nonumber\\ &\hspace{1.8in}{\color{blue}+ (\beta^2+1) ||\vec{k}||^2  \ft{\delta E}^r(\vec{k},t)\,  \ft{\delta E}^r(-\vec{k},t)\Big) }\nonumber\\
&{\color{blue}+\kappa \int_{\mathbb{R}^3} d^3k \;\delta_{b}^i \ft{\varphi} {}_a^b(\vec{k},t) \, \ft{\delta \mathcal{E}}\t{}{^a_i}(-\vec{k},t)}\nonumber\\
&{\color{blue}+\kappa \int d^3k  \frac{ik_b}{2} \kappa\ft{\delta \sigma}^{\prime a} \left[ \ft{\varphi}{}_a^b - \delta_a^b \left( \ft{\varphi}{}_c^c +m^2 \ft{\widetilde{V}}-\ft{\epsilon} \right)\right] +\kappa \int d^3k \left[2i \kappa ||\vec k|| \ft{\delta \tau} \hat{k}^b \ft{p}'{}_b \right]}\nonumber\\
&{\color{blue}-\kappa\int_{\mathbb{R}^3} \frac{d^3k}{||\vec{k}||^2} \Bigg( -\ft{p}\t{}{_c}(\vec{k},t) \,\ft{p}\t{}{_d}(-\vec{k},t) \left[ 4\delta^{cd}- \hat{k}^c \hat{k}^d \right] + \ft{\epsilon}(\vec{k},t)\,\ft{\epsilon}(-\vec{k},t) \left[\frac{1}{4}+\frac{3}{2}\right]} \nonumber\\
&\hspace{1in}{\color{blue}-\frac{3}{2} m^2\,\ft{\widetilde{V}}(\vec{k},t) \,\ft{\epsilon}(-\vec{k},t)-\ft{\varphi} {}_a^a(\vec{k},t) \,\ft{\epsilon}(-\vec{k},t) \Bigg)}\nonumber\\
    &+ O(\delta^3,\kappa^2)
\end{align}
\endgroup
As expected the physical Hamiltonian differs from $\mathbf{\delta H_{can}}$ only by terms that involve the linearised constraints at least linearly and $\delta \mathbf{H}$ agrees exactly with the expression for the physical Hamiltonian shown in \eqref{eq:PhysHam} in the main text. As discussed in the main text in section \ref{sec:ConstrObs} below \eqref{eq:Choicetausigma} as well as \eqref{eq:linHamredps} the fall-off behaviour of the physical Hamiltonian is suitable for the choices of $\delta\tau$ and $\delta\sigma^c$ in \eqref{eq:Choicetausigma} where we also discuss the fall-off behaviour of the gauge invariant variables for these choices.

\section{Two possible choices for a gauge fixing}

\subsection{Gauge fixing following from the clocks introduced in section \ref{sec:ConstrObs}}
\label{app:GaugeFixing}
In this appendix we present a possible gauge fixing that naturally arises from the choice of the clocks introduced in section \ref{sec:ConstrObs} in the context of constructing Dirac observables. The induced gauge fixing is obtained by setting the gauge fixing conditions ${\cal G}^I$ shown in \eqref{eq:GFCondClocks} in section \ref{sec:ConstrObs} equal to zero. Evaluating the constraints in momentum space, using the basis introduced in appendix \ref{app:Basis}, yields the following seven relations among the elementary phase space variables at every point $(\vec k,t)$:
\begin{align}
||\vec{k}|| \ft{\delta E}^{\overline{1}} &= \ft{\delta A}^{\overline{1}} - \ft{\delta A}^{\overline{2}} & i ||\vec{k}|| (\ft{\delta A}^4 + \ft{\delta A}^5) &= - 2\beta \hat{k}^a \ft{p}\t{}{_a}
\\||\vec{k}|| \ft{\delta E}^{1} &= \ft{\delta A}^{2} - \ft{\delta A}^{1} & i ||\vec{k}|| \ft{\delta A}^{\overline{2}} &=  2\beta m^a(\vec{k}) \ft{p}\t{}{_a}
\\||\vec{k}|| \ft{\delta E}^{3} &= \ft{\delta A}^{5} - \ft{\delta A}^{4} & i ||\vec{k}|| \ft{\delta A}^{2} &= 2\beta \overline{m}^a(\vec{k}) \ft{p}\t{}{_a}\\ ||\vec{k}|| (\ft{\delta A}^4 - \ft{\delta A}^5) &= -\ft{\epsilon}\,,
\end{align}
where $\ft{\epsilon}$ and $\ft{p}{}_a$ denote the three-dimensional Fourier transforms of the energy and momentum density of the scalar field. The gauge fixing conditions, obtained by requiring ${\cal G}^I$ to vanish and in addition choosing $\tau(t,\vec{x})=t, \sigma^a=x^a_\sigma={\rm const}$ and $\xi^j=0$, hence $\delta \tau=\frac{t}{\kappa}$, $\delta\sigma^a = \frac{x_\sigma^a}{\kappa}={\rm const}$, give seven additional conditions in momentum space for every $\vec{k}$ and $t$:
\begin{align}
    \ft{\delta A}^3 &=0 & \ft{\delta E}^2 &= -\ft{\delta E}^1 \\
    \ft{\delta A}^1 &= 0 & \ft{\delta E}^{\overline{2}} &= -\ft{\delta E}^{\overline{1}} \\
    \ft{\delta A}^{\overline{1}} &= 0 & 3 \ft{\delta E}^3 - \ft{\delta E}^4 - \ft{\delta E}^5 &= \frac{1}{||\vec k||^2} \ft{\epsilon} \\
    \ft{\delta E}^5- \ft{\delta E}^4 &=  \frac{2}{||\vec{k}||} \left( \ft{\delta A}^3 + \ft{\delta A}^4 + \ft{\delta A}^5 \right)\,.
\end{align}
Note that $\delta \tau$ or $\delta\sigma$ were chosen to be constant in position and hence vanish in all appearing combinations in the gauge fixing conditions. Substituting these results into \eqref{eq:EinMomSpace} and \eqref{eq:AinMomSpace} yields an expression for $\ft{\delta E}$ and $\ft{\delta A}$ in terms of the physical degrees of freedom:
\begin{align}\label{eq:Eredps}
\ft{\delta E}\t{}{^a_i} =& \ft{\delta E}^{+} m^a m_i + \ft{\delta E}^{-} \overline{m}^a \overline{m}_i + \frac{\ft{\epsilon}}{||\vec{k}||^2}\delta^a_i\nonumber\\ &+ \frac{2\beta i}{||\vec{k}||^2} \ft{p}\t{}{_c} \left[ m^c \left(\hat{k}^a \overline{m}_i - \overline{m}^a \hat{k}_i\right) - \overline{m}^c \left( \hat{k}^a m_i- m^a \hat{k}_i \right) - \hat{k}^c \left(m^a \overline{m}_i - \overline{m}^a m_i \right) \right]  \nonumber\\
=& \ft{\delta E}^{+} m^a m_i + \ft{\delta E}^{-} \overline{m}^a \overline{m}_i + \frac{\ft{\epsilon}}{||\vec{k}||^2}\delta^a_i -\frac{2\beta}{||\vec k||^2} \ft{p}{}_c \t{\epsilon}{^c^a_i}
\end{align}
\begin{align}\label{eq:Aredps}
\ft{\delta A}\t{}{_a^i} =& \ft{\delta A}^{+} m_a m^i + \ft{\delta A}^{-} \overline{m}_a \overline{m}^i - \frac{2\beta i}{||\vec{k}||} \ft{p}\t{}{_c} \left[ \overline{m}^c m_a \hat{k}^i + m^c \overline{m}_a \hat{k}^i - \frac{1}{2} \hat{k}^c \left( m_a \overline{m}^i + \overline{m}_a m^i \right) \right] \nonumber\\&+ \frac{\ft{\epsilon}}{2 ||\vec{k}||} \left(\overline{m}_a m^i - m_a \overline{m}^i \right)\nonumber\\
=& \ft{\delta A}^{+} m_a m^i + \ft{\delta A}^{-} \overline{m}_a \overline{m}^i - \frac{2\beta i}{||\vec{k}||} \ft{p}\t{}{_c} \left[ \hat{k}^i \delta_a^c -\frac{1}{2} \hat{k}^c \delta_a^i  -\frac{1}{2}\hat{k}^c\hat{k}_a\hat{k}^i\right] + \frac{i\ft{\epsilon}}{2||\vec k||} \t{\epsilon}{_b_a^i}\hat{k}^b\,.
\end{align}
Note that one can also obtain setting the constraints equal to zero in the expressions shown in \eqref{eq:obsnvarE} and \eqref{eq:obsnvarA}. The independent physical degrees of freedom are encoded in the matter fields $(\phi,\pi)$ as well as the symmetric transverse-traceless parts of the connection and densitised triads that we denote as
\begin{align}
\ft{\delta \mathcal{A}}\t{}{_a^i}(\vec{k},t) &:= \ft{\delta A}^{+}(\vec{k},t) \, m_a(\vec{k}) \, m^i(\vec{k}) + \ft{\delta A}^{-}(\vec{k},t) \, \overline{m}_a(\vec{k}) \, \overline{m}^i(\vec{k}) = \ft{P}\t{}{_a^i_j^b}(\vec{k}) \ft{\delta A}\t{}{_b^j}(\vec{k},t)\\
    \ft{\delta \mathcal{E}}\t{}{^a_i}(\vec{k},t) &:= \ft{\delta E}^{+}(\vec{k},t) \, m^a(\vec{k}) \, m_i(\vec{k}) + \ft{\delta E}^{-}(\vec{k},t) \, \overline{m}^a(\vec{k}) \, \overline{m}_i(\vec{k}) = \ft{P}\t{}{^a_i^j_b}(\vec{k}) \ft{\delta E}\t{}{^b_j}(\vec{k},t) 
\end{align}
with the projectors $\ft{P}\t{}{^a_i^j_b}(\vec{k})$ and $ \ft{P}\t{}{_a^i_j^b}(\vec{k})$ onto the symmetric transverse-traceless parts that were defined in \eqref{TTprojectors}.
As expected, with the components of $\ft{\delta \mathcal{A}}\t{}{_a^i},\ft{\delta \mathcal{E}}\t{}{^a_i}(\vec{k},t)$ we end up with the six physical components given by
\begin{equation}
\ft{\delta E}^{+}, \ft{\delta E}^{-}, \ft{\delta A}^{+}, \ft{\delta A}^{-}, \phi, \pi\,.
\end{equation}
Without the matter fields, one gets the well-known four phase space field degrees of freedom of gravitational waves in vacuum, corresponding to two field degrees of freedom in the Lagrangian framework, and the real scalar matter field leads to two additional phase space field degrees of freedom. Reinserting \eqref{eq:Eredps} and \eqref{eq:Aredps} into the Hamiltonian \eqref{eq:totalhamlin} yields the total Hamiltonian of the linearised theory on the reduced phase space, \eqref{eq:linHamredps}. For this, we defined the transverse-traceless projectors
\begin{align}\label{transverseprojector}
    [P^\pm(\vec{k})]_a^b &:= \overline{m}_a(\pm\vec{k})\, \overline{m}^b(\pm\vec{k})\,,
\end{align}
such that $[P^\pm(\vec{k})]_a^b \delta_b^i \,  \ft{\delta E}\t{}{^a_i} = \ft{\delta E}^\pm$, $[P^\pm(\vec{k})]_b^a \,  \delta^b_i \ft{\delta A}\t{}{_a^i} = \ft{\delta A}^\pm$ and $\ft{P}\t{}{^a_i^j_b}(\vec{k}) = \sum\limits_{r\in\{\pm\}} [P^r(\vec{k})]^a_i [P^{-r}(\vec{k})]^j_b$.\\

\subsection{A further gauge fixing often used for ADM variables introduced in \cite{arnowitt1962dynamics}}
\label{app:GFADMclocks}
At this point, we would like to make a short comparison to the clocks introduced in \cite{Dittrich:2006ee} based on seminal work in \cite{arnowitt1962dynamics} and \cite{kuchavr1970ground}. Likewise to \cite{Dittrich:2006ee} we denotes this set of clocks as ADM clocks. These clocks, similarly to the ones we use in this work, also do not contain the symmetric transverse-traceless gravitational degrees of freedom $\delta A^\pm$ and $\delta E^\pm$. Hence, these four phase space degrees of freedom are again identified as the physical degrees of freedom in the gravitational sector, as in our case. However, the gauge fixing induced by setting the corresponding gauge fixing conditions $\delta{\cal G}^I=\delta T^I-\tau^I$ used in \cite{Dittrich:2006ee} to zero is different from the one used by in section \ref{sec:ConstrObs} in this work here, hence this set of clocks is not equivalent to the one discussed in section \ref{sec:ConstrObs}. This can be easily seen by considering the clock $^GT^i$ associated with the Gau\ss{} constraint from \cite{Dittrich:2006ee} in momentum space. Requiring that the components $m_i(\vec k) ^G\ft{T}^i(\vec k)$ and $\overline{m}_i(\vec k) ^G\ft{T}^i(\vec k)$ vanish yields $\ft{\delta E}^1 = \ft{\delta E}^{\overline{1}}=0$. 

In contrast the gauge fixing obtained by setting the constraints as well as gauge fixing conditions $\delta{\cal G}^I$ used in section \ref{sec:ConstrObs} equal to zero, we can read off from the transformation introduced in appendix \ref{app:Trafo} that $\ft{\delta E}^1 = -\frac{2i\beta}{||\vec{k}||^2} \overline{m}^a \ft{p}{}_a$ and $\ft{\delta E}^{\overline{1}}= \frac{2i\beta}{||\vec{k}||^2} m^a \ft{p}{}_a$. This gauge fixing condition would be equivalent to the Lorenz-like condition for the Gau\ss{}constraint chosen in this work in section \ref{sec:ConstrObs} in \ref{eq:GaussandHamclock} 
only for the vacuum case, as can be seen in \cite{ashtekar1991gravitons}\footnote{Here only the vacuum case of linearised gravity is considered and the gauge fixing employed is not equivalent to the one used in this work and discussed in appendix \ref{app:GaugeFixing}. In \cite{ashtekar1991gravitons} one condition is that $\delta \t{E}{^a_i}$ is traceless, that is $\delta \t{E}{^a_i} \delta_a^i=0$, which is not satisfied for the gauge fixing discussed in appendix \ref{app:GaugeFixing} where we have $\ft{\delta E}\t{}{^a_i}\delta_a^i = \frac{3\ft{\epsilon}}{||\vec k||^2}$.}. Setting the gauge fixing conditions ${\cal G}^I_{\rm ADM}$ induced by the ADM clocks used in \cite{Dittrich:2006ee} equal to zero imposes the following set of conditions for the phase space variables:
\begin{align}
    \ft{\delta E}^1 = \ft{\delta E}^{\overline{1}} = \ft{\delta E}^2 =\ft{\delta E}^{\overline{2}} = 2\ft{\delta A}^3+\ft{\delta A}^4 +\ft{\delta A}^5 = \ft{\delta E}^5-\ft{\delta E}^4 = \ft{\delta E}^3-\ft{\delta E}^4-\ft{\delta E}^5=0\,,
\end{align}
which implies on the constraint hypersurface that
\begin{align}
    {\ft{\delta E}}\t{}{^a_i} &= {\ft{\delta \mathcal{E}}}\t{}{^a_i} + \frac{\ft{\epsilon}}{2||\vec{k}||^2} (\delta^a_i + \hat{k}^a\hat{k}_i)\\
    {\ft{\delta A}}\t{}{_a^i} &= {\ft{\delta \mathcal{A}}}\t{}{_a^i}-\frac{2i\beta}{||\vec k||} \ft{p}{}_b \left( \hat{k}^i \delta^b_a + \hat{k}_a \delta^{ib} -\frac{1}{2} \hat{k}^b \delta_a^i +\hat{k}_a\hat{k}^i \hat{k}^b\right) +\frac{i\ft{\epsilon}}{2||\vec k||} \hat{k}^b \t{\epsilon}{_b_a^i}
\end{align}
In this particular gauge, the physical Hamiltonian becomes
\begingroup
\allowdisplaybreaks
\begin{align}
\delta \mathbf{H} =
    &\int_{\mathbb{R}^3} d^3x \;\epsilon(\vec{x},t) \nonumber\\
&+\kappa \int_{\mathbb{R}^3}d^3k\; \frac{1}{2\beta^2}\, \sum_{r\in\{\pm\}}\Big( \ft{\delta A}^r(\vec{k},t) \,\ft{\delta A}^r (-\vec{k},t)+  2r\,||\vec{k}|| \ft{\delta E}^r(\vec{k},t) \, \ft{\delta A}^r(-\vec{k},t) \nonumber\\ &\hspace{1.8in}+ (\beta^2+1) ||\vec{k}||^2  \ft{\delta E}^r(\vec{k},t)\,  \ft{\delta E}^r(-\vec{k},t)\Big) \nonumber\\
&+\kappa \int_{\mathbb{R}^3} d^3k \;\delta_{b}^i \ft{\varphi} {}_a^b(\vec{k},t) \, \ft{\delta \mathcal{E}}\t{}{^a_i}(-\vec{k},t)\nonumber\\ &-\kappa\int_{\mathbb{R}^3} \frac{d^3k}{||\vec{k}||^2} \Bigg( -\ft{p}\t{}{_c}(\vec{k},t) \,\ft{p}\t{}{_d}(-\vec{k},t) \left[ 4\delta^{cd}- 3\hat{k}^c \hat{k}^d \right] + \ft{\epsilon}(\vec{k},t)\,\ft{\epsilon}(-\vec{k},t) \left[\frac{1}{4}+1\right] \nonumber\\
&\hspace{1in}- m^2\,\ft{\widetilde{V}}(\vec{k},t) \,\ft{\epsilon}(-\vec{k},t)-\ft{\varphi} {}_a^b(\vec{k},t) \,\ft{\epsilon}(-\vec{k},t) \left[ \frac{1}{2}\delta^a_b +\frac{1}{2}\hat{k}^a\hat{k}_b \right] \Bigg)\nonumber\\
    &+ O(\delta^3,\kappa^2)\,.\label{eq:gfhamditadm}
\end{align}
\endgroup
As can be readily seen, it differs from the Hamiltonian in the gauge used in this work in \eqref{eq:deltaHNewVar} only in the term encoding self-interaction of the scalar field. The reason is that by choosing the clocks, different choices for physical temporal and spatial coordinates were established and these are determined by the choice of different sets of clocks that on the gauge fixed surface are equal to $\delta\tau$ and $\delta\sigma^a$ respectively. As discussed in section \ref{sec:ConstrObs} by an abuse of notation we understand the $\vec x$ and $t$ arguments of the fields to be the associated values of the diffeomorphism and Hamiltonian clock, $\delta \tau$ and $\delta \sigma^a$ respectively. By inspection of the clocks, it turns out that the ADM clocks used in this section, corresponding to the parameters $x^a_{ADM}$ and $t_{ADM}$ are related to the parameters used in this works, $t$ and $x^a$, by
\begin{align}
    t_{ADM} &= -\frac{1}{2}t +\frac{1}{2}\partial^a (p_a  * G^{\Delta\Delta})\\
    x^a_{ADM} &= \frac{1}{2} x^a+ \partial^a (\epsilon * G^{\Delta\Delta})\,,
\end{align}
where we neglected factors of $\kappa$ and $\beta$, as the notation regarding these factors  in \cite{Dittrich:2006ee} partly differs from the one used in this work. 

We realise that in the case of vacuum gravity the choices for physical temporal and spatial coordinates agree but differ for non-vanishing momentum and energy density of the scalar field 
~\\
\\
Compared to the work in \cite{anastopoulos2013master}, where for the ADM constraints the same gauge fixing as in this section was used, their gauge fixed Hamiltonian appears with different prefactors in the self-interaction part compared to the one in \eqref{eq:gfhamditadm}. A reason might be, that we were not able to reproduce the expression for $V$ in their equation (6) in \cite{anastopoulos2013master} which consists of the second order terms of the expansion of $\sqrt{q_{ab}}\: ^{(3)}R$, where $q_{ab}$ denotes the spatial ADM metric and $^{(3)}R$ the three-dimensional Ricci scalar. However, in contrast the result for this expansion given by the same authors in \cite{lagouvardos2021gravitational} could be reproduced by our computations. 

\section{Local invertibility of $\big(\mathds{1}-\Sigma(t,t_0) \big)$ in the TCL master equation} \label{sec:sigma_inv}

An essential ingredient in the TCL master equation was the assumption that the superoperator $\big(\mathds{1}-\Sigma(t,t_0) \big)$ was invertible in a suitable neighbourhood of $t_0$ and we used that the inverse can be written in terms of a (perturbative) geometric series akin to

\begin{align}
    \big( \mathds{1} - \Sigma(t,t_0) \big)^{-1} = \sum\limits_{n=0}^\infty \big( \Sigma(t,t_0)\big)^n.
\end{align}

In order to represent the inverse in this way we need to ensure that the series actually converges which we will prove in this part of the appendix. The first thing to notice is the fact that $\Sigma(t_0,t_0) = 0$, which directly follows from the very definition (\ref{eq:Sigmat}). Secondly, all constituents of $\Sigma(t,t_0)$ are continuous in $t$, which renders $\Sigma(t,t_0)$ itself continuous. Consequently, there exists an interval $[t_0, t]$ with $t > t_0$ such that $\big(\mathds{1}-\Sigma(t,t_0) \big)$ is invertible on the entire interval. We shall denote with $\Sigma_R(t,t_0) \in \mathcal{B}(\mathcal{B}(\mathcal{H}))$ the restriction to precisely that neighbourhood, while the latter indicates that it is a bounded superoperator on the base Hilbert space. Hence, we can immediately define a norm of $\Sigma_R(t,t_0)$ which is essentially inherited from the Hilbert space:

\begin{align*}
    \lVert \Sigma_R(t,t_0) \rVert_{\mathcal{B}(\mathcal{B}(\mathcal{H}))} := \sup_{A \in \mathcal{B}(\mathcal{H})} \lVert \Sigma_R(t,t_0) A \rVert_{\mathcal{B}(\mathcal{H})}, \quad  \lVert A \rVert_{\mathcal{B}(\mathcal{H})} := \sup_{\Psi \in \mathcal{H}} \lVert A \Psi \rVert_{\mathcal{H}} = 1, \quad \lVert \Psi \rVert_{\mathcal{H}} =1.
\end{align*}

By virtue of the definition of the restriction and the fact that $( \mathds{1} - \Sigma(t,t_0))^{-1}$ is invertible for all times in $[t_0, t]$ we have $\lVert \Sigma_R(t,t_0) \rVert := \lambda < 1$, where we omit the explicit $\mathcal{B}(\mathcal{B}(\mathcal{H}))$-label since it follows from context. The time interval for which $\big(\mathds{1}-\Sigma(t,t_0) \big)$ is invertible depends on the coupling strength of the interaction, which is immediately evident from the definition (\ref{eq:Sigmat}), which linearly depends on $\alpha$ in the lowest orders of the expansion of the involved propagators. Heuristically speaking this means that the time interval $[t_0, t]$ is longer the smaller the coupling constant is. In order to prove this, we need to establish the fact that the inversion can be indeed written in the form of a geometric series. Firstly, the series is absolutely convergent:

\begin{align*}
    \lVert \sum\limits_{n=0}^\infty \big( \Sigma_R(t,t_0)\big)^n \rVert \leq \sum\limits_{n=0}^\infty \lVert \big( \Sigma_R(t,t_0)\big)^n \rVert \leq \sum\limits_{n=0}^\infty \lVert \Sigma_R(t,t_0) \rVert^n = \sum\limits_{n=0}^\infty \lambda^n < \infty.
\end{align*}

Since $\mathcal{B}(\mathcal{B}(\mathcal{H})$ is a Banach space by construction and hence complete with respect to the induced norm, the series $\sum\limits_{n} \big( \Sigma_R(t,t_0)\big)^n$ converges to an operator in $\mathcal{B}(\mathcal{B}(\mathcal{H}))$. In full analogy to the geometric series from standard calculus, let us introduce the following expression:

\begin{align}
\label{eq:DefRN}
    \mathfrak{R}(N) := \big( \Sigma_R(t,t_0)\big)^N \big(\mathds{1} - \Sigma_R(t,t_0) \big)^{-1}.
\end{align}
With the help of this auxiliary quantity, we then establish a telescoping series:

\begin{align*}
    \big( \Sigma_R(t,t_0)\big)^M = \big( \Sigma_R(t,t_0)\big)^M \big( \mathds{1} - \Sigma_R(t,t_0) \big) \big(\mathds{1}-\Sigma_R(t,t_0) \big)^{-1} = \mathfrak{R}(M) - \mathfrak{R}(M+1).
\end{align*}

Consequently, the finite sum can be written in terms of two contributions, that is

\begin{align*}
    \sum\limits_{n=0}^N \big( \Sigma_R(t,t_0)\big)^n = \mathfrak{R}(0) - \mathfrak{R}(N+1),
\end{align*}

where now it is  possible to take the limit of $N \to \infty$ and complete the proof:

\begin{align*}
    \lim_{N \to \infty} \sum\limits_{n=0}^N \big( \Sigma_R(t,t_0)\big)^n = \sum\limits_{n=0}^\infty \big( \Sigma_R(t,t_0)\big)^n = \mathfrak{R}(0) - \lim_{N \to \infty} \mathfrak{R}(N+1) = \mathfrak{R}(0) = \big(\mathds{1}-\Sigma_R(t,t_0) \big)^{-1},
\end{align*}

where we used the definition of $\mathfrak{R}(N)$ in \eqref{eq:DefRN} in the last step and moreover made use of the previous estimate:

\begin{equation*}
    \lim_{N \to \infty} \lVert \mathfrak{R}(N+1) \rVert \leq \lim_{N \to \infty} \lVert \big( \Sigma_R(t,t_0)\big)^N \rVert \; \lVert \big(\mathds{1}-\Sigma_R(t,t_0) \big)^{-1} \rVert = \lim_{N \to \infty} \lambda^{N+1} \; \lVert \big(\mathds{1}-\Sigma_R(t,t_0) \big)^{-1} \rVert = 0,
\end{equation*}

since $\lambda < 1$ and $\lVert \big(1-\Sigma_R(t,t_0) \big)^{-1} \rVert$ is finite on the interval $[t_0, t]$ as we have shown already above.

\section{Evaluation of the correlation functions}\label{AppendixCorrelationFunctions}
In this appendix we will derive the correlation functions and their spectral representation, working with a regularisation established by formulating the theory in a box of finite volume introduced in equation \eqref{BoxIntro}. Given that the two number operators $n_k^+ := (b_k^+)^\dagger b_k^+$ and $n_k^- := (b_k^-)^\dagger b_k^-$ mutually commute, as they live on different Hilbert spaces, the density matrix of the environment can be written in terms of the following tensor product:
\begin{equation}\label{9:rhosplit}
\rho_\mathcal{E} = \rho_{\mathcal{E}+} \otimes \rho_{\mathcal{E}-},
\end{equation} 
where 
\begin{equation}
\rho_{\mathcal{E}r} = \frac{1}{Z_{\mathcal{E}r}}\exp\left\{-\beta \sum\limits_{\vec{k}\in \mathds{K}} {\Omega}_k n_k^r\right\},
\end{equation} 
with $r\in\{+,-\}$, $Z_{\mathcal{E}r} := tr_{\mathcal{E}r}\left\{\rho_{\mathcal{E}r}\right\}$ and $tr_{\mathcal{E}r}$ denoting the partial trace over the $r$-part of the environmental Hilbert space $\mathcal{H}_\mathcal{E} = \mathcal{H}_{\mathcal{E}+} \otimes \mathcal{H}_{\mathcal{E}-}$. 

At this point we introduce an alternative notation for the summation over the discrete $\vec{k}$-vectors. Instead of summing over all elements of the set $\mathds{K}$ that contains the discrete, permitted $\vec{k}$-vectors, we want to sum over an index running over the natural numbers. Such a bijection exists, because each possible $\vec{k}$ consists of three components $k_x,k_y,k_z \in \frac{\pi\mathbb{Z}}{L}$ with $L$ denoting the length of the box. Thus, we can identify an element $\vec{k}\in\mathds{K}$ uniquely by providing three numbers. A bijection between $\mathbb{Z}^3$ and $\mathbb{N}$ can be constructed  therefore we can sum over $j\in \mathbb{N}$ instead of $\vec{k}\in\mathds{K}$.\\
We use this notation and rewrite the number operator in terms of the occupation number basis of the corresponding part of the Hilbert space:
\begin{equation}
n_i^r = \sum\limits_{n_1^r=0}^{\infty} \sum\limits_{n_2^r=0}^{\infty} ... \sum\limits_{n_i^r=1}^{\infty} ... \;  n_i^r \ket{n_1^r,n_2^r,...,n_i^r,...}\bra{n_1^r,n_2^r,...,n_i^r,...}\,,
\end{equation}
where the individual $n_i^r \in \mathbb{N}_0$ on the right hand side are the eigenvalues of the occupation number operator and $\ket{n_1^r,n_2^r,...,n_i^r,...}$ the corresponding eigenstates.
Due to the orthonormality of different Fock states $\ket{n_1^r, n_2^r,...}$ we obtain
\begin{equation}
\exp\left\{-\beta \sum\limits_{j\in \mathbb{N}} {\Omega}_j n_j^r\right\}= \sum\limits_{n_1^r=0}^{\infty} \sum\limits_{n_2^r=0}^{\infty} ...  \prod_{i\in\mathbb{N}} e^{-\beta \, n_i^r {\Omega}_i} \ket{n_1^r,n_2^r,...}\bra{n_1^r,n_1^r,...}.
\end{equation}
Due to the regularisation established by considering the system to be in a finite box, we can express the partial trace in terms of Fock states and find:
\begin{equation}
tr_{\mathcal{E}r}\left\{.\right\} = \sum\limits_{n_1^r=0}^{\infty} \sum\limits_{n_2^r=0}^{\infty} ...  \bra{n_1^r,n_2^r,...} . \ket{n_1^r,n_2^r,...}\,. 
\end{equation}
We can use this expression to evaluate the partition sum of the Gibbs state:
\begin{align}\label{9:psr}
Z_{\mathcal{E}r} &= tr_{\mathcal{E}r} \left(\exp\left\{-\beta \sum\limits_{j\in \mathbb{N}} {\Omega}_j n_j^r\right\} \right) =\sum\limits_{n_1^r=0}^{\infty} \sum\limits_{n_2^r=0}^{\infty} ...  \prod_{j\in\mathbb{N}} e^{-\beta \, n_j^r {\Omega}_j} = \sum\limits_{n_1^r=0}^{\infty} \sum\limits_{n_2^r=0}^{\infty} ...  \prod_{j\in\mathbb{N}} \left[e^{-\beta \,{\Omega}_j}\right]^{ n_j^r}\nonumber\\
&= \left(\sum\limits_{n_1^r=0}^\infty\left[e^{-\beta \,{\Omega}_1}\right]^{ n_1^r}\right) \cdot \left(\sum\limits_{n_2^r=0}^\infty\left[e^{-\beta \,{\Omega}_2}\right]^{ n_2^r}\right) \cdot ...= \prod_{i\in\mathbb{N}} \left(\sum\limits_{n_i^r=0}^\infty\left[e^{-\beta \,{\Omega}_i}\right]^{ n_i^r}\right)= \prod_{i\in\mathbb{N}} \frac{1}{1-e^{-\beta\, \Omega_i}} \,.
\end{align}
In the last step we assumed that $\vec{k}=0$, which corresponds to $\Omega_k=0$, is not contained in the set $\mathds{K}$, which is the usual infrared divergence present in quantum field theory. Then $e^{-\beta \,\Omega_i}<1$ always holds and the expression is  a geometric series.\\
In actual applications it is important that the partition sum is finite. This is indeed the case, a proof can be found for instance in \cite{fahn2020masterthesis}. The final partition sum for part $r$ of the environment in \eqref{9:psr} is independent of this label $r$, so we get the same result for $r=+$ and $r=-$ respectively. From \eqref{9:rhosplit} one can then see that the partition sum of the total environmental Gibbs state is thus just given by the square of \eqref{9:psr}:
\begin{equation}
Z_{\mathcal{E}} = \left[ \prod_{j\in\mathbb{N}} \frac{1}{1-e^{-\beta\, \Omega_j}} \right]^2\,.
\end{equation}
Now with this result the thermal Wightman functions can be computed explicitly:
\begin{align}
\label{eq:WightmanFctns}
    G^>&\t{}{^a_i^b_j}(\vec{x}-\vec{y},t-s) = <\delta \mathcal{E}\t{}{^a_i}(\vec{x},t) \delta \mathcal{E}\t{}{^b_j}(\vec{y},s)> =  tr_\mathcal{E} \left\{ \delta \mathcal{E}\t{}{^a_i}(\vec{x},t) \delta \mathcal{E}\t{}{^b_j}(\vec{y},s) \rho_\mathcal{E} \right\} \nonumber\\
    &= \int \frac{d^3k d^3p}{2(2\pi)^3\sqrt{\Omega_k \Omega_p}} \sum\limits_{r,u\in\{\pm\}} [P^{-r}(\vec{k})]^a_i [P^{-u}(\vec{p})]^b_j \nonumber\\&\hspace{1in} \cdot tr_\mathcal{E}\left\{ \left[ b_k^r e^{-i\Omega_k t +i\vec{k}\vec{x}} + (b_{-k}^r)^\dagger e^{i\Omega_k t +i\vec{k}\vec{x}} \right] \left[ b_p^u e^{-i\Omega_p s +i\vec{p}\vec{y}} + (b_{-p}^u)^\dagger e^{i\Omega_p s +i\vec{p}\vec{y}} \right] \rho_\mathcal{E}\right\}\,,
\end{align}
where $[P^{-r}(\vec{k})]^a_i := [P^{-r}(\vec{k})]^a_b \delta^b_i$.
Using the explicit expressions given above for the Gibbs state and the trace in the occupation number basis, we obtain the following results:
\begin{align}
    tr_\mathcal{E}\left\{ b_k^r b_p^u \rho_\mathcal{E}\right\} &= 0\\
    tr_\mathcal{E}\left\{ b_k^r (b_{-p}^u)^\dagger \rho_\mathcal{E}\right\} &= \delta^{ru} \delta(\vec{k}+\vec{p}) tr_\mathcal{E}\left\{ b_k^r (b_k^r)^\dagger \rho_\mathcal{E}\right\} = \delta^{ru} \delta(\vec{k}+\vec{p}) tr_\mathcal{E}\left\{ [ n_k^r+1] \rho_\mathcal{E}\right\} \\
    tr_\mathcal{E}\left\{ (b_{-k}^r)^\dagger b_p^u \rho_\mathcal{E}\right\} &= \delta^{ru} \delta(\vec{k}+\vec{p}) tr_\mathcal{E}\left\{ (b_k^r)^\dagger b_k^r \rho_\mathcal{E}\right\} = \delta^{ru} \delta(\vec{k}+\vec{p}) tr_\mathcal{E}\left\{ n_k^r \rho_\mathcal{E}\right\} \\
    tr_\mathcal{E}\left\{ (b_{-k}^r)^\dagger (b_{-p}^u)^\dagger \rho_\mathcal{E}\right\} &= 0\,.
\end{align}
Reinserting this back into  the Wightman functions in  \eqref{eq:WightmanFctns} they simplify to 
\begin{align}
    G^>\t{}{^a_i^b_j}(\vec{x}-\vec{y},t-s)= &\int \frac{d^3k}{2(2\pi)^3 \Omega_k} \sum\limits_{r\in\{\pm\}} [P^{-r}(\vec{k})]^a_i [P^{-r}(-\vec{k})]^b_j \nonumber\\&\times \left[ (tr_\mathcal{E}\left\{ n_k^r \rho_\mathcal{E}\right\}+1) e^{-i\Omega_k (t-s) +i\vec{k}(\vec{x}-\vec{y})} + tr_\mathcal{E}\left\{ n_k^r \rho_\mathcal{E}\right\}e^{-i\Omega_k (s-t) +i\vec{k}(\vec{y}-\vec{x})} \right]\,.
\end{align}
Now the remaining trace can  be computed with respect to the occupation number basis and yields the expected Bose-Einstein distribution:
\begingroup
\allowdisplaybreaks
\begin{align}
 N(\Omega_k):=tr_\mathcal{E} \left\{ n_{k}^r \, \rho_{\mathcal{E}} \right\} &= \frac{1}{Z_{\mathcal{E}r}} \sum\limits_{\substack{n_1^r=0\\\tilde{n}_1^r=0}}^{\infty} ... \sum\limits_{\substack{n_K^r=0\\\tilde{n}_K^r=0}}^{\infty} \bra{n_1^r,...,n_K^r} n_k^r e^{-\beta \sum\limits_{j=1}^K \tilde{n}_j^r {\Omega}_j} \ket{\tilde{n}_1^r,...,\tilde{n}_K^r}\braket{\tilde{n}_1^r,...,\tilde{n}_K^r|n_1^r,...,n_K^r}\nonumber\\
&= \frac{1}{Z_{\mathcal{E}r}} \sum\limits_{n_1^r=0}^{\infty} ... \sum\limits_{n_K^r=0}^{\infty} \bra{n_1^r,...,n_K^r} n_k^r e^{-\beta \sum\limits_{j=1}^K n_j^r {\Omega}_j} \ket{n_1^r,...,n_K^r}\nonumber\\
&=\frac{1}{-\beta}\frac{1}{Z_{\mathcal{E}r}} \frac{\partial}{\partial \Omega_k}  \sum\limits_{n_1^r=0}^{\infty} ... \sum\limits_{n_K^r=0}^{\infty} \bra{n_1^r,...,n_K^r} e^{-\beta \sum\limits_{j=1}^K n_j^r {\Omega}_j} \ket{n_1^r,...,n_K^r}\nonumber\\
&= \frac{1}{-\beta Z_{\mathcal{E}r}} \frac{\partial Z_{\mathcal{E}r}}{\partial \Omega_k} =\frac{1}{-\beta} \frac{\partial}{\partial \Omega_k} \ln(Z_{\mathcal{E}r}) = \frac{1}{-\beta}\frac{\partial}{\partial \Omega_k} \left[ -\sum\limits_{j=1}^K \ln\left( 1-e^{-\beta \Omega_j} \right)\right] \nonumber\\
&= \frac{e^{-\beta \Omega_k}}{1-e^{-\beta \Omega_k}} =\frac{1}{e^{\beta \Omega_k}-1} \,.
\end{align}
\endgroup
This result is independent of the polarisation, hence we can evaluate the sum over polarisations separately and obtain the transverse-traceless projector
\begin{equation}
    \sum\limits_{r\in\{\pm\}} [P^{-r}(\vec{k})]^a_i [P^{-r}(-\vec{k})]^b_j = \ft{P}\t{}{^a_i^b_j}(\vec{k})\,.
\end{equation}
Hence we can express the Wightman function in the following manner:
\begin{equation}
    G^>\t{}{^a_i^b_j}(\vec{x}-\vec{y},t-s) =: \t{P}{^a_i^b_j} G^>(\vec{x}-\vec{y},t-s)
\end{equation}
with
\begin{align}
    G^>(\vec{x}-\vec{y},t-s) &= \int \frac{d^3k}{2(2\pi)^{\frac{3}{2}} \Omega_k} \left[ (N(\Omega_k)+1) e^{-i\Omega_k (t-s) +i\vec{k}(\vec{x}-\vec{y})} + N(\Omega_k) \, e^{-i\Omega_k (s-t) +i\vec{k}(\vec{y}-\vec{x})} \right] \nonumber\\
    &=\int \frac{d^4k}{2(2\pi)^{\frac{3}{2}} \Omega_k} \left[ (N(\Omega_k)+1)  \delta(k^0-\Omega_k)+ N(\Omega_k)  \delta(k^0+\Omega_k) \right] e^{-i k^0 (t-s) +i\vec{k}(\vec{x}-\vec{y})} \nonumber\\
    &=\int \frac{d^4k}{2(2\pi)^{\frac{3}{2}} \Omega_k} [N(k^0)+1] \left[ \delta(k^0-\Omega_k)+  \delta(k^0+\Omega_k) \right] e^{-i k^0 (t-s) +i\vec{k}(\vec{x}-\vec{y})}\,,
\end{align}
where we used that $N(-\Omega_k) = -[N(\Omega_k)+1]$. Defining 
\begin{equation*}
 \rho^>(k^0,||\vec{k}||) = [1+N(k^0)] \rho(k^0,||\vec{k}||)\quad{\rm and}\quad  
 \rho(k^0,||\vec{k}||) = \sqrt{\frac{\pi}{2}}\frac{1}{\Omega_k} \left[ \delta(k^0-\Omega_k)- \delta(k^0+\Omega_k) \right]\, ,
\end{equation*}
we can write down the Wightman function in its spectral decomposition:
\begin{equation}
     G^>(\vec{x}-\vec{y},t-s) = \int \frac{d^4k}{(2\pi)^{\frac{4}{2}} } \rho^>(k^0,||\vec{k}||) e^{-i k^0 (t-s) +i\vec{k}(\vec{x}-\vec{y})}\,.
\end{equation}
For $G^<(\vec{x}-\vec{y},t-s)$, defined as
\begin{equation}
    G^<\t{}{^a_i^b_j}(\vec{x}-\vec{y},t-s) =: \t{P}{^a_i^b_j} G^<(\vec{x}-\vec{y},t-s)
\end{equation}
a similar derivation yields
\begin{equation}
     G^<(\vec{x}-\vec{y},t-s) = \int \frac{d^4k}{(2\pi)^{\frac{4}{2}} } \rho^<(k^0,||\vec{k}||) e^{-i k^0 (t-s) +i\vec{k}(\vec{x}-\vec{y})}
\end{equation}
with 
\begin{equation*}
\rho^<(k^0,||\vec{k}||) = N(k^0) \rho(k^0,||\vec{k}||)\, .    
\end{equation*}

From the equalities $G^<(\vec{x}-\vec{y},t-s) = G^>(\vec{y}-\vec{x},s-t)$ and $\t{P}{^b_j^a_i}=\t{P}{^a_i^b_j}$, one can directly read off $G^<(\vec{x}-\vec{y},t-s)$ and construct the thermal Green's function
\begin{align}
    G^{(F)}\t{}{^a_i^b_j}(\vec{x}-\vec{y},t-s) :=& \t{P}{^a_i^b_j}G^>(\vec{x}-\vec{y},t-s) \theta(t-s) + \t{P}{^a_i^b_j}G^<(\vec{x}-\vec{y},t-s) \theta(t-s)\nonumber\\
    =& \t{P}{^a_i^b_j} \left[\theta(t-s )\int \frac{d^3k}{2(2\pi)^{\frac{3}{2}} \Omega_k}  e^{-i\Omega_k (t-s) +i\vec{k}(\vec{x}-\vec{y})} +  \theta(s-t)\int \frac{d^3k}{2(2\pi)^{\frac{3}{2}} \Omega_k}  e^{-i\Omega_k (s-t) +i\vec{k}(\vec{y}-\vec{x})}\right] \nonumber\\ +& \t{P}{^a_i^b_j}\int \frac{d^3k}{(2\pi)^{\frac{3}{2}} \Omega_k} N(\Omega_k) \cos[(\Omega_k (t-s) -\vec{k}(\vec{x}-\vec{y})]\nonumber\\
    =& G^{(F)}_{\Theta=0}\t{}{^a_i^b_j}(\vec{x}-\vec{y},t-s) + \t{P}{^a_i^b_j}\int \frac{d^3k}{(2\pi)^{\frac{3}{2}} \Omega_k} N(\Omega_k) \cos[(\Omega_k (t-s) -\vec{k}(\vec{x}-\vec{y})]
\end{align}
with the Heaviside function $\theta(\tau)$. The first part is the usual Green's function one obtains for a temperature parameter $\Theta$ that vanishes and a second thermal contribution present in the finite temperature case. In case of vanishing temperature parameter $\Theta=0$ , we obtain $N(\Omega_k)=0$ and find the  Green's function of the zero temperature case that takes the following form:
\begin{align}
    G^{(F)}_{\Theta=0}&\t{}{^a_i^b_j}(\vec{x}-\vec{y},t-s)\nonumber\\ &= \t{P}{^a_i^b_j} \left[\theta(t-s )\int \frac{d^3k}{2(2\pi)^{\frac{3}{2}} \Omega_k}  e^{-i\Omega_k (t-s) +i\vec{k}(\vec{x}-\vec{y})} +  \theta(s-t)\int \frac{d^3k}{2(2\pi)^{\frac{3}{2}} \Omega_k}  e^{-i\Omega_k (s-t) +i\vec{k}(\vec{y}-\vec{x})}\right] \nonumber\\
    &=\t{P}{^a_i^b_j} \int_{\mathbb{R}^4} \frac{d^4k}{(2\pi)^{\frac{4}{2}}} \frac{i}{k^2+i\epsilon} e^{-ik^0(t-s)+i\vec{k}(\vec{x}-\vec{y})} \nonumber\\
    &= \t{P}{^a_i^c_l} \t{P}{^b_j^e_m} \delta^{dl} \delta^{fm} \left[ \delta_{ce} \delta_{df} + \delta_{cf} \delta_{de} - \delta_{cd} \delta_{ef} \right] \frac{1}{2}\int_{\mathbb{R}^4} \frac{d^4k}{(2\pi)^{\frac{4}{2}}} \frac{i}{k^2+i\epsilon} e^{-ik^0(t-s)+i\vec{k}(\vec{x}-\vec{y})}\,.
\end{align}
Thus, the results is just given by the transverse-traceless projection of the spatial part of the graviton propagator in harmonic gauge, see for instance \cite{donoghue2017epfl} for the explicit form of the graviton propagator. Note that we chose the integration contour in the second step such that we obtain the Feynman propagator.
 After a four dimensional Fourier transformation of the entire thermal Green's function and again choosing the Feynman prescription we obtain:
\begin{align}
    \ft{G}^{(F)}\t{}{^a_i^b_j}(p) = \ft{P}\t{}{^a_i^b_j}(\vec{p}) \Bigg\{ \frac{i}{p^\mu p_\mu+i\epsilon} + 2\pi N(\Omega_p) \delta(p^\mu p_\mu) \Bigg\}\,,
\end{align}
which indeed has the expected form, see for instance \cite{koksma2011decoherence}.

\end{appendices}
\bibliographystyle{unsrt}
\bibliography{GravDec.bib}
\end{document}